\pdfoutput=1
\documentclass[11pt,twoside,a4paper,cmspaper,final,collab]{cms-tdr}
\def\svnVersion{4255406}\def\svnDate{2023/07/18}\def\cmsCernNoTag{CERN-EP-2022-053}\def\cmsCernDate{\today}\def\cmsMessage{Published in the European Physical Journal C as \href{http://dx.doi.org/10.1140/epjc/s10052-023-11631-7}{\doi{10.1140/epjc/s10052-023-11631-7}.}}
\begin{document}\cmsNoteHeader{SMP-20-003}

\newcommand{\artemide}{\textsc{arTeMiDe}\xspace}
\newcommand{\MGaMC}{\textsc{mg5}\_a\textsc{mc}\xspace}
\newcommand{\GE}{\textsc{Geneva}\xspace}
\newcommand{\cascade}{\textsc{Cascade}\xspace}

\newcommand{\MGPE}{\MGaMC + \PYTHIA8\xspace}
\newcommand{\MGCAS}{\MGaMC + PB(\cascade)\xspace}
\newcommand{\MGCASzero}{\MGaMC (0 jet at NLO)+ PB (\cascade)\xspace}
\newcommand{\MGCASone}{\MGaMC (1 jet at NLO)+ PB (\cascade)\xspace}

\newcommand{\NqLL}{\ensuremath{\mathrm{N^3LL}}\xspace}
\newcommand{\minnlo}{\textsc{MiNNLO$_\mathrm{PS}$}\xspace}
\newcommand{\phistar}{\ensuremath{\varphi^{*}_{\eta}}\xspace}
\newcommand{\Mll}{\ensuremath{m_{\Pell\Pell}}\xspace}
\newcommand{\PTll}{\ensuremath{\pt(\Pell\Pell)}\xspace}
\newcommand{\qt}{\ensuremath{q_\mathrm{T}}\xspace}
\newcommand{\mrangea}{\ensuremath{50< m_{\Pell\Pell} <76\GeV}\xspace}
\newcommand{\mrangeb}{\ensuremath{76< m_{\Pell\Pell} <106\GeV}\xspace}
\newcommand{\mrangec}{\ensuremath{106< m_{\Pell\Pell} <170\GeV}\xspace}
\newcommand{\mranged}{\ensuremath{170< m_{\Pell\Pell} <350\GeV}\xspace}
\newcommand{\mrangee}{\ensuremath{350< m_{\Pell\Pell} <1000\GeV}\xspace}
\newenvironment{tolerant}[1]{\par\tolerance=#1\relax}{ \par }
\newcommand{\ggll}{\ensuremath{\gamma \gamma \to \Pell^+ \Pell^-}}
\cmsNoteHeader{SMP-20-003}

\title{Measurement of the mass dependence of the transverse momentum of lepton pairs in Drell--Yan production in proton-proton collisions at \texorpdfstring{$\sqrt{s} = 13\TeV$}{sqrt(s) = 13 TeV}}
\titlerunning{Mass dependence of the transverse momentum of Drell--Yan pairs at 13\TeV}
\date{\today}
\abstract{The double differential cross sections of the Drell--Yan lepton pair ($\ell^+\ell^-$, dielectron or dimuon) production are measured as functions of the invariant mass \Mll, transverse momentum \PTll, and \phistar. The \phistar observable, derived from angular measurements of the leptons and highly correlated with \PTll, is used to probe the low-\PTll region in a complementary way. Dilepton masses up to 1\TeV are investigated. Additionally, a measurement is performed requiring at least one jet in the final state. To benefit from partial cancellation of the systematic uncertainty, the ratios of the differential cross sections for various \Mll ranges to those in the \PZ mass peak interval are presented. The collected data correspond to an integrated luminosity of 36.3\fbinv of proton--proton collisions recorded with the CMS detector at the LHC at a centre-of-mass energy of 13\TeV. Measurements are compared with predictions based on perturbative quantum chromodynamics, including soft-gluon resummation. }

\hypersetup{
pdfauthor={CMS Collaboration},
pdftitle={Measurement of the mass dependence of the transverse momentum of lepton pairs in Drell-Yan production in proton-proton collisions at sqrt(s) = 13 TeV},
pdfsubject={CMS},
pdfkeywords={CMS, SMP, Drell-Yan, cross section, transverse-momentum dependence}
}

\maketitle 

\section{Introduction}

The Drell--Yan (DY) production of charged-lepton pairs in hadronic collisions~\cite{Drell:1970wh}
provides important insights into the partonic structure of hadrons and the evolution of the parton distribution functions (PDFs).
At leading order (LO) in perturbative quantum chromodynamics (pQCD), the DY process is described in terms of an 
$s$-channel $Z/\gamma^*$ exchange process convolved with collinear quark and antiquark parton distribution functions of the proton. At 
LO, the lepton pair transverse momentum \PTll, corresponding to the exchanged boson transverse
momentum, is equal to zero. At higher orders, initial-state QCD radiation gives rise to a sizable \PTll.
Whereas the spectrum for large \PTll values is expected to be described through fixed-order
calculations in pQCD, at small values ($\pt < \mathcal{O}(\Mll)$), where \Mll is the invariant mass of the lepton pair, soft-gluon resummation
to all orders is required~\cite{Dokshitzer:1978yd,Collins:1984kg}.
In addition, the low-\PTll region also includes the effects of the intrinsic transverse motion of the partons in the
colliding hadrons that has to be extracted from data and parameterized.
The resummation functions are universal and obey renormalisation group equations, predicting a simple scale dependence in the leading logarithmic approximation, where the scale is given by \Mll. 
Therefore, measuring the \PTll spectrum in a wide \Mll range tests the validity of 
the resummation approach and the precision of different predictions. 
Calculations for inclusive DY production as a function of \Mll and \PTll 
are available up to next-to-next-to-leading order (NNLO) in pQCD~\cite{Hamberg:1990np,Catani:2009sm,Catani:2007vq,Melnikov:2006kv}. 
Soft-gluon resummation can be computed analytically, 
either in transverse momentum dependent parton distributions (TMD)
or in parton showers of Monte Carlo (MC) 
event generators matched with matrix element
calculations~\cite{Bacchetta:2017gcc,Bacchetta:2019sam,Camarda:2019zyx,
Bizon:2018foh,Ebert:2020dfc,Becher:2020ugp,Hautmann:2017xtx, Hautmann:2017fcj}.

\begin{tolerant}{5000}
The \PTll resolution is dominated by the uncertainties in the magnitude of the transverse momenta of the
leptons, whereas the measurement precision of the lepton angle does not contribute significantly.
The kinematic quantity \phistar~\cite{Banfi:2010cf,Banfi:2012du,Marzani:2013nza}, derived from these lepton angles, is defined
by the equation:
\begin{equation}
\phistar \equiv \tan\left(\frac{\pi - \Delta\varphi}{2}\right) \, \sin(\theta^*_{\eta}).
\end{equation}
The variable $\Delta\varphi$ is the opening angle between the leptons in the plane transverse to the beam
axis. The variable $\theta^*_{\eta}$ is the scattering angle of the dileptons with respect to the beam 
in the longitudinally boosted frame where the leptons are back to back. It is related to the pseudorapidities of the oppositely
charged leptons by the relation $\cos(\theta^*_{\eta}) = \tanh \left[(\eta^{-} - \eta^{+})/2\right]$. 
The variable \phistar, by construction greater than zero, is closely related to the normalized transverse momentum $\PTll/\Mll$~\cite{Banfi:2010cf}.
Since \phistar depends only on angular variables,
its resolution is significantly better than that of the transverse momentum, especially at low-\PTll values, but its
interpretation in terms of initial-state radiation (ISR) is not as direct as that of \PTll.
\end{tolerant}

The DY process in the presence of one jet is a complementary way to investigate the initial-state QCD radiations.
The requirement of a minimal transverse momentum associated with this jet is reflected in the \PTll distribution by momentum
conservation. When more hadronic activity than a single jet is present in the events, the transverse momentum balance between
the leading jet and the lepton pair has a broad distribution. As a consequence, the full \PTll spectrum in the
presence of jets brings additional information, since at small values it is sensitive to
numerous hard QCD radiations. Furthermore, DY production in association with at least one jet also brings up contributions where
virtual partons acquire transverse momentum, whose collinear radiations will have a significant
angle with respect to the beam, which contributes as a component of the final \PTll.

This paper presents a DY differential cross section measurement in bins of \Mll, over
the range of 50\GeV to 1\TeV, as functions of \PTll 
and \phistar for inclusive DY production, and in events with at least one jet as a function of \PTll.
The data were collected in 2016 with the CMS detector at the CERN LHC,
corresponding to an integrated luminosity of 36.3\fbinv of proton-proton (pp) collisions at a centre-of-mass energy of $\sqrt{s} = 13\TeV$.
To reduce the uncertainties, the measured cross sections combine measurements of separately extracted cross sections for the
electron and the muon channels. The measurements presented in this paper are extensively discussed in Ref.~\cite{Moureaux:2783681}.

Complementary measurements of the DY process have been performed recently by 
the CMS, ATLAS, and LHCb Collaborations at the CERN LHC ~\cite{CMS:2011aa, Chatrchyan:2011cm, Chatrchyan:2013tia, CMS:2014jea, Khachatryan:2015oaa, Sirunyan:2018owv, Sirunyan:2019bzr,Aad:2011dm, Aad:2013iua, Aad:2014xaa, Aad:2014qja, Aad:2015auj, Aaboud:2016btc,Aad:2019wmn,Aaij:2012vn, Aaij:2012mda, Aaij:2015gna, Aaij:2015zlq, Aaij:2016mgv} and by the CDF and D0 Collaborations at the Fermilab Tevatron~\cite{Affolder:1999jh,Abbott:1999yd,TevatronWZ:D0PhysRevLett2008_100,TevatronWZ:D0PhysLettB2010_693,TevatronWZ:D0PhysRevLett2011_106,CDF:2012brb,PhysRevD.91.072002}.
The cross section measurements presented in this paper extend the mass range below and above the \PZ boson resonance 
with respect to the previous CMS measurements of \PTll dependence.

The outline of this paper is the following: in Section~\ref{sec:detector} a brief description of the 
CMS detector is given. 
In Section~\ref{sec:eventselection} the selection criteria of the measurement are described. 
The simulation samples used in the measurement are described in Section~\ref{samples}. 
Section~\ref{sec:unfolding} explains the details of the unfolding procedure and the systematic 
uncertainties are given in Section~\ref{sec:systematics}. Theory predictions used for comparison with the measurements are described in Section~\ref{sec:theory}. The results are presented in 
Section~\ref{sec:results} and a summary of the paper is given in Section~\ref{sec:conclusion}.

\section{The CMS detector}
\label{sec:detector}

The central feature of the CMS apparatus is a superconducting solenoid of 6\unit{m} internal diameter, 
providing a magnetic field of 3.8\unit{T}. Within the solenoid volume are a silicon pixel and strip tracker, 
a lead tungstate crystal electromagnetic calorimeter (ECAL), and a brass and scintillator hadron calorimeter (HCAL), 
each composed of a barrel and two endcap sections. Forward calorimeters extend the $\eta$ coverage provided 
by the barrel and endcap detectors.
Muons are detected in gas-ionization chambers made of detection planes using three technologies: drift tubes, 
cathode strip chambers, and resistive plate chambers, embedded in the steel flux-return yoke outside the solenoid.

The global event reconstruction (also called particle-flow event reconstruction~\cite{Sirunyan:2017ulk}) 
reconstructs and identifies each individual particle in an event, with an optimized combination of all subdetector 
information.
In this process, the identification of the particle type (photon, electron, muon, charged or neutral hadron) 
plays an important role in the determination of the particle direction and energy.

\begin{tolerant}{5000}
 Electrons 
are identified as a primary charged particle track and potentially many ECAL energy clusters corresponding to this track extrapolation to the ECAL and to possible bremsstrahlung photons emitted along the way.
The electron momenta are estimated by combining energy measurements in the ECAL with momentum measurements in the tracker~\cite{Sirunyan:2020ycc}.
The momentum resolution for electrons with $\pt \approx 45\GeV$ from $\PZ \to \Pe \Pe$ decays ranges from 1.7 to 4.5\%.
It is better in the barrel region than in the endcaps, and also depends on the bremsstrahlung energy emitted by the electron as it traverses the material in front of the ECAL.
\end{tolerant}

Muons 
 are identified as tracks in the central tracker consistent with either a track or several hits in the muon system,
and associated with calorimeter deposits compatible with the muon hypothesis. 
The reconstructed muon global track, for muons with $20 <\pt < 100\GeV$,  has a relative transverse momentum resolution of 1.3--2.0\% in the barrel and better than 6\% in the endcaps. The \pt resolution in the barrel is better than 10\% for muons with \pt up to 1\TeV~\cite{Sirunyan:2018fpa}.
The resolution is further improved with corrections derived from the \PZ mass distribution~\cite{Bodek:2012id}.
 
Charged hadrons are identified as charged particle tracks not identified as electrons or as muons. Finally, neutral hadrons are identified as HCAL energy clusters not linked to any charged-hadron trajectory, or as a combined ECAL and HCAL energy excess with respect to the expected charged-hadron energy deposit. For each event, hadronic jets are clustered from these reconstructed particles using the infrared- and collinear-safe anti-\kt algorithm~\cite{Cacciari:2008gp, Cacciari:2011ma} with a distance parameter of 0.4.
 Jet momentum is determined as the vectorial sum of all particle momenta in the jet, typically within 5 to 10\% of the true momentum over the entire \pt spectrum and detector acceptance.

The primary vertex (PV) is taken to be the vertex corresponding to the hardest scattering in the event, evaluated using tracking information alone, as described in Section 9.4.1 of Ref.~\cite{CMS-TDR-15-02}.
 
 Events of interest are selected using a two-tiered trigger system. The first level (L1), composed of custom hardware processors, uses information from the calorimeters and muon detectors to select events at a rate of around 100\unit{kHz} within a fixed latency of about 4\mus~\cite{Sirunyan:2020zal}. The second level, known as the high-level trigger, consists of a farm of processors running a version of the full event reconstruction software optimized for fast processing, and reduces the event rate to around 1\unit{kHz} before data storage~\cite{Khachatryan:2016bia}.

A more detailed description of the CMS detector is reported in Ref.~\cite{Chatrchyan:2008zzk}, together with a definition of the coordinate system used and the relevant kinematic variables.

\section{Event selection}
\label{sec:eventselection}

The initial event selection requires a dielectron trigger with a \pt threshold of 23 and
12\GeV on the two leading electrons in the electron channel. In the muon channel we require a dimuon trigger with \pt thresholds of 18 and 7\GeV or a
single-muon trigger with a \pt threshold of 24\GeV.
The final selection is restricted to the region where the triggers are fully efficient: $\pt > 25\GeV$ for the leading lepton, $\pt > 20\GeV$ for the
subleading lepton and $\abs {\eta} < 2.4$ for both channels.

An event must contain exactly two isolated leptons of the same flavour (with the
isolation criteria as detailed in Ref.~\cite{Sirunyan:2019bzr}). In addition
the two leptons must have opposite
charges. Events with a third lepton with \pt greater than 10\GeV and $\abs{\eta}<2.4$
are vetoed.

Due to the high instantaneous luminosity of the LHC, additional proton--proton
interactions occur during
the same bunch crossing (pileup) that contribute additional overlapping tracks
and energy deposits in the event, and result in an apparent increase
of jet momenta.
To mitigate this effect, tracks identified as originating from pileup
vertices are discarded and an offset correction is applied to correct for the
remaining neutral pileup contributions~\cite{CMS-PAPERS-JME-18-001}.
The two identified leptons can be reconstructed as jets. Those jets are disregarded by requiring 
a separation, $\Delta R = \sqrt{\smash[b]{(\Delta \eta)^2 + (\Delta \phi)^2}}$, between the
reconstructed jets and these lepton candidates to be larger than 0.4.

To suppress the contamination of jets coming from pileup, a multivariate discriminant is used.
The pileup contamination is also reduced by the choice of the final selection:
jets are required to have a minimum
transverse momentum of 30\GeV and, to ensure high-quality track information, they 
are limited to a rapidity range of $\abs {y}< 2.4$.

To reduce the $\PQt\PAQt$ background, events containing one or more
\PQb tagged jets are vetoed.
The medium discrimination working point of the combined secondary vertex
\PQb tagging algorithm~\cite{Sirunyan:2017ezt} is used.
The effect on the signal is small and is corrected for in the unfolding procedure.

The effects of finite detector resolution and selection efficiency are
corrected by using the unfolding procedure described in
Section~\ref{sec:unfolding}.
Scale factors are applied to the simulation used for the unfolding, to correct
for differences with respect to the data in the efficiencies of the different
selections:
trigger, lepton identification, lepton isolation, and b-tagged jet veto.
For the trigger, the factor is given as a function of $\abs {\eta}$ of the two
leptons and is applied once per lepton pair. The value of the scale factor is
close to one. When dealing with the identification and isolation efficiencies,
the scale factor is given
per lepton as a function of its \pt and $\abs{\eta}$, and applied to each of
the two selected leptons~\cite{Sirunyan:2019bzr}. 

\section{Simulated samples and backgrounds}
\label{samples}
\begin{tolerant}{5000}
For the simulation of the $\PZ/\gamma^*$ process (including the $\PGt^+\PGt^-$ background),
a sample is generated with \MGvATNLO~\cite{Alwall:2014hca} version 5.2.2.2 (shortened to \MGaMC)
using the FxFx jet merging scheme~\cite{Frederix:2012ps}. The parton
shower, hadronization, and QED final-state radiation (FSR) are calculated with
\PYTHIA8.212~\cite{Sjostrand:2014zea}
using the CUETP8M1 tune~\cite{Khachatryan:2015pea}. The matrix element
calculations include $\PZ/\gamma^* + 0,1,2$ jets at next-to-leading order (NLO),
giving an LO accuracy for $\PZ/\gamma^* + 3$ jets. The NLO NNPDF~3.0~\cite{Ball:2014uwa} is used for the matrix element
calculation. In control plots and when comparing to the measurement, this
prediction is normalized to the cross section obtained directly from the
generator, 1977\,pb per lepton channel (for $\Mll>50\GeV$).
\end{tolerant}

Other processes that can give a final state with two oppositely charged same-flavour leptons are $\PW\PW$, 
$\PW\PZ$, $\PZ\PZ$, $\gamma \gamma$, \ttbar pairs, and single top quark production. 
The \ttbar and single top backgrounds are generated at NLO using the
\POWHEG
version~2~\cite{Nason:2004rx,Frixione:2007vw,Alioli:2010xd,Frixione:2007nw}
interfaced to \PYTHIA8.
Background samples corresponding to diboson electroweak production (denoted VV in the figure 
legends)~\cite{Nason:2013ydw} are generated at NLO with \POWHEG
interfaced to \PYTHIA8 ($\PW\PW$) or at LO with \PYTHIA8 alone
($\PW\PZ$ and $\PZ\PZ$). These samples are generated using NLO NNPDF~3.0 for the matrix element
calculation.
The $\gamma \gamma$ background process leading to two charged leptons in the final state,
\ggll, is simulated using \textsc{LPair}~\cite{Vermaseren:1982cz,Baranov:1991yq} interfaced with \PYTHIA6 and using
the default $\gamma$-PDF of Suri--Yennie~\cite{Suri:1971yx}.
This contribution is split into three components, since the interaction at each
proton vertex process can be elastic or inelastic.

\begin{tolerant}{5000}
The total cross sections of $\PW\PZ$ and $\PZ\PZ$ diboson samples are
normalized to the NLO prediction calculated
with \MCFM~v6.6~\cite{Campbell:2015qma}, whereas the cross sections of the $\PW\PW$
samples are normalized to the NNLO prediction~\cite{Gehrmann:2014fva}. The total cross section of the $\PQt\PAQt$
production is normalized to the prediction with NNLO accuracy in QCD and
next-to-next-to-leading logarithmic (NNLL) accuracy in soft gluon
resummation calculated with \TOPpp~2.0~\cite{Czakon:2011xx}. The single top and $\gamma \gamma$ background distributions are normalized to the cross sections calculated by their respective event generators.
\end{tolerant}

It is possible for hadrons to mimic the signature of an electron in the
detector. The main processes that contribute to this background are
\PW + jet production, when the \PW decays leptonically, and
QCD multijet events. Such backgrounds are nonnegligible
only in the electron channel.

The contamination of the signal region by events containing hadrons
misidentified as electrons is estimated using a control region where two
electrons of the same sign are required. This control region mainly contains
events with hadrons misidentified as electrons and events originating from the
DY process when the charge of one electron is incorrectly attributed.
The probability of charge misidentification is obtained as a function of \pt and $\eta$ of each electron in the \PZ peak region
($81 < \Mll <101\GeV$), where the hadron contamination is negligible even in
the control region.
These probabilities are then used to estimate the charge
misidentification rate for other values of \Mll.
The difference between the observed number of events in the control region and
the estimated charge misidentification rate is assumed to be the contamination
from hadron background.
We observe that the numbers of
misidentified-lepton events in the same-sign electron sample and in the
signal (opposite-sign electron) sample are compatible.

\begin{figure}[htbp!]
    \centering

    \includegraphics[width=0.49\textwidth]{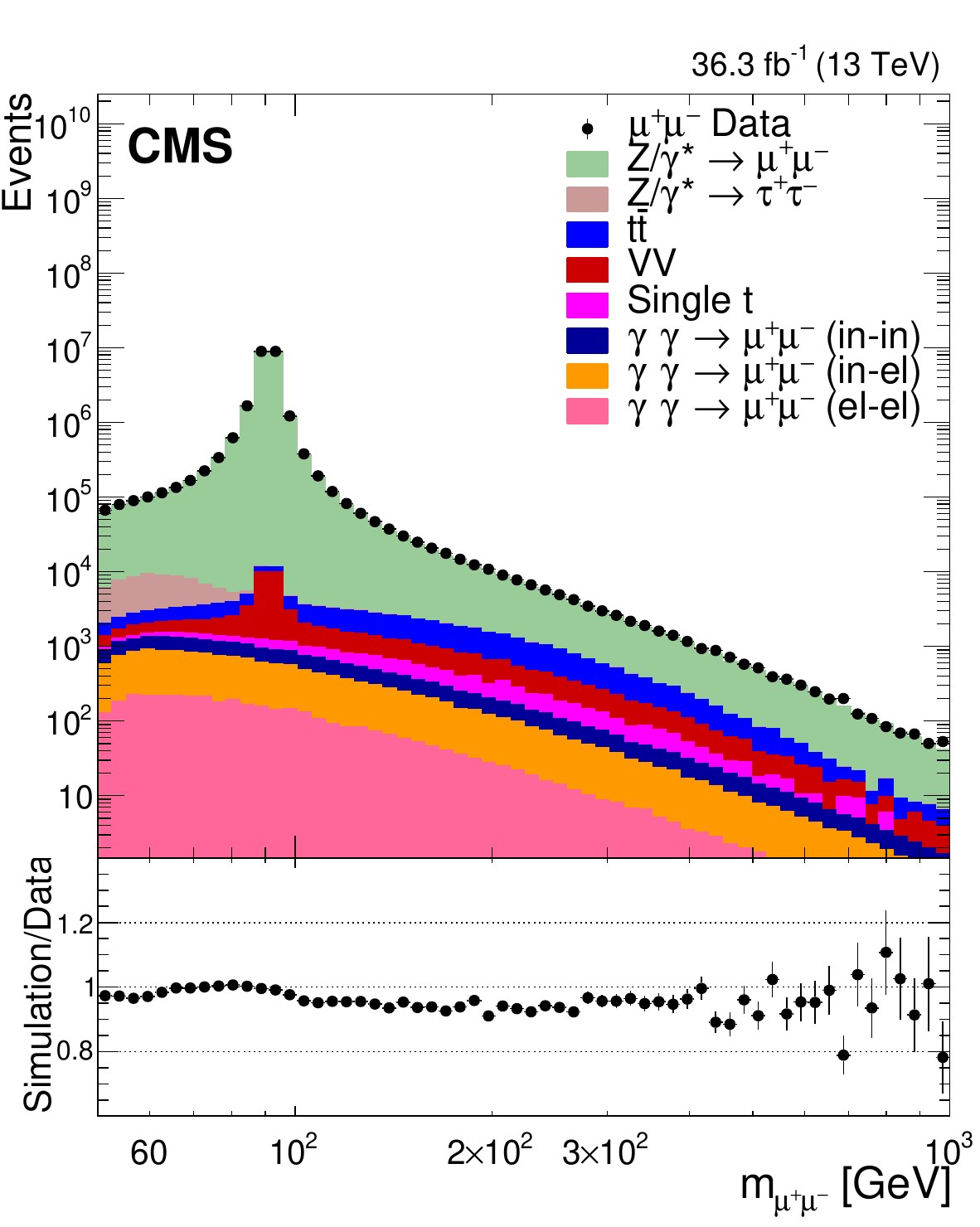}
    \includegraphics[width=0.49\textwidth]{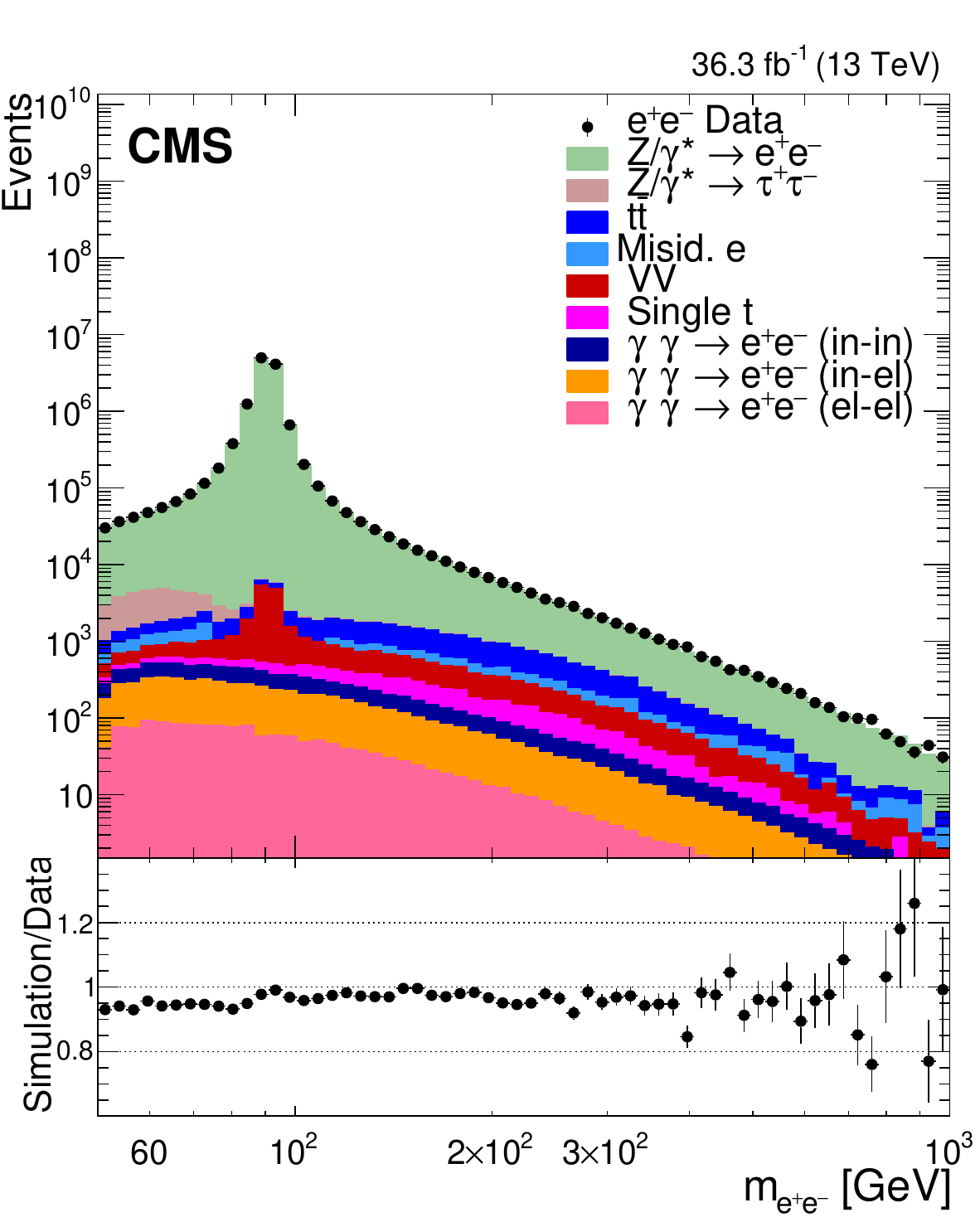}

    \caption{
      Distributions of events passing the selection requirements in the muon (left) and electron channels
      (right). Each plot also presents in the lower part a ratio of simulation over data.
      Only statistical uncertainties are shown as error bars on the data points, whereas the ratio presents the statistical uncertainty in the simulation and the data. The plots show the number of events without normalization to the bin width. The different background contributions are discussed in the text.
    }
    \label{fig:reco-level-leptons}
\end{figure}

\begin{figure*}[htbp!]
    \centering
    \includegraphics[width=0.37\textwidth]{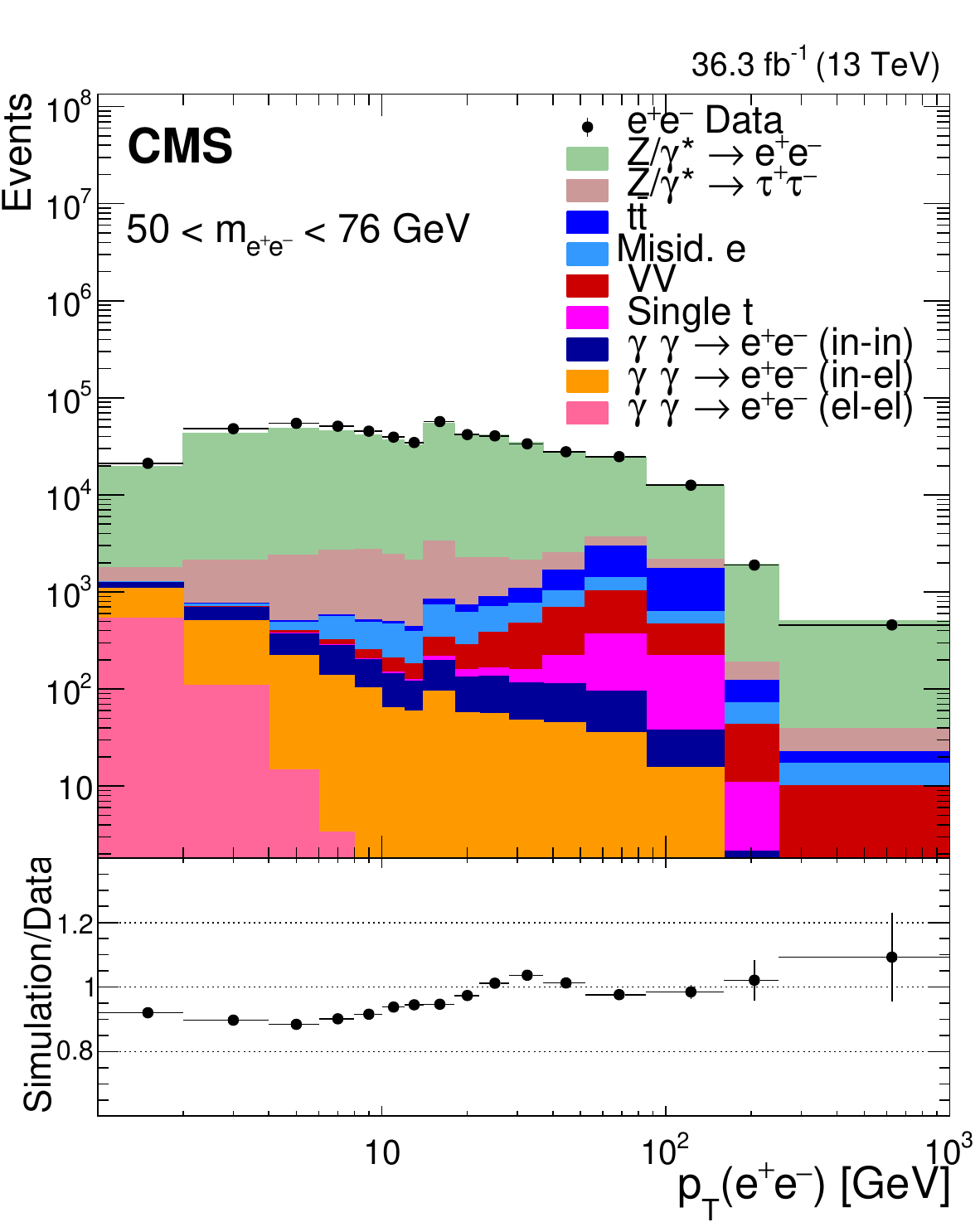}
    \includegraphics[width=0.37\textwidth]{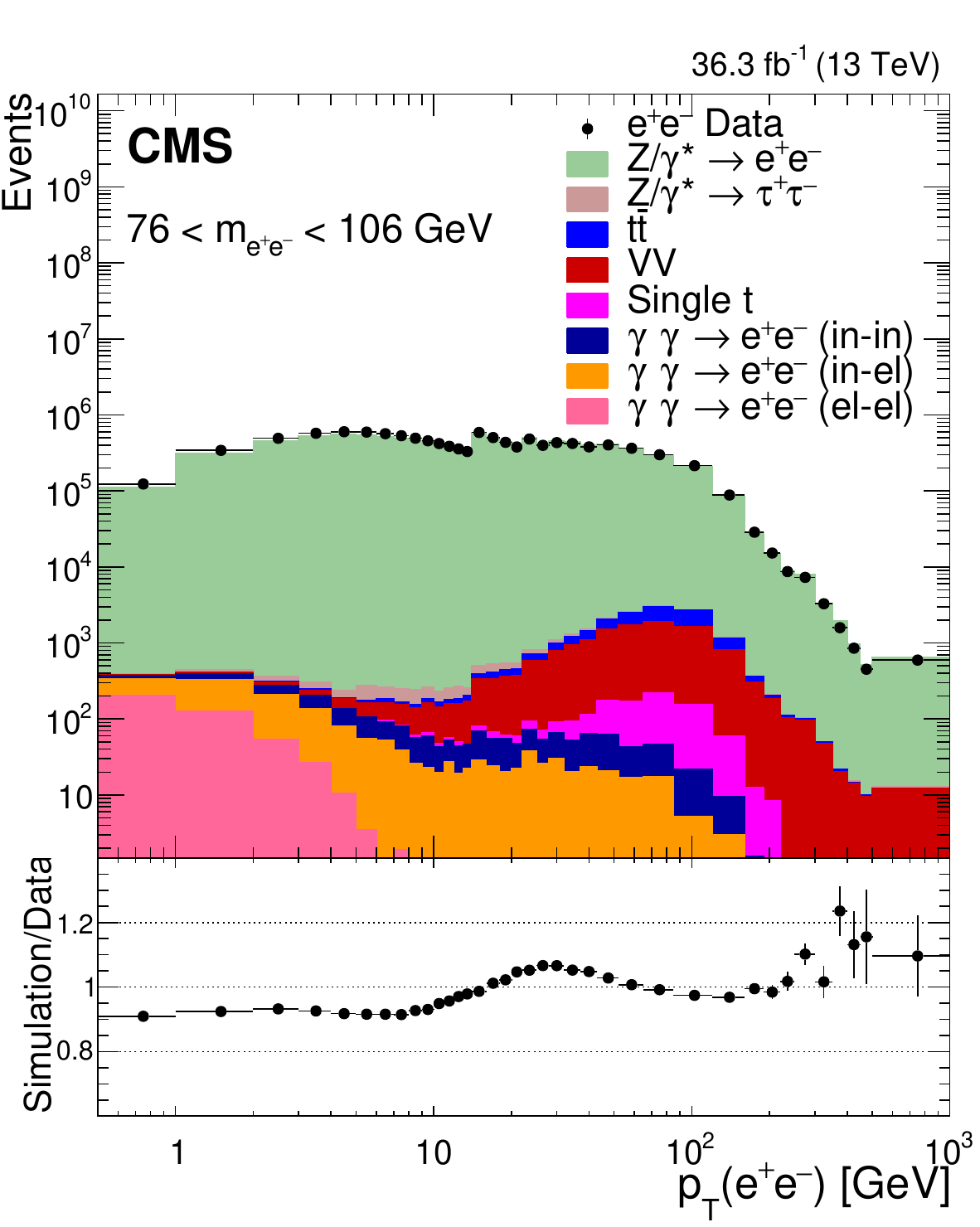}
    \includegraphics[width=0.37\textwidth]{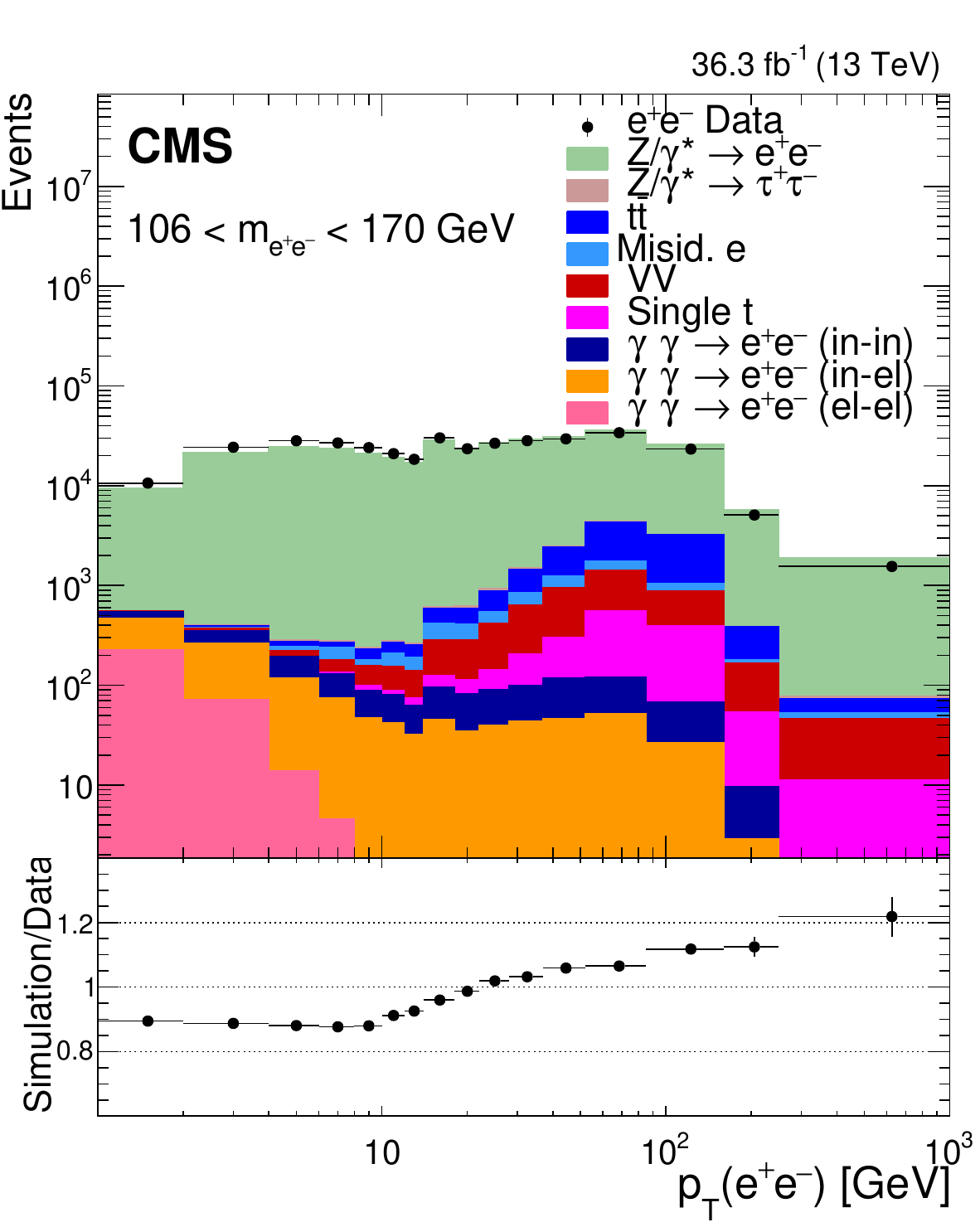}
    \includegraphics[width=0.37\textwidth]{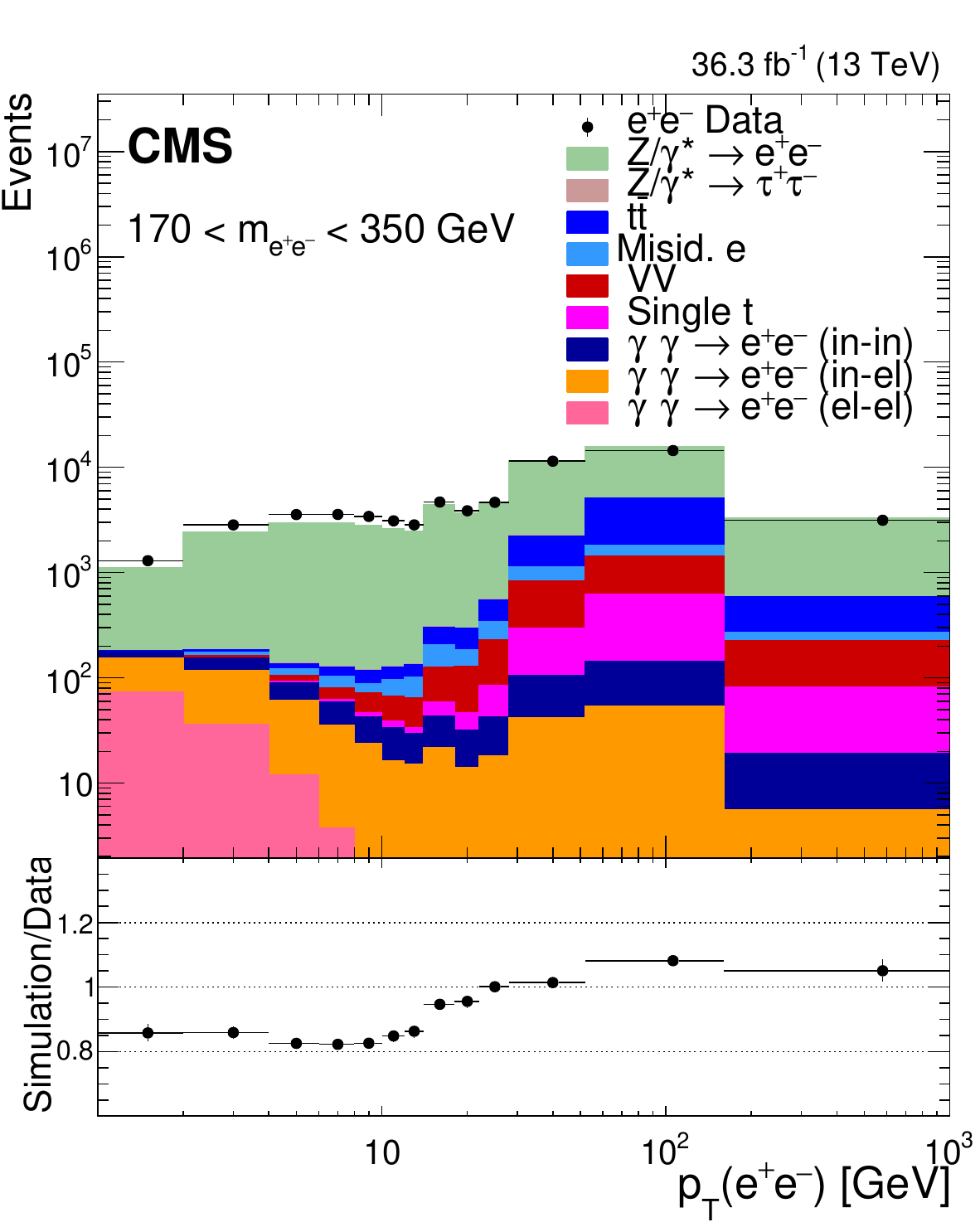}
    \includegraphics[width=0.37\textwidth]{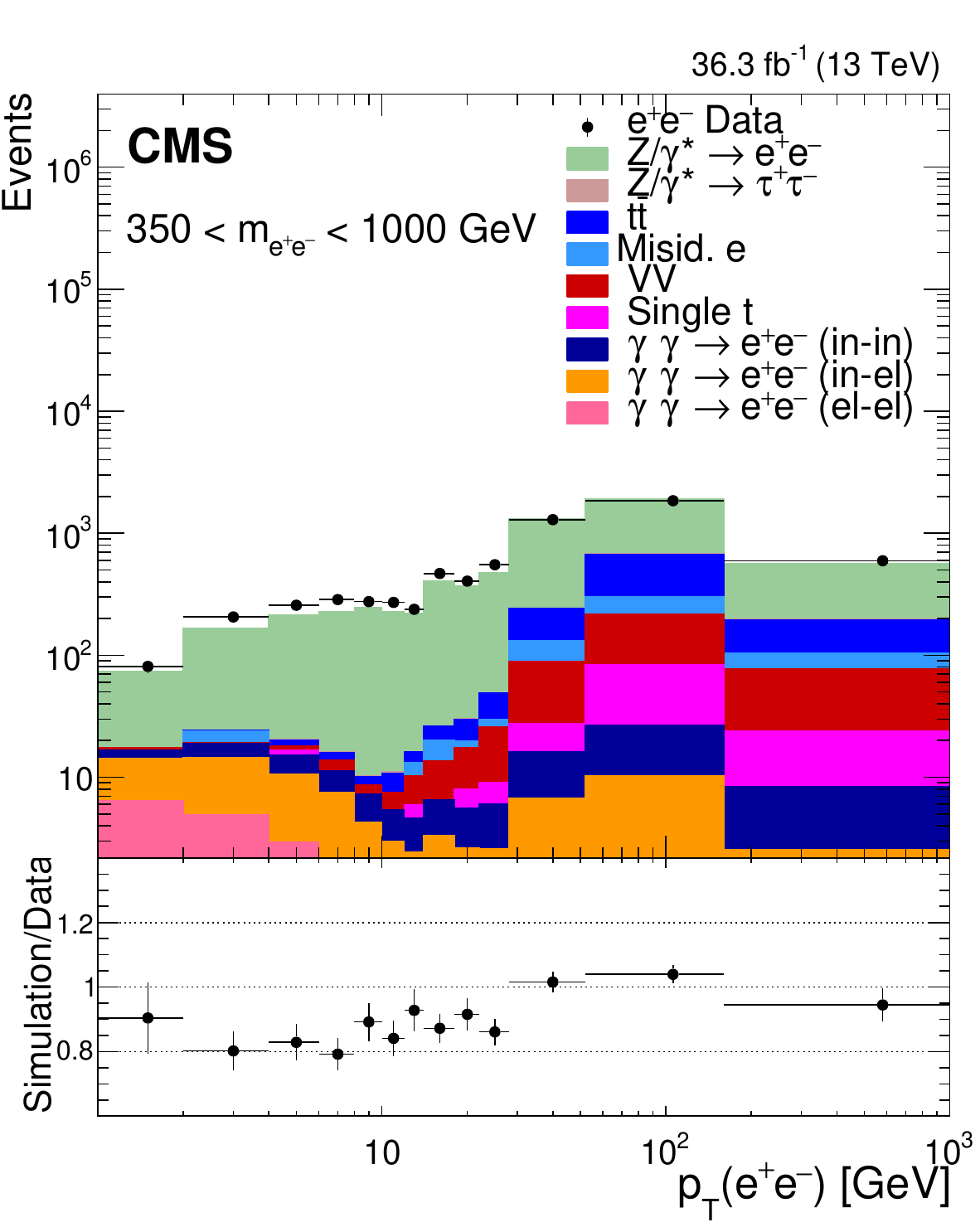}

    \caption{
      Distributions of events passing the selection requirements in the electron channel
      as a function of the dilepton \pt in five ranges of invariant mass:
       50 to
      76\GeV (upper left), 76 to 106\GeV (upper right), 106 to 170\GeV (middle left), 170 to 350\GeV (middle right), and 350 to 1000\GeV (lower).
      More details are given in Fig.~\ref{fig:reco-level-leptons}.
    }
    \label{fig:reco-level-pt}
\end{figure*}

The number of events at the reconstructed level is compared with the sum of the
contributions from signal and backgrounds.
In Fig.~\ref{fig:reco-level-leptons}, the dilepton mass spectrum is shown for both the electron and and the muon channels, whereas
Fig.~\ref{fig:reco-level-pt} shows the \PTll distributions in various \Mll bins for the electron channel only.
Globally, the background contamination is lower than 1\%. The background becomes
around 10\% for \Mll
outside of the \PZ boson mass peak and up to 30\% in some bins.
The simulated samples are processed through a \GEANTfour~\cite{GEANT4:2002zbu} based simulation of the CMS detector, with the same reconstruction algorithms as of data. They also include a pileup profile that is reweighed to match the profile of the data.

\section{Measured observables and unfolding procedure}
\label{sec:unfolding}
The measurement of the DY cross section is carried out with respect to the \pt and \phistar of the 
dilepton pairs produced inclusively, and with respect to \pt for pairs produced in association 
with at least one jet. For the inclusive case, the measurement is divided into five invariant mass 
bins: 50--76, 76--106, 106--170, 170--350, 350--1000\GeV; the last bin is not included when requiring 
at least one jet because of the small number of events available. The measurement of the ratio of cross 
section in mass bins 50--76, 106--170, 170--350, 350--1000\GeV to the cross section around the \PZ mass peak(76--106\GeV) 
is also performed. The bin widths are chosen to be as small as possible, based on the detector resolution and the number of events.

To correct for the detector resolution and the efficiency of the
selection, an unfolding procedure is applied to the measured distributions one dimensionally in each mass bin.
To obtain the particle-level distributions from the reconstructed distributions, the unfolding uses a response matrix based
on the simulated signal sample.
To unfold, the D'Agostini iterative method with early stopping is used as
implemented in \textsc{RooUnfold}~\cite{D'Agostini:1994zf}.
The result, converging towards the maximum likelihood estimate, is affected by fluctuations increasing with the
number of iterations.
The fluctuations are studied using pseudo-experiments for each number of
iterations following the method used in Ref.~\cite{Sirunyan:2018cpw}. The procedure is stopped just before the fluctuations become
significant with respect to the statistical uncertainty. The number of iterations ranges between 4 and 25.

The particle level refers to stable particles ($c\tau>1\unit{cm}$), other than neutrinos, in the final state.
To correct for energy losses due to QED FSR, leptons are ``dressed",
\ie, all the prompt photons with a distance smaller than $\Delta R=0.1$ to the
lepton axis are added to the lepton momentum.
The cross section is extracted in the following phase space: leading and subleading dressed leptons satisfying $\pt> 25$
and 20\GeV and $\abs{\eta}<2.4$. When at least one additional jet is required, it
must satisfy $\pt>30\GeV$, $\abs{y}<2.4$, and be spatially separated from the
dressed leptons by $\Delta R>0.4$.

The cross sections are first extracted separately for the electron and muon channels. They are compatible for all studied distributions and the two channels are combined
to reduce the statistical uncertainties.
The combined differential cross sections are calculated bin-by-bin as the weighted mean values of the differential cross sections of the two channels.
The systematic and statistical uncertainties are obtained using the linear
combination method described in Ref.~\cite{Khachatryan:2016crw}, considering as
fully correlated the uncertainties in
the jet energy scale and resolution, the pileup, the background
subtraction, \PQb tagging, and the integrated luminosity.
Other uncertainties are considered as uncorrelated.
\section{Systematic uncertainties}
\label{sec:systematics}

Several sources of uncertainties in the measurement are considered.
The integrated luminosity is measured with a precision of
$1.2\%$~\cite{CMS:2021xjt}, which results in a relative uncertainty
of almost the same value in the measurement. Small variations are caused by the
subtraction of the background contributions estimated from the simulation.

The uncertainties coming from the lepton trigger efficiencies are estimated by varying the applied scale factors up and down by one standard
deviation.
The uncertainties from identification and reconstruction efficiencies are
estimated for various sources including QED FSR, resolution, background
modeling, and the tag object selection in the tag-and-probe procedure, as well as the statistical component treated
separately for each scale factor in \pt and $\eta$ of the
lepton~\cite{Sirunyan:2019bzr}.
The efficiency uncertainties include a one percent effect in the L1 trigger
caused by a timing problem in ECAL endcaps.
The lepton energy scale uncertainties are estimated by varying the lepton energy and \pt by $\pm 1$ standard deviation (reach 0.75\% (0.5\%) for electrons (muons) depending on $\eta$ and \pt).
Uncertainties coming from the lepton energy resolution are estimated by spreading the lepton energy using the generator-level information.

The uncertainty in the jet energy scale is estimated by varying the jet momenta in data
by 2.5\% to 5\%, depending on the energy and pseudorapidity of the jet.
The uncertainty in the jet energy resolution is estimated by varying the smearing factor used to match the simulated jet energy resolution to data by $\pm$1 standard deviation around its central value.

\begin{tolerant}{5000}
A systematic uncertainty is attributed to the normalisation of the background
samples estimated by Monte Carlo event generators.
The theoretical uncertainty in the cross section of the dominant \ttbar background
is$\approx$6\%, using the \TOPpp~2.0 program and including scale and PDF
variations.
The uncertainties in the other background cross sections are smaller.
In particular, it has been verified that 6\% covers the differences of the \ggll samples generated
using Suri--Yennie and LuxQED~\cite{Manohar:2016nzj,Manohar:2017eqh} photon PDFs.
In a conservative way,
the uncertainties in all other Monte Carlo based background estimates are also estimated to be 6\%. This uncertainty is applied to fully-elastic, semi- and fully-inelastic cases.
\end{tolerant}

The uncertainty in the misidentified electron background estimation using same-sign events is obtained using 
an uncertainty in the charge misidentification estimation of about 10\% per
electron at $\pt (\Pe)=150\GeV$, rising with $\pt(\Pe)$.
A 20\% total uncertainty in the charge misidentification is used and
propagated to the estimate of this background.

Alternative pileup profiles are generated by varying the amount of pileup
events by 5\%, and the difference to the nominal sample is propagated to the final results.

The unfolding model uncertainty is estimated by reweighting the simulated sample to match the data shape for each distribution, and using this as an alternate model for unfolding. The difference with respect to the results obtained with the simulated sample is assigned as the uncertainty. 
The statistical uncertainty coming from the limited sample size is also included, provided by the \textsc{RooUnfold} package.

The systematic uncertainties are propagated to the measurement through the unfolding procedure by computing new 
response matrices varying the quantities by one standard deviation up and down.
All the experimental uncertainties are symmetrized by taking the average of the deviations from the central value. 
The uncertainty sources are independent and the resulting uncertainties are added
in quadrature. 

\begin{figure*}[htbp!]
    \centering
    \includegraphics[height=0.4\textwidth]{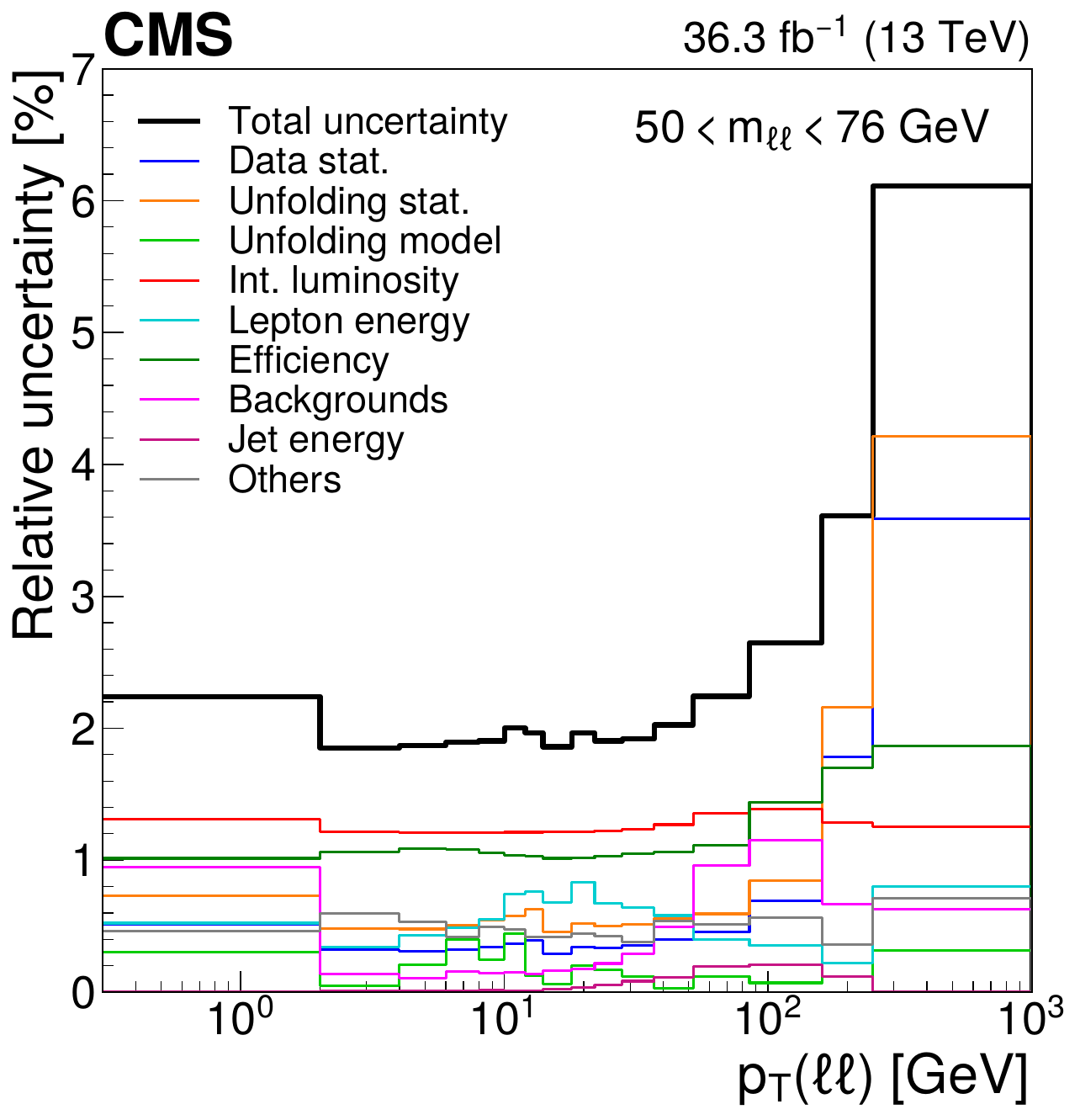}
    \hspace{1.5mm}
    \includegraphics[height=0.4\textwidth]{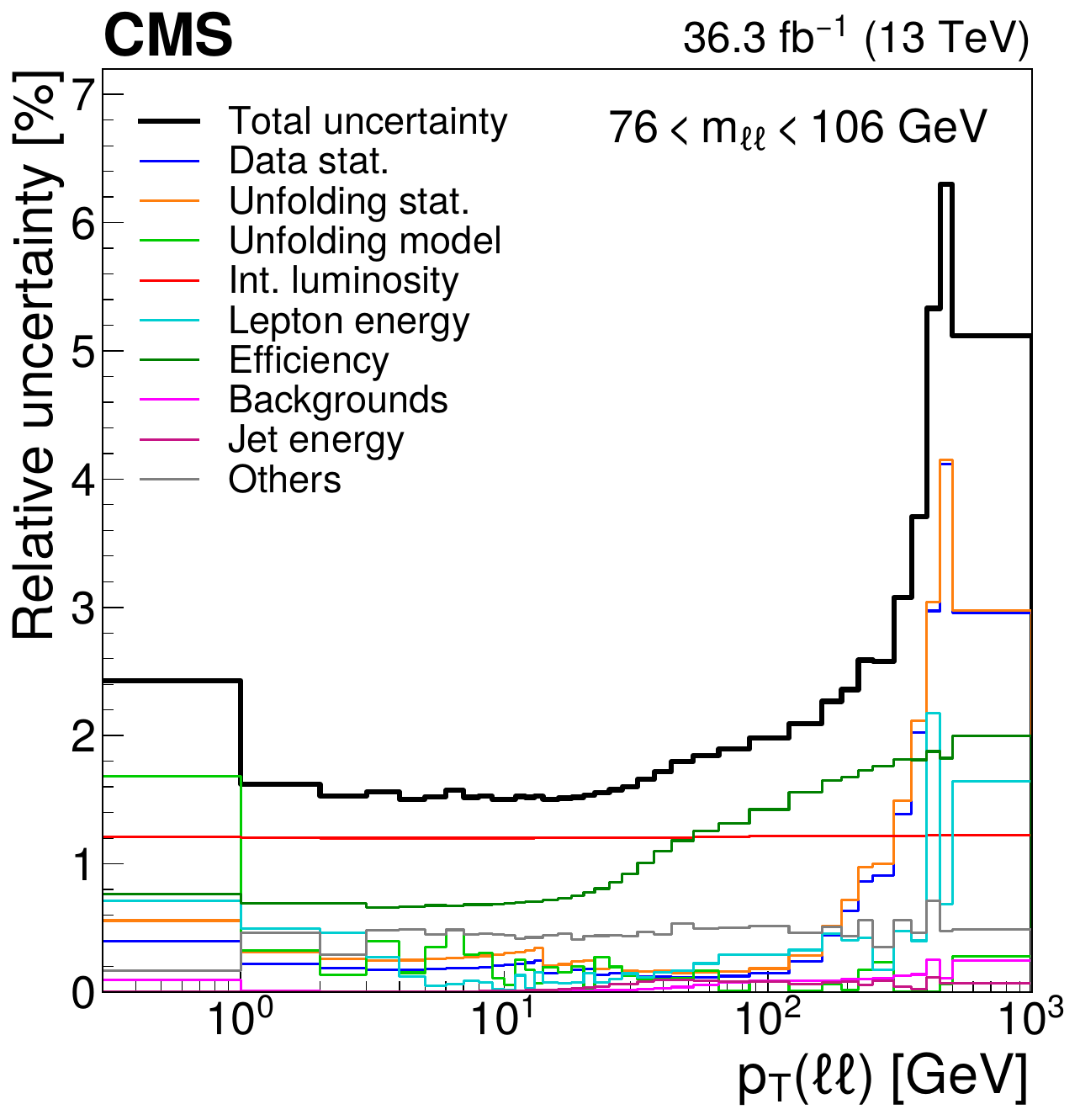}
    \includegraphics[height=0.4\textwidth]{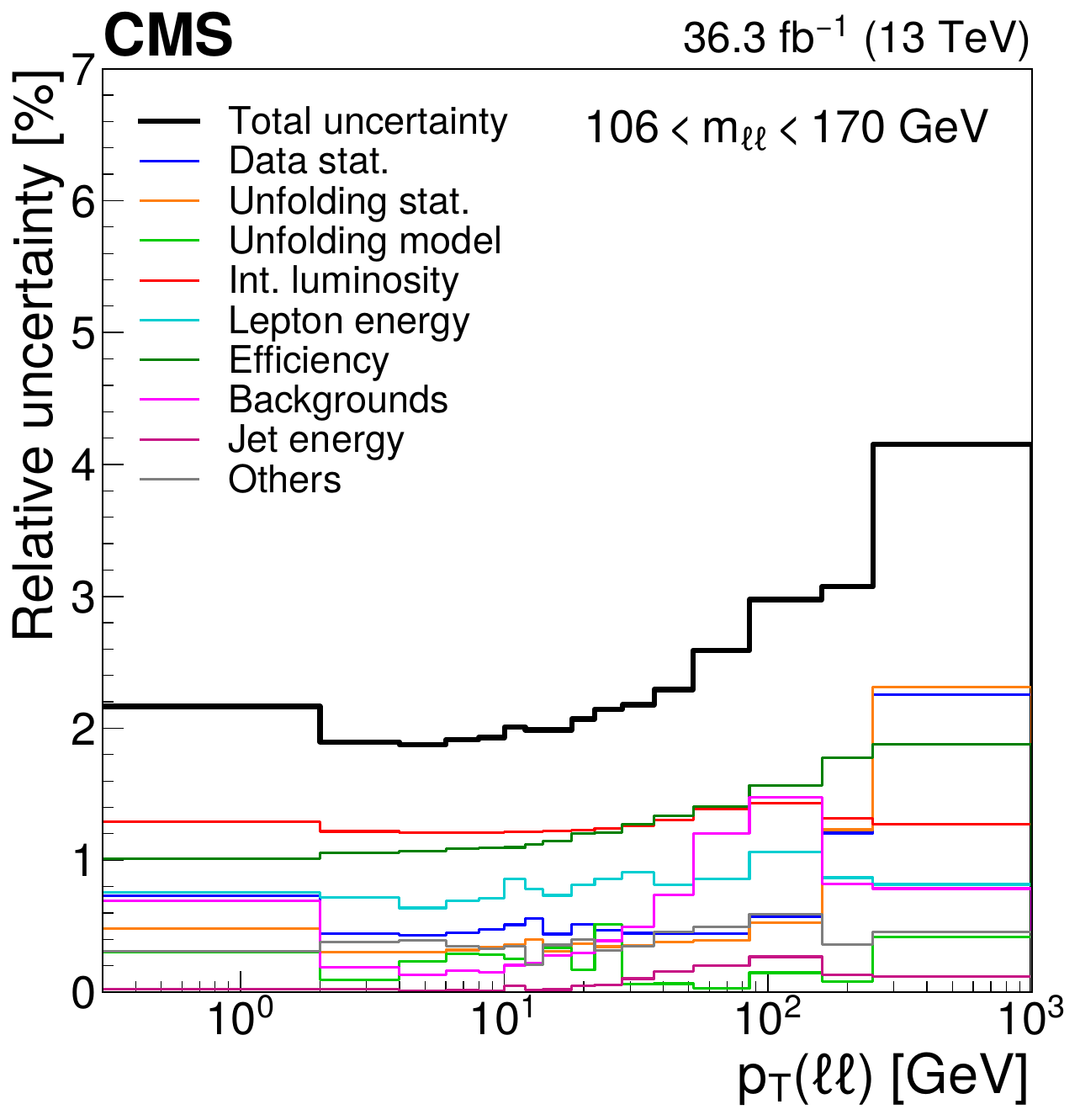}
    \includegraphics[height=0.4\textwidth]{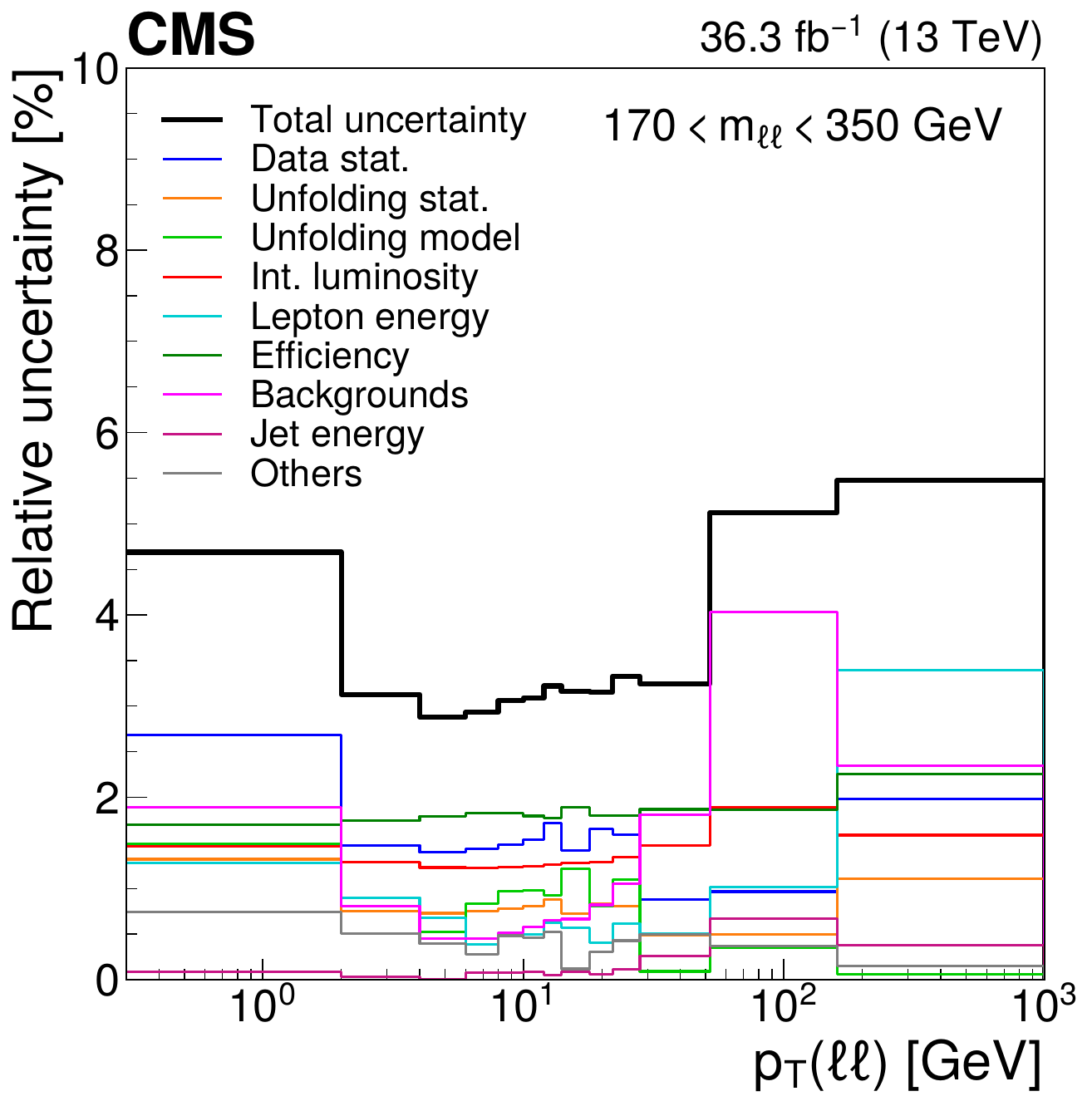}
    \includegraphics[height=0.4\textwidth]{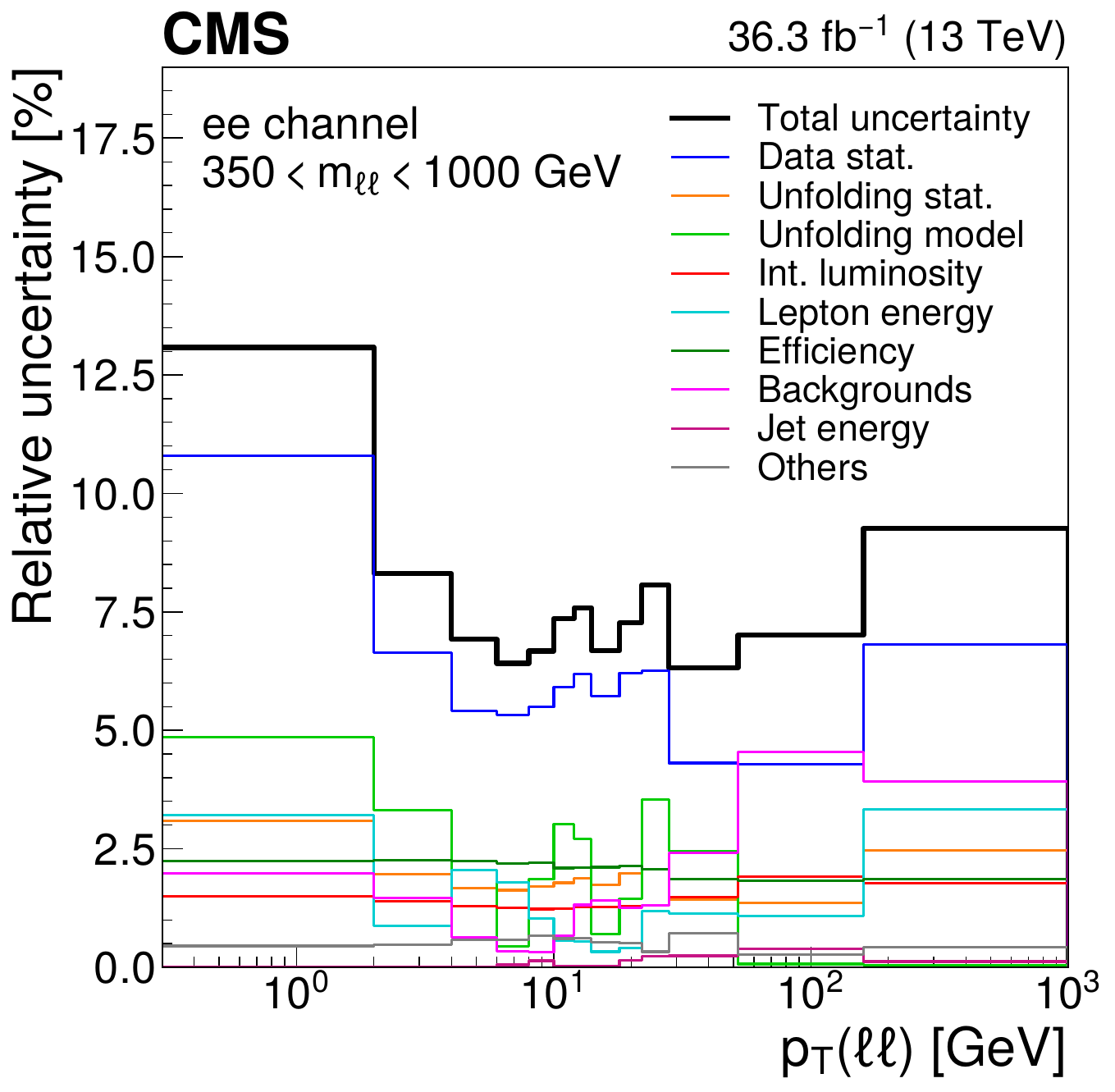}

        \caption{
            Estimates of the uncertainties in inclusive differential cross sections in \PTll in
            various invariant mass ranges: \mrangea (upper left),
            \mrangeb (upper right), \mrangec (middle left),
            \mranged (middle right), and \mrangee (lower).
            The black line is the quadratic sum of the colored lines.
           }
	\label{fig:UnfCombPt0err}
\end{figure*}

For the inclusive measurement the main sources of uncertainties are the
integrated luminosity measurement,
the identification and trigger efficiency corrections of the leptons, and the energy scale of the leptons.
For the DY + $\geq$1 jet case, the major uncertainties come from the jet energy scale and the unfolding model.
The estimates of systematic uncertainties for the inclusive differential cross sections in \PTll 
for various \Mll ranges are shown in Fig.~\ref{fig:UnfCombPt0err}.

\begin{figure*}[htbp!]
    \centering

    \includegraphics[height=0.45\textwidth]{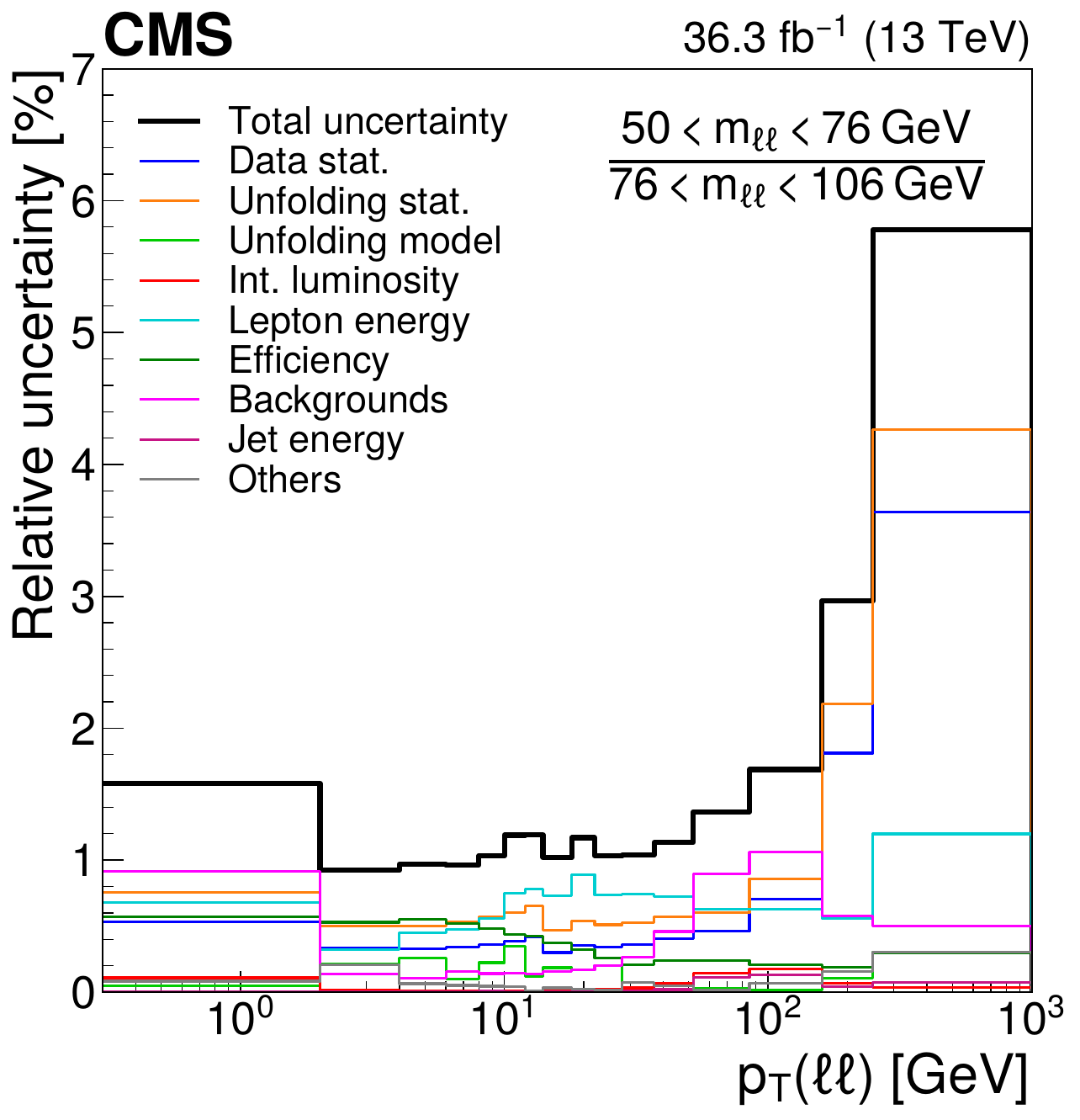}
    \hspace{1.6mm}
    \includegraphics[height=0.45\textwidth]{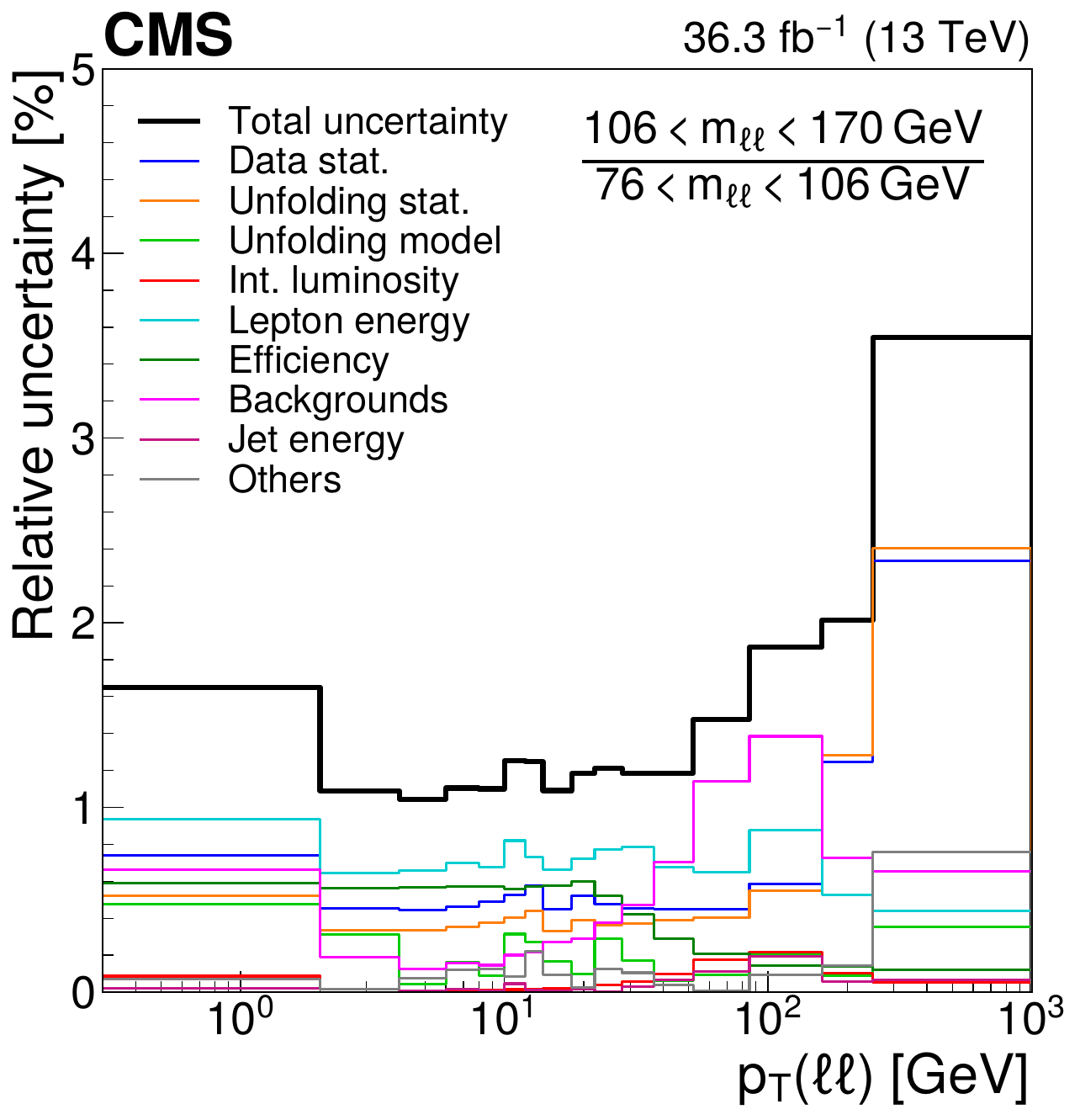}
    \includegraphics[height=0.45\textwidth]{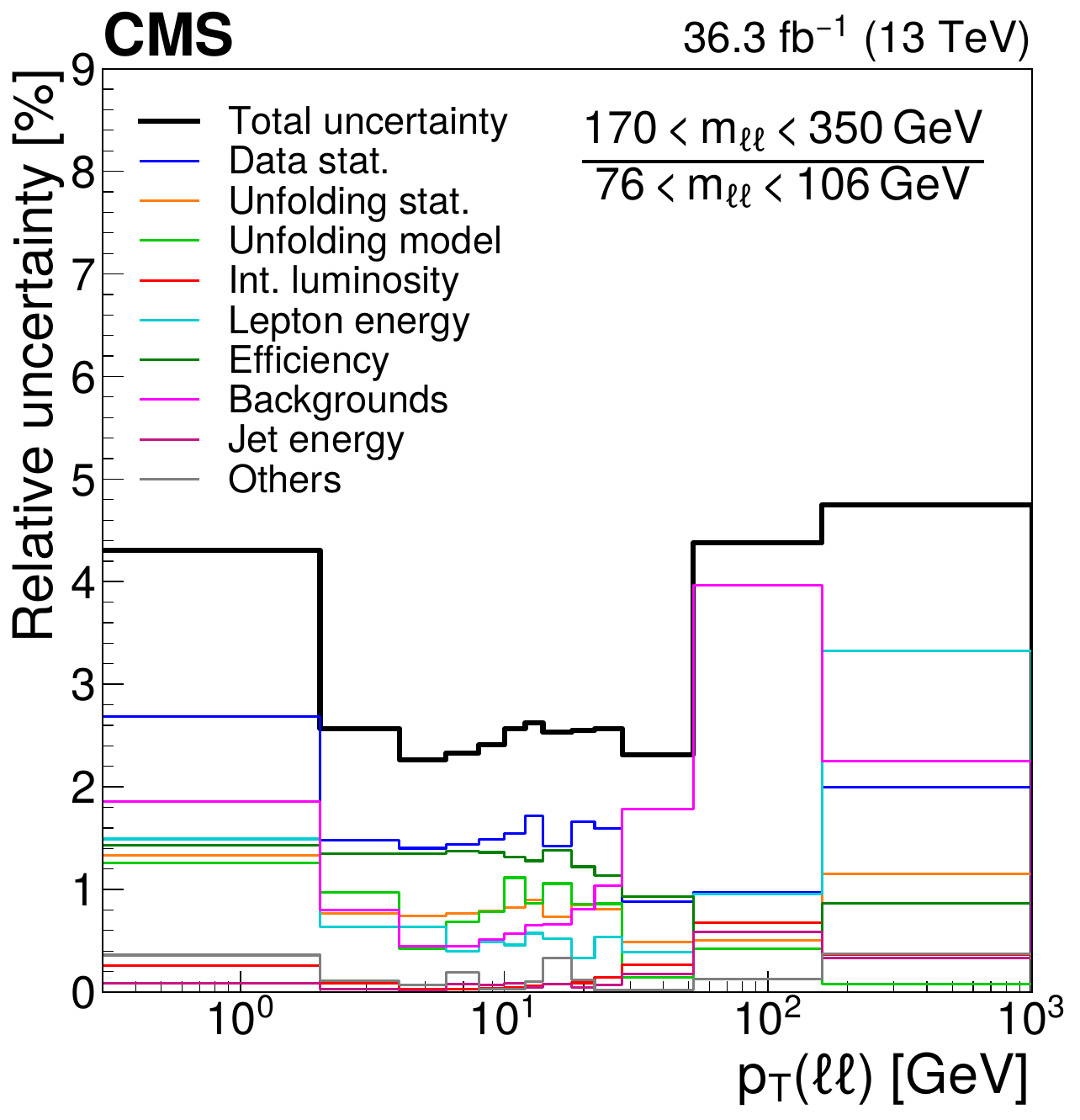}
    \includegraphics[height=0.45\textwidth]{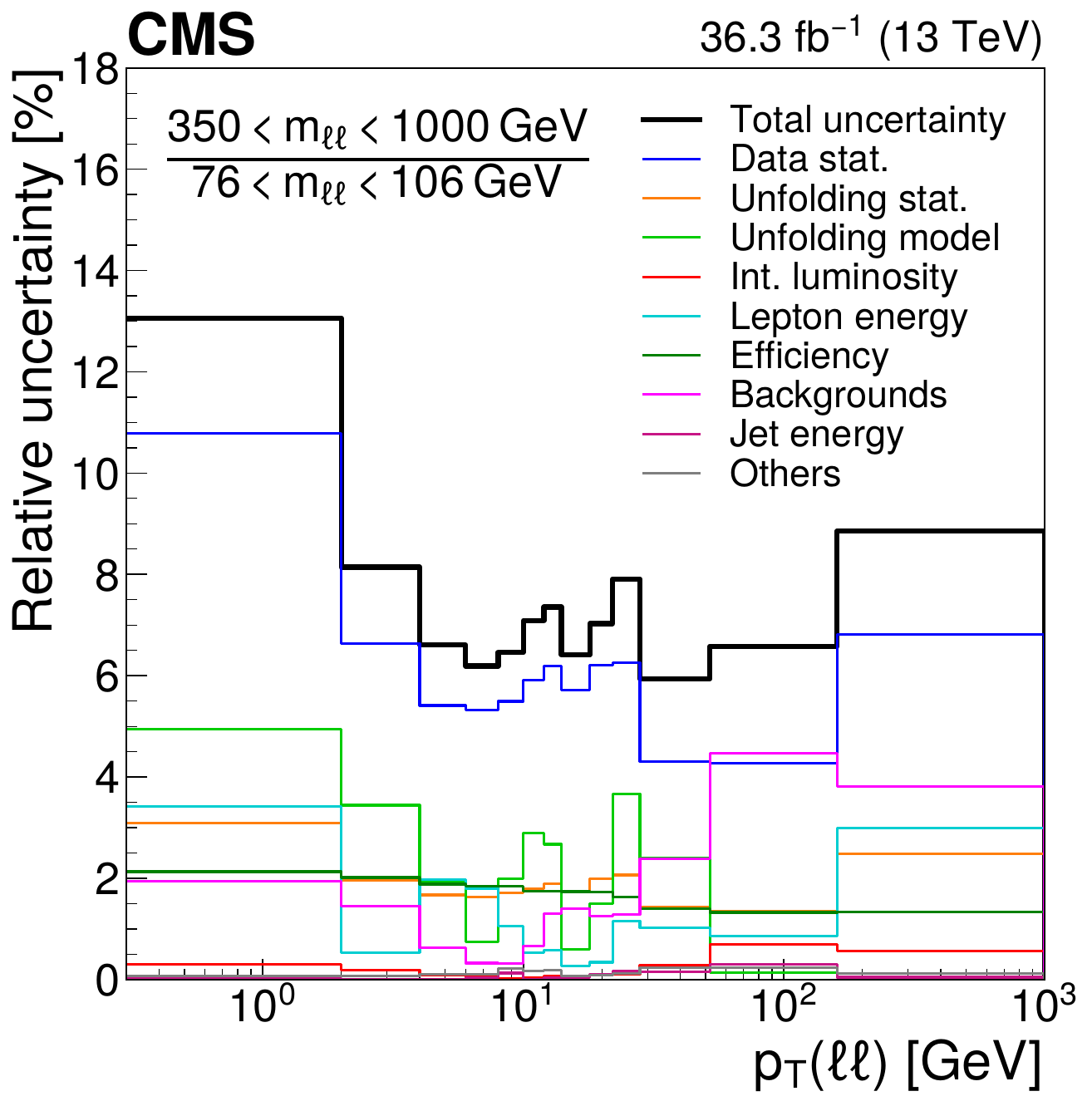}

    \caption{
        Estimates of the uncertainties in inclusive differential cross section ratios in \PTll for invariant mass ranges
        with respect to the peak region \mrangeb: \mrangea (upper left), \mrangec (upper right), \mranged (lower left),
        and \mrangee (lower right). The black line is the quadratic sum of the colored lines.
    }
    \label{fig:UnfCombPt0Ratioerr}
\end{figure*}

When calculating the cross section ratios, for each \PTll or \phistar bin, all
uncertainties are taken as fully
correlated between the numerator and the denominator, except the data and Monte
Carlo statistical uncertainties. The total uncertainty corresponds to the quadratic sum of the sources.
The estimates of systematic uncertainties for the ratios of the inclusive
\PTll distributions are shown in  Fig.~\ref{fig:UnfCombPt0Ratioerr}.

\section{Theory predictions}
\label{sec:theory}
The measured data are compared with the \MGPE baseline sample described
in Section~\ref{samples}.
The QCD scale uncertainties are estimated by varying the renormalisation and factorisation scales simultaneously by factors of 2 and 1/2 (omitting the variations in opposite direction and taking the envelope). The strong coupling (\alpS) and PDF uncertainties are estimated as the standard deviation of weights from the replicas
provided in the NNPDF~3.0 PDF set~\cite{Ball:2014uwa}.

An event sample at NNLO with a jet merging method is generated with \minnlo~\cite{Monni:2019whf, Monni:2020nks}.
The coupling  \alpS is evaluated independently at each vertex at a scale that depends 
on the kinematic configuration. Sudakov form factors are used to interpolate between the scales.
The NNLO version of the NNPDF~3.1 PDF set~\cite{NNPDF:2017mvq} is used along with the 
\PYTHIA version 8~\cite{Sjostrand:2014zea}
for the parton showers based on the CP5 tune~\cite{CMS:2019csb} and multiparton interactions (MPI), but including a harder primordial \kt of 2.2 \GeV obtained from tuning the \kt parameter to describe the observed \phistar distribution of Ref.~\cite{Sirunyan:2019bzr}.

The results are also compared with a third prediction from the parton branching (PB) TMD
method~\cite{Hautmann:2017xtx, Hautmann:2017fcj} obtained from
\cascade3~\cite{baranov2021cascade3}.
This prediction is of particular interest since the initial-state parton showers
are fully determined by
TMD and their backward PB evolution, and therefore are free of tuning parameters.
The matrix element calculation is performed at NLO for $\PZ + 0$~jet using
\MGaMC for the inclusive distributions (labelled \MGCASzero), and for $\PZ + 1$~jet for the distributions
where one jet is required in the final state (labelled \MGCASone).
Initial-state parton showers, provided by the PB TMD method are matched to the NLO matrix element~\cite{Martinez:2019mwt}, using the latest
TMD PB set:
PB-NLO-HERAI+II-2018-set2~\cite{PhysRevD.99.074008}.
The final parton shower, hadronization, and QED FSR steps are performed with \PYTHIA6~\cite{Sjostrand:2006za}. 
This approach is equivalent to the inclusion of the next-to-leading logarithmic
soft-gluon resummation on top of the fixed-order NLO calculations.
The theoretical uncertainties in the cross section are estimated by variation of scales and from TMD
uncertainties.
This approach is expected to describe the inclusive cross section at low \PTll
($< 20\GeV$) well,
and to fail for larger \PTll, since higher-order matrix element contributions are missing, as already observed for
the \PZ boson mass peak range~\cite{Sirunyan:2019bzr}.
Recently, this approach has been developed to include multi-jet merging~\cite{Martinez:2021chk} at LO, 
which allows a larger \PTll\  region to be described as well.

A fourth prediction is based on an independent approach relying on TMDs obtained from 
fits to DY and \PZ boson measurements at different energies~\cite{Scimemi:2017etj, Scimemi:2019cmh} 
using an NNLO evolution. The corresponding numerical evaluations are provided
by the \artemide2.02 code~\cite{arTeMiDe-web}. The resummation corresponds to an
N$^3$LL approximation. The uncertainty is obtained from scale variations. 
Due to the approximation of ordering among the scales, the prediction has a
limited range of validity for the calculation of: $\PTll < 0.2 \ \Mll$.
Predictions for the \phistar cross section dependence as well as the 1 jet case are not 
provided by \artemide. The \artemide sample does not include the QED FSR;
a correction is derived from the \PYTHIA8 shower in the \MGaMC sample.
The uncertainty is derived by taking the difference with respect to corrections
derived from the \POWHEG sample described in Ref.~\cite{Sirunyan:2019bzr}.
This uncertainty is smaller than 1\% for $\PTll<0.2\ \Mll$

Two more predictions are obtained from the \GE 1.0-RC3 program~\cite{Alioli_2013, Alioli:2015toa, PhysRevD.104.094020} 
combining higher-order resummation with a DY calculation at NNLO. 
Originally, the resummation was carried out at NNLL including partially N$^{3}$LL 
on the 0-jettiness variable $\tau_0$~\cite{PhysRevLett.105.092002}. 
More recently it includes the \qt 
resummation at \NqLL in the Radish formalism~\cite{Monni:2016ktx,Bizon:2017rah} for the 0 jet
case, whereas it keeps the 1-jettiness resummation for the 1 jet case.
Two samples are generated, one in the 0-jettiness approach and one in the \qt resummation
approach. The calculation uses the PDF4LHC15 NNLO~\cite{Butterworth:2015oua} PDF 
set with $\alpS(m_{\PZ})=0.118$, the world average. 
The events are showered using a specially modified
version of \PYTHIA8, which is also used for nonperturbative effects and
QED radiation in the initial and final states using a modified tune based on
CUETP8M1.
The theoretical uncertainties are estimated by variation of scales and from the resummation 
as described in Ref.~\cite{Alioli:2015toa}. No uncertainty is assigned to the jetiness resummation.

\section{Results and discussion}
\label{sec:results}

\subsection{\texorpdfstring{\PTll}{pt(ll)} results}

\begin{figure*}[htbp!]
 \centering
 \includegraphics[height=0.40\textwidth]{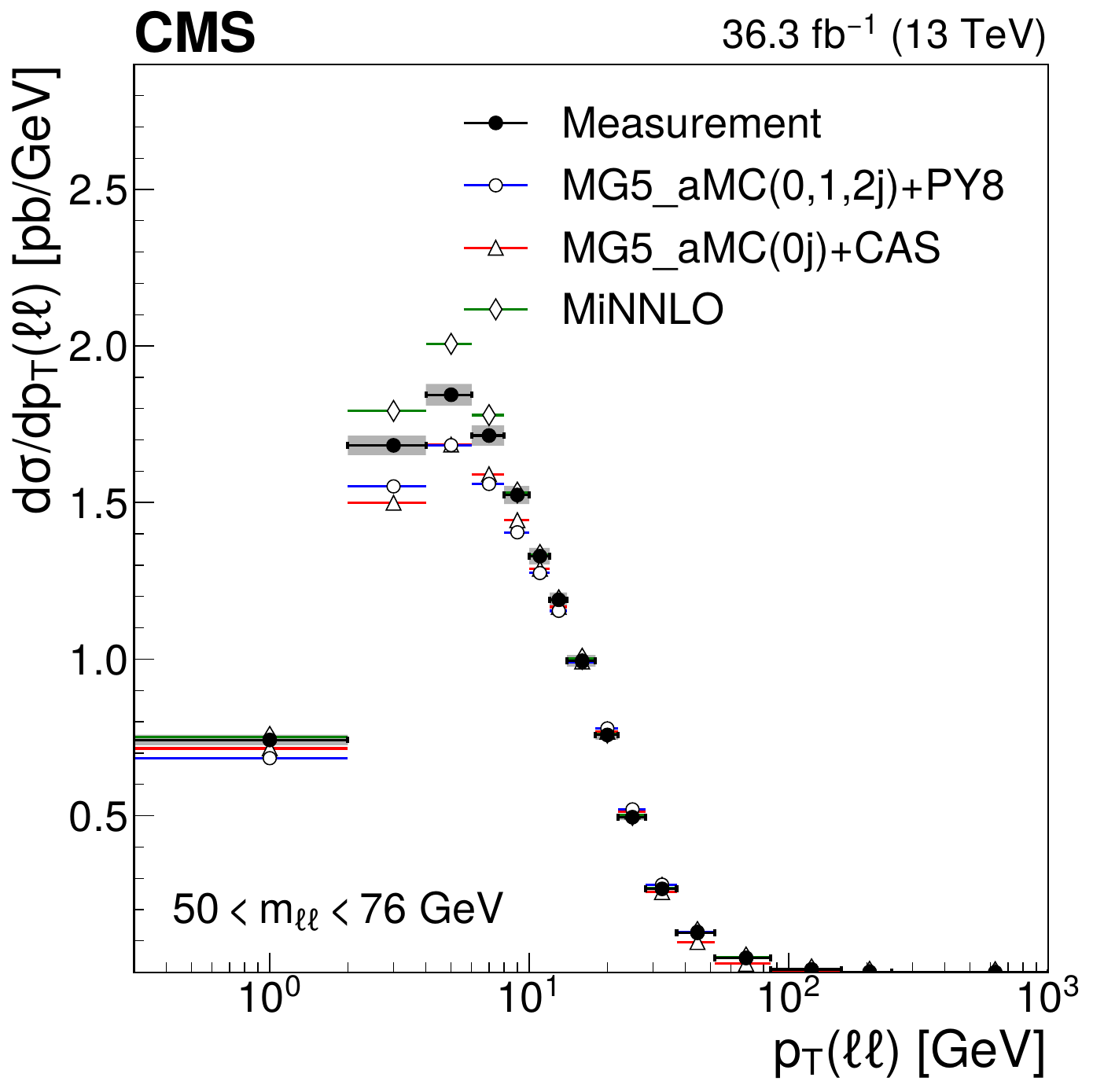}
 \hspace{2.2mm}
 \includegraphics[height=0.40\textwidth]{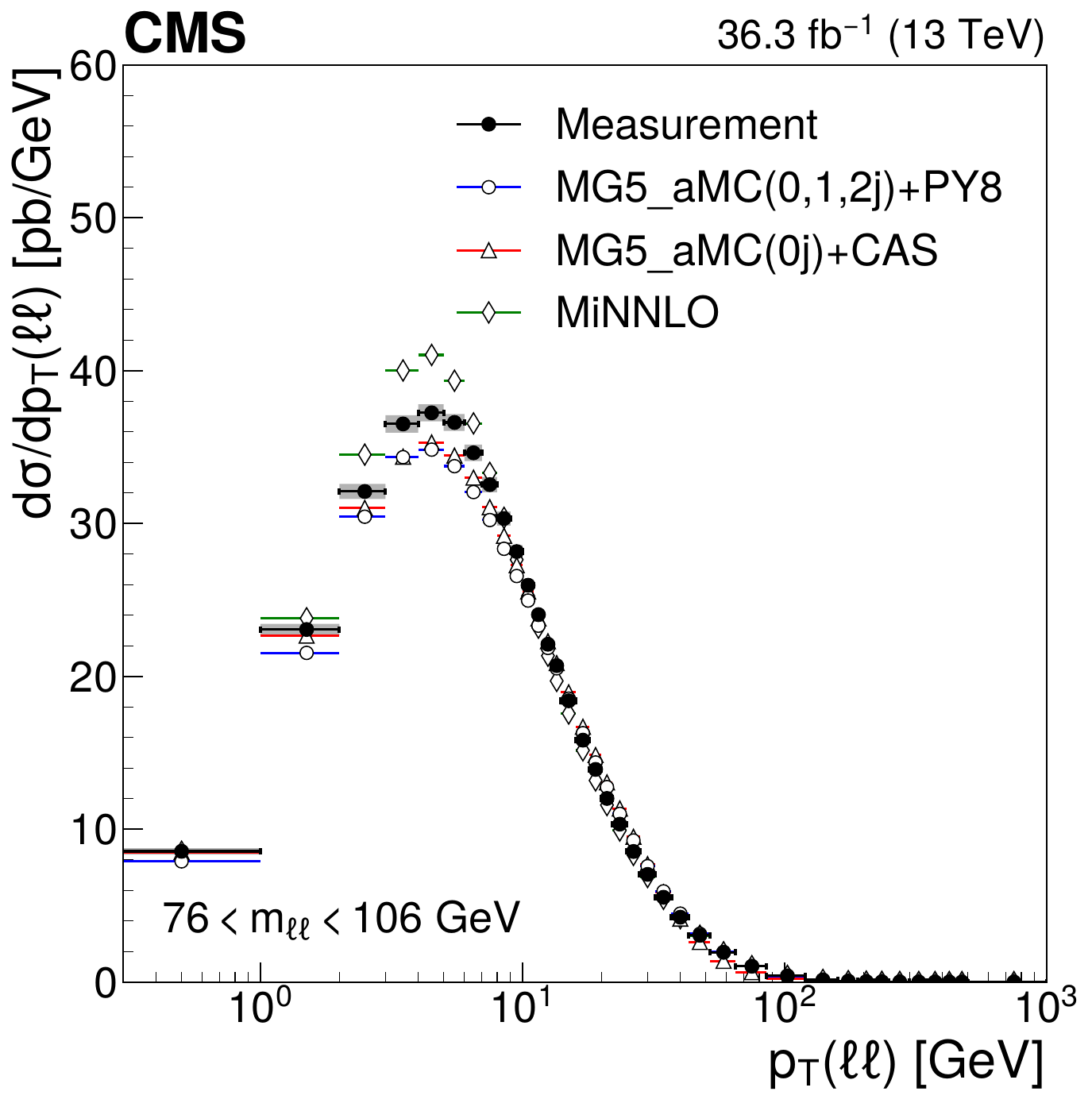}
 \includegraphics[height=0.40\textwidth]{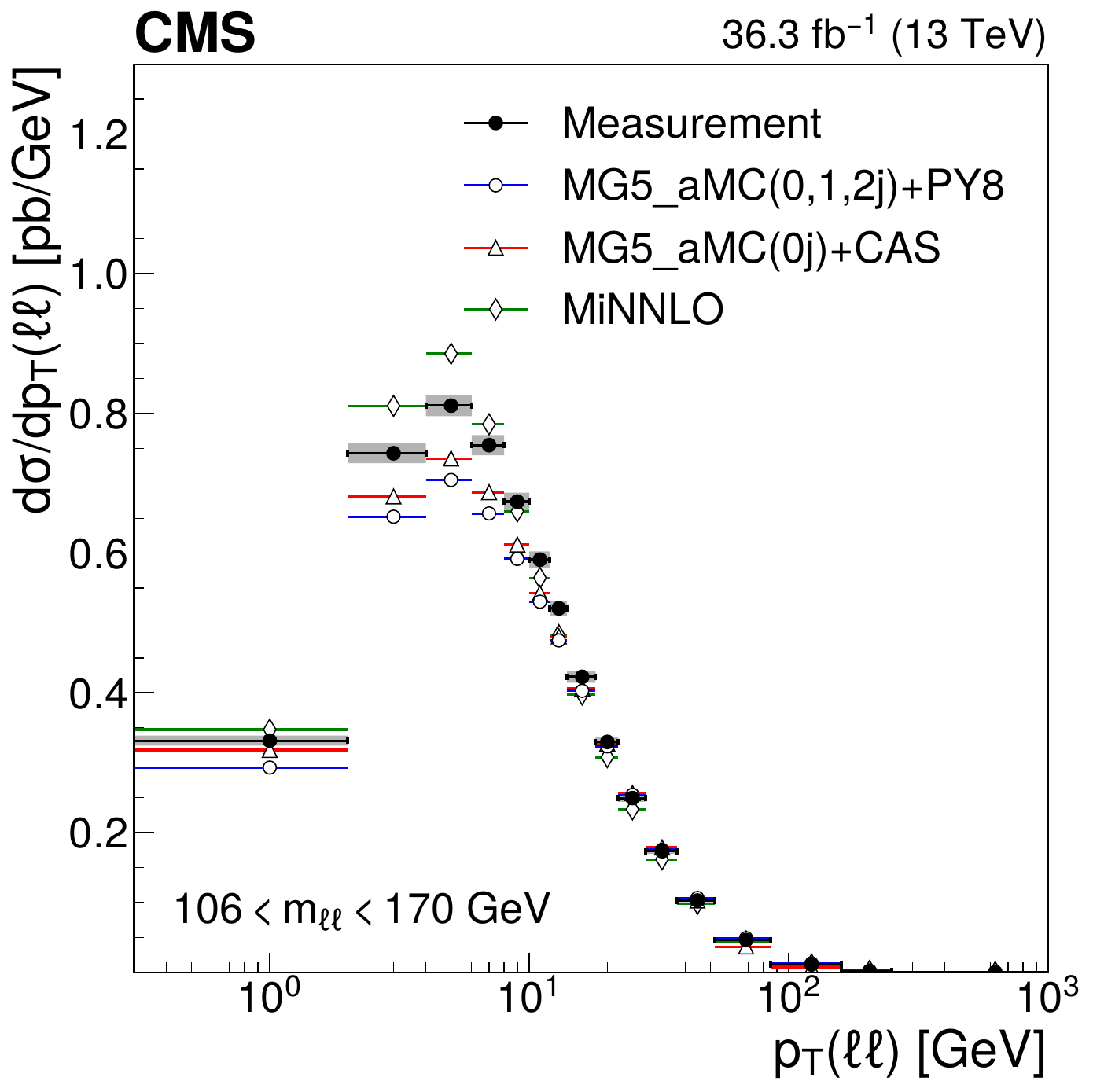}
 \includegraphics[height=0.40\textwidth]{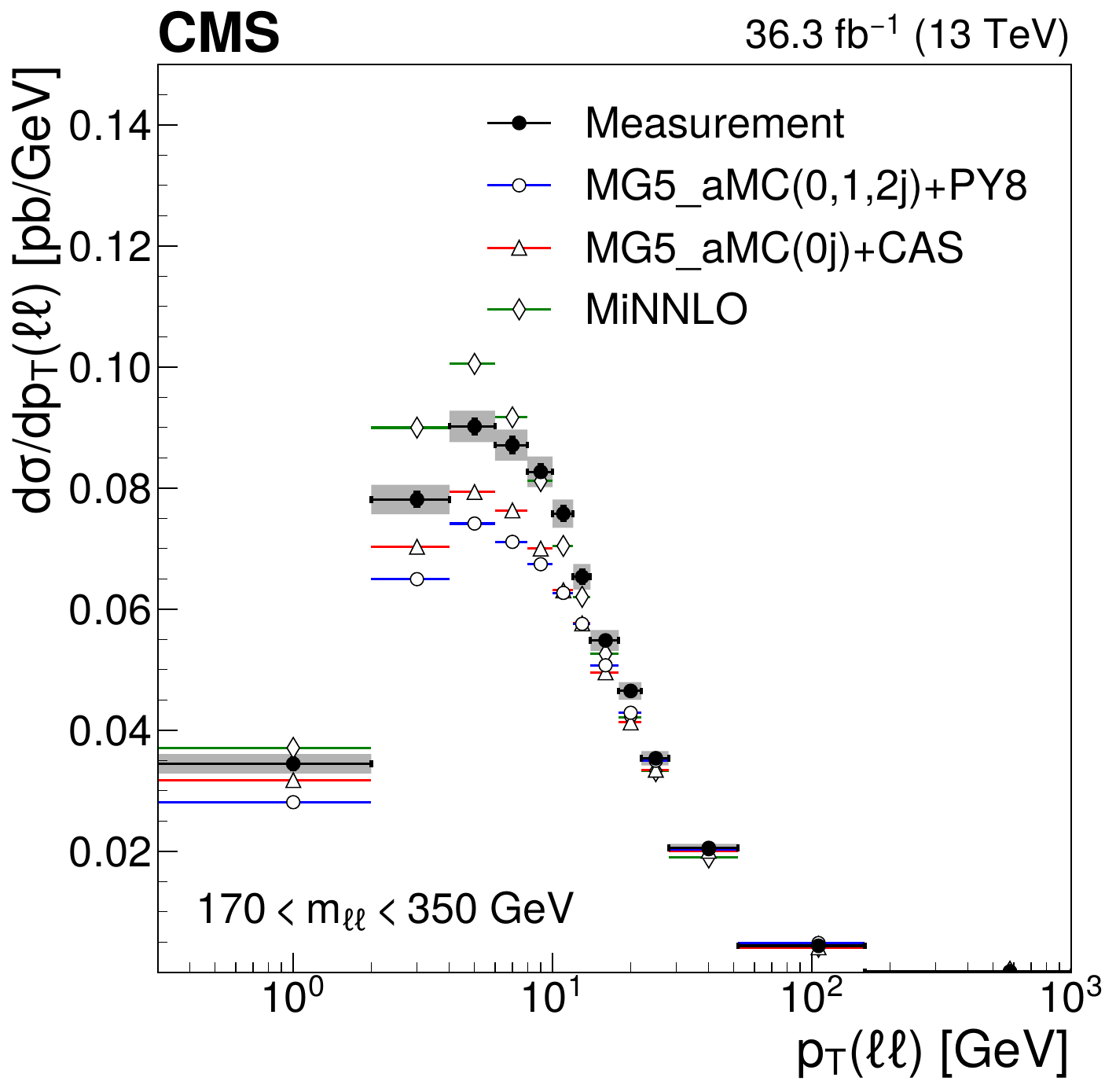}
 \includegraphics[height=0.40\textwidth]{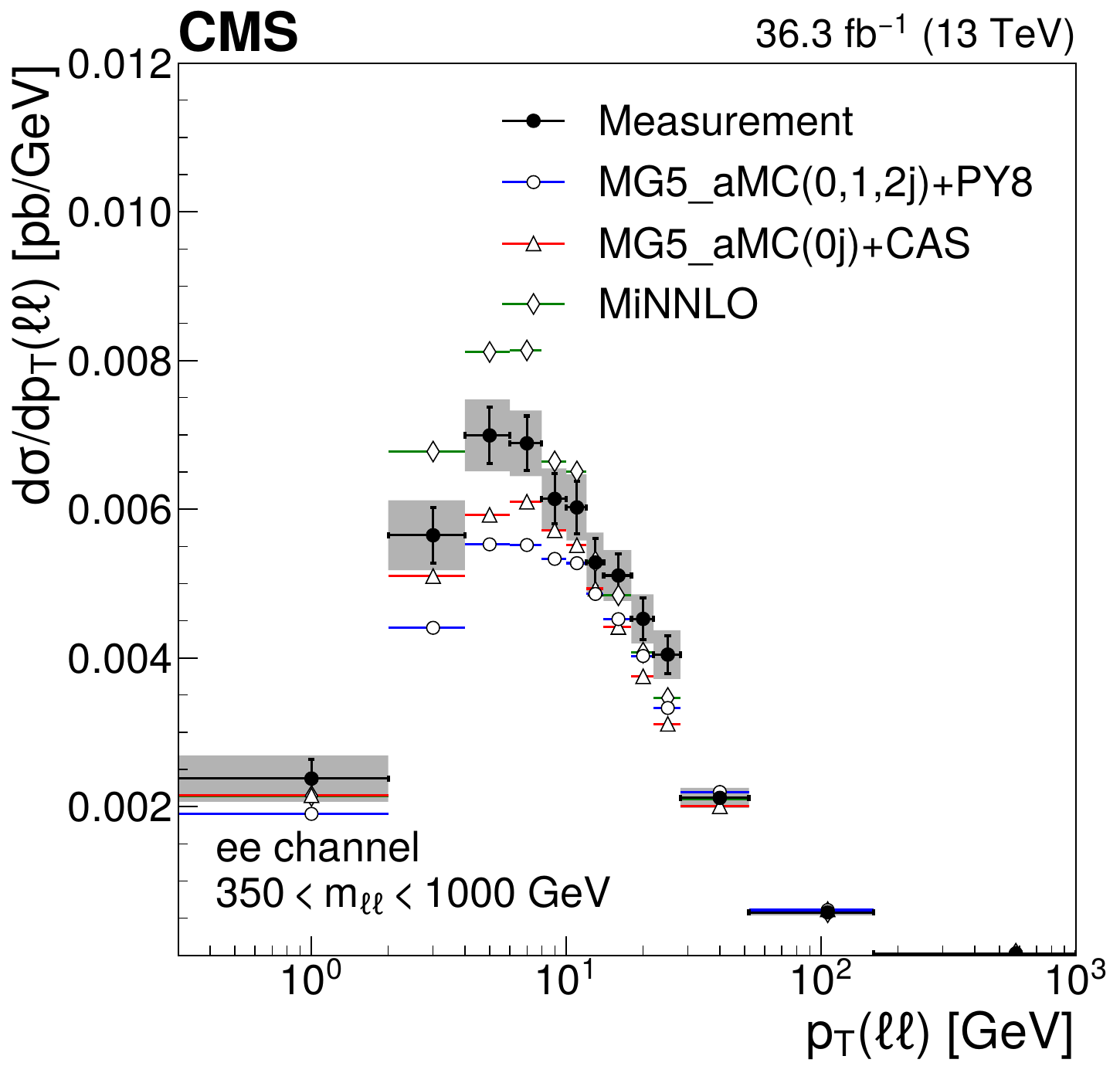}

 \caption{Differential cross sections in \PTll in various invariant mass ranges: 
         \mrangea (upper left), \mrangeb (upper right), \mrangec 
         (middle left), \mranged (middle right), and \mrangee (lower).
         The error bars on data points (black dots) correspond to the statistical uncertainty of the measurement and the 
         shaded bands around the data points correspond to the total experimental uncertainty.
         The measurement is compared with \MGaMC (0, 1, and 2 jets at NLO) + \PYTHIA8 (blue dots), \minnlo (green diamonds) 
         and \MGaMC (0 jet at NLO)+ PB (\cascade) (red triangles). 
        }
 \label{fig:UnfCombPt0}
\end{figure*}

\begin{figure*}[htbp!]
  \centering
  \includegraphics[width=0.48\textwidth]{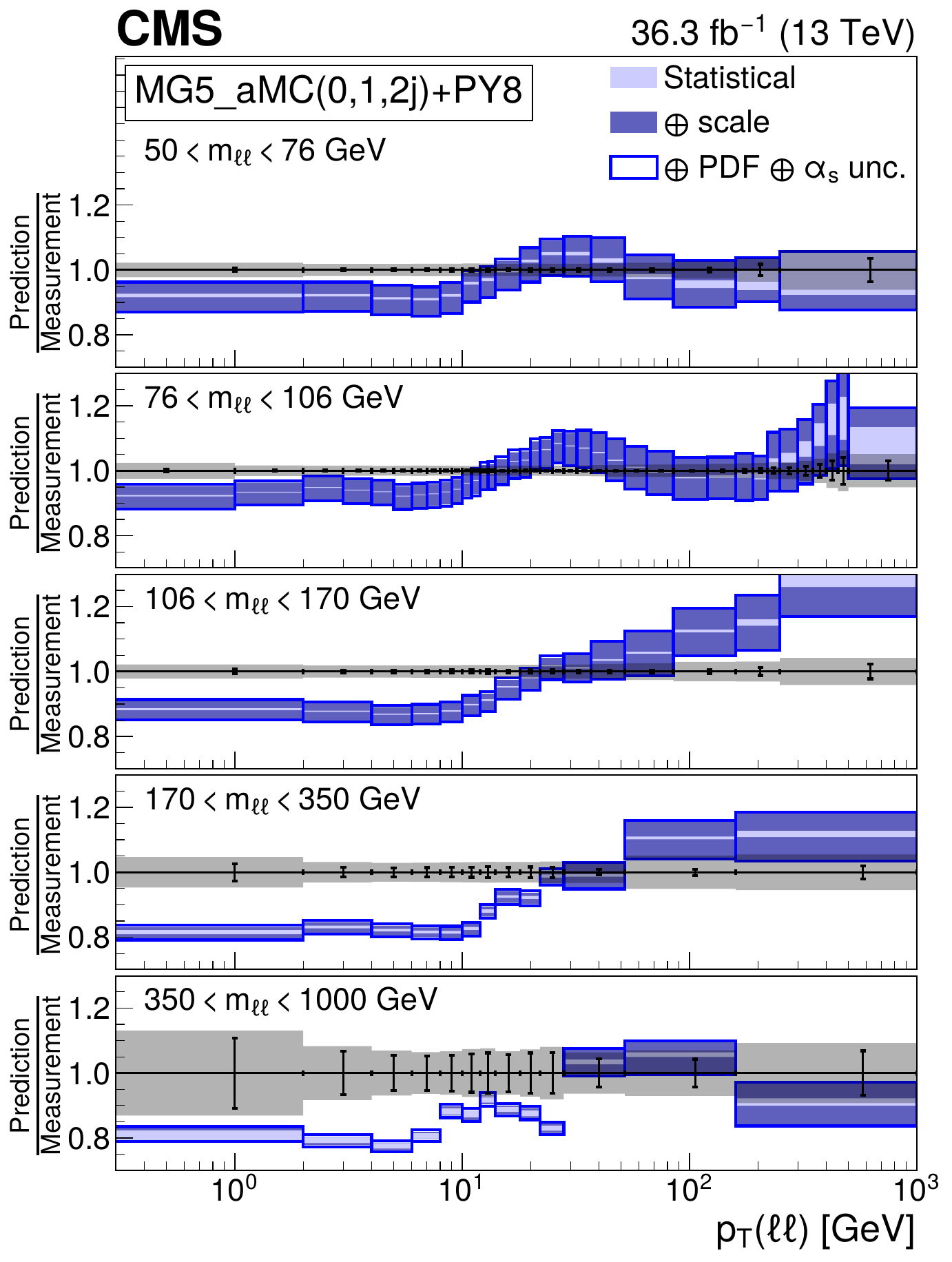}
  \includegraphics[width=0.48\textwidth]{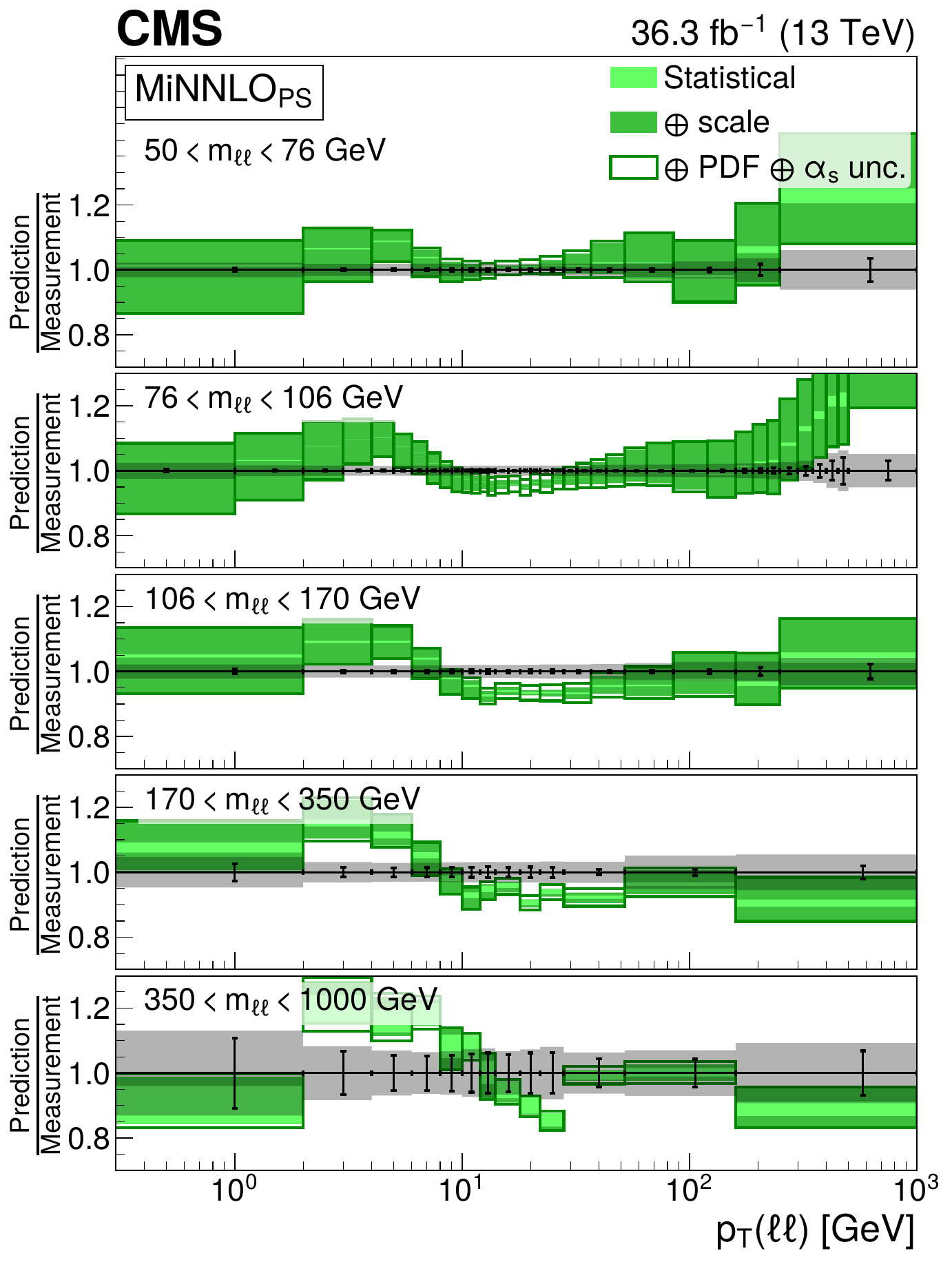}
  \caption{Comparison to Monte Carlo predictions based on a matrix element with parton shower merging. 
           The ratio of \MGaMC (0, 1, and 2 jets at NLO) + \PYTHIA8 (left) and \minnlo (right) predictions to the measured differential cross sections
           in \PTll are presented for various \Mll ranges.
           The error bars correspond to the statistical uncertainty of the measurement and the shaded bands to the total experimental uncertainty.
           The light color band corresponds to the statistical uncertainty of the simulation and the dark color band includes 
           the scale uncertainty. The largest bands include PDF and \alpS uncertainties, added in quadrature.  
          }
 \label{fig:UnfCombPt0a}
\end{figure*}

\begin{figure*}[htbp!]
  \centering
  \includegraphics[width=0.48\textwidth]{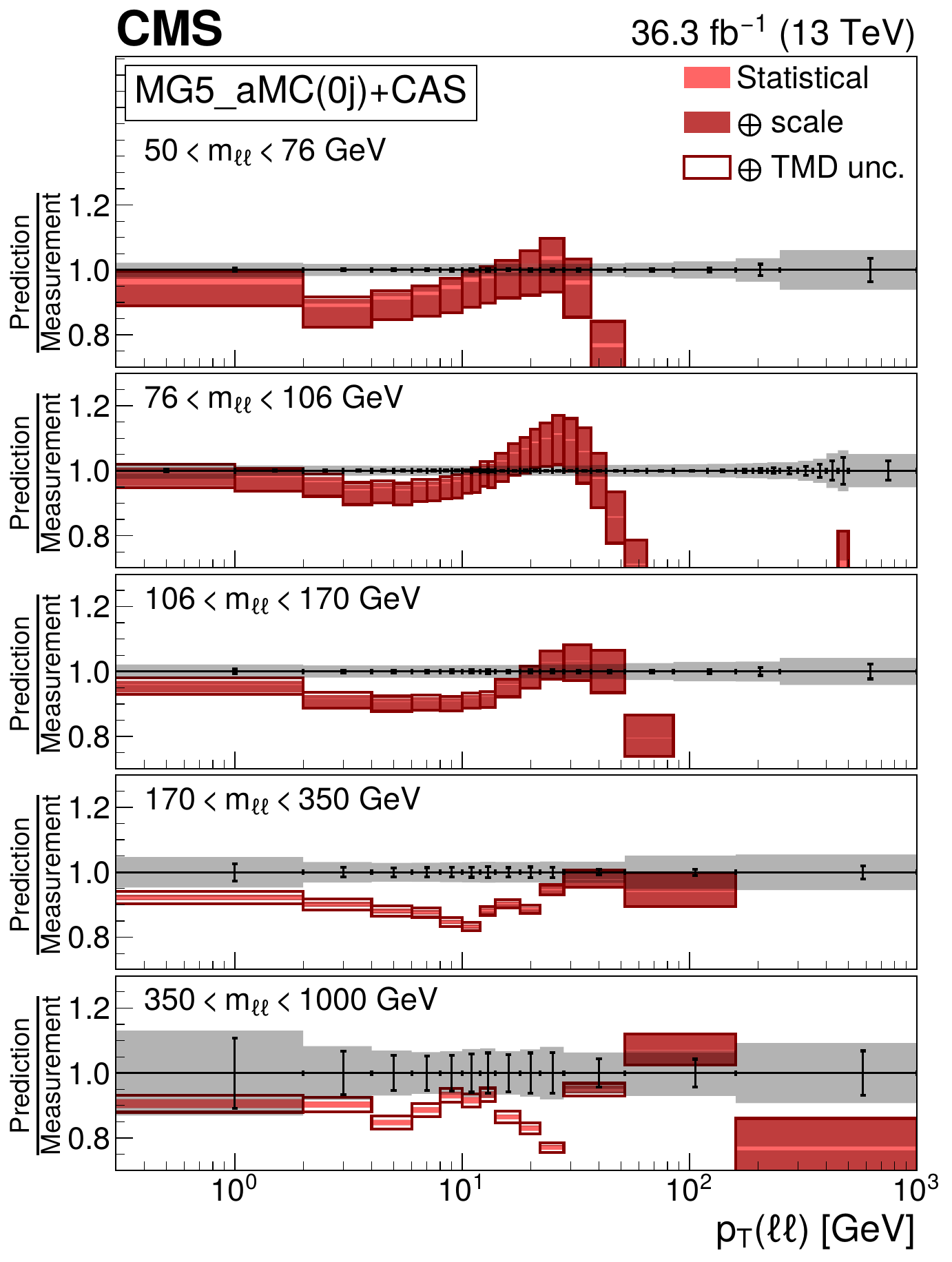}
  \includegraphics[width=0.48\textwidth]{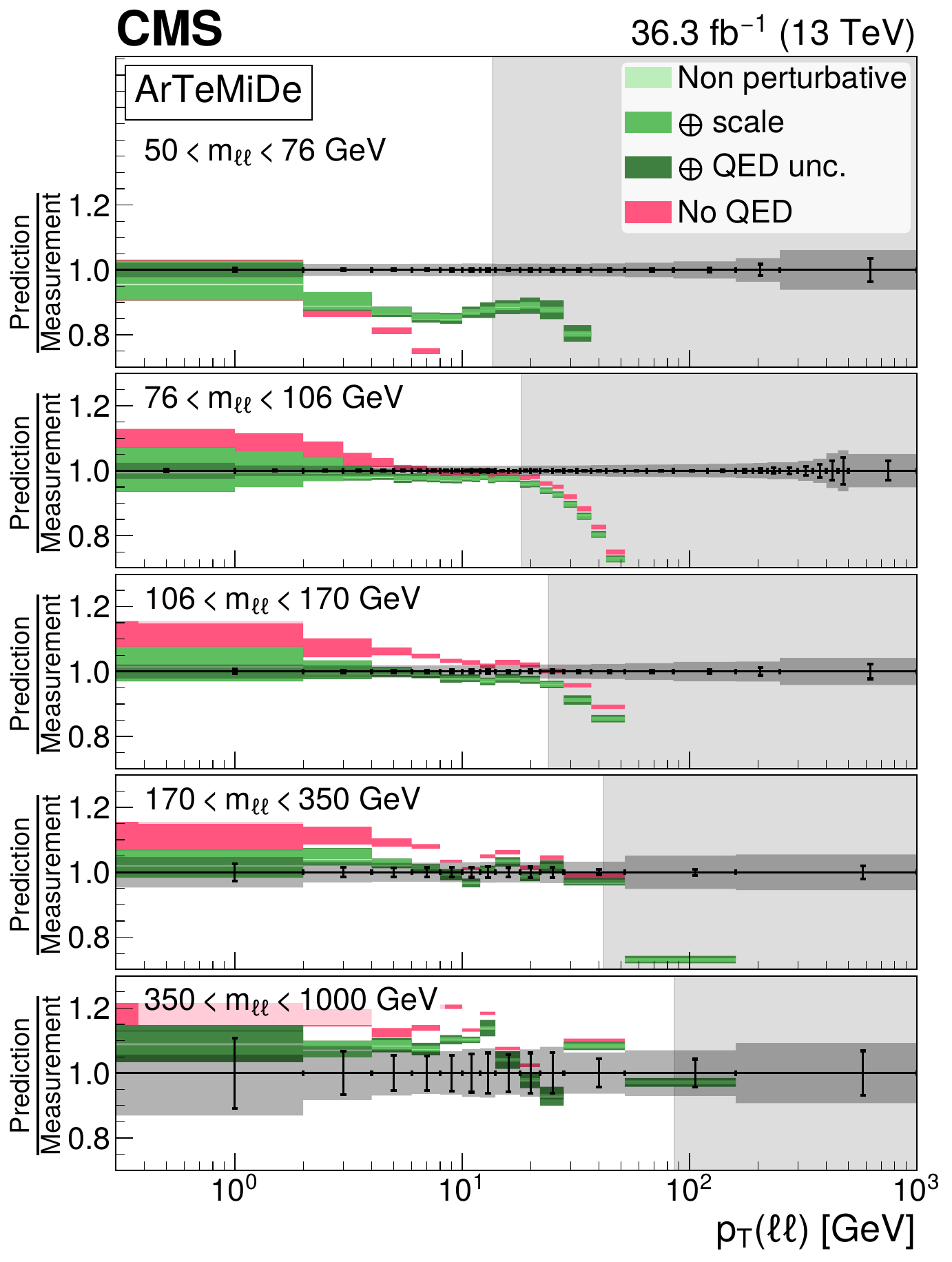}
  \caption{Comparison to TMD based predictions.
           The ratio of \MGaMC (0 jet at NLO) + PB (\cascade) (left) and \artemide (right) predictions to the measured differential cross sections
           in \PTll are presented for various \Mll ranges.
           The error bars correspond to the statistical uncertainty of the measurement and the shaded bands to the total experimental uncertainty.
           The light (dark) green band around \artemide predictions represent the nonperturbative (QCD scale) uncertainties, 
           the darker green representing the QED FSR correction uncertainties. 
           The range of invalidity is shaded with a gray band.
           The light color band around \cascade prediction corresponds to the statistical uncertainty of the simulation
           and the dark color band includes the scale uncertainty. The largest bands include TMD uncertainty, 
           added in quadrature.  
        }
 \label{fig:UnfCombPt0b}
\end{figure*}

\begin{figure*}[htbp!]
  \centering
  \includegraphics[width=0.48\textwidth]{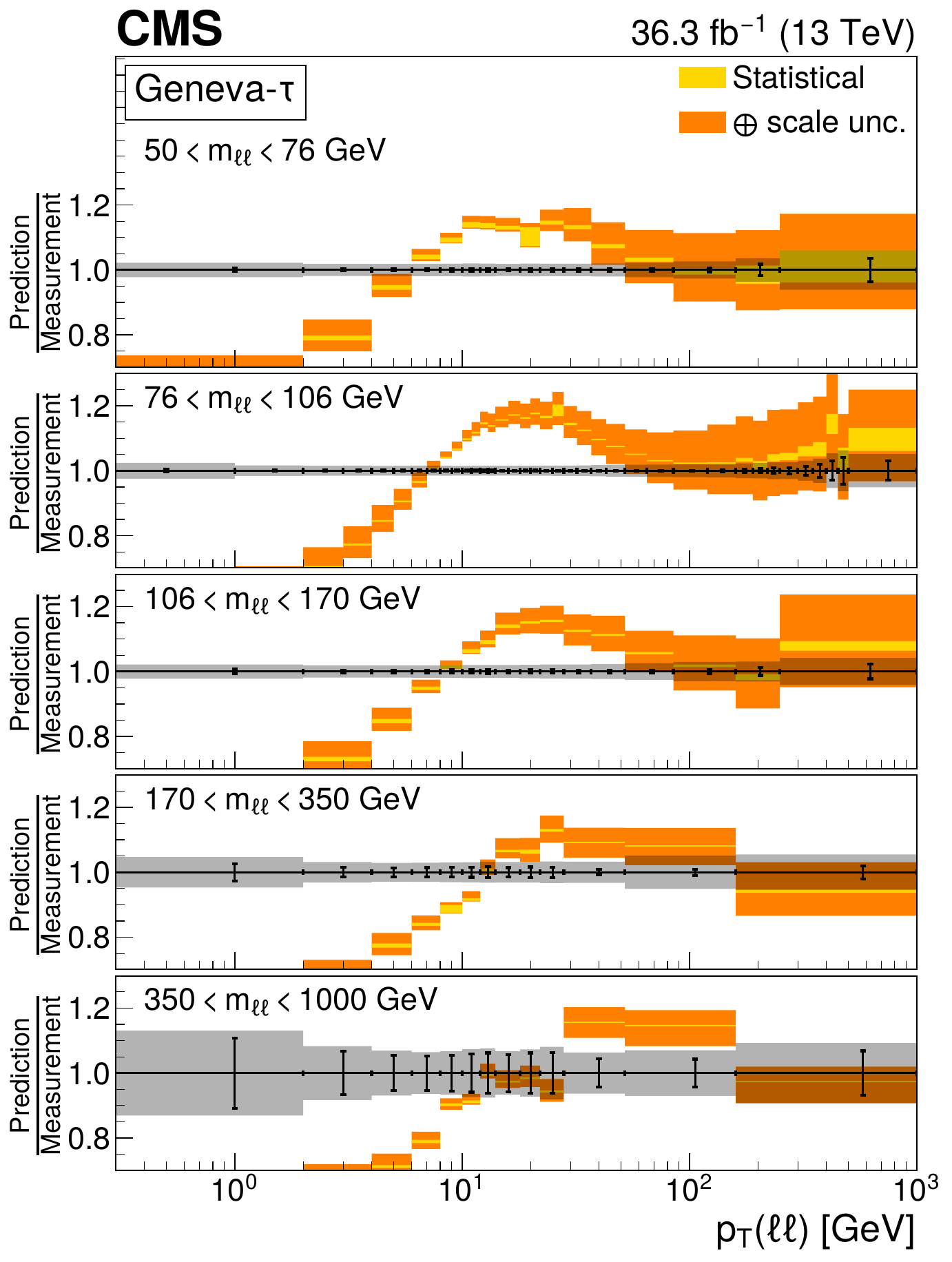}
  \includegraphics[width=0.48\textwidth]{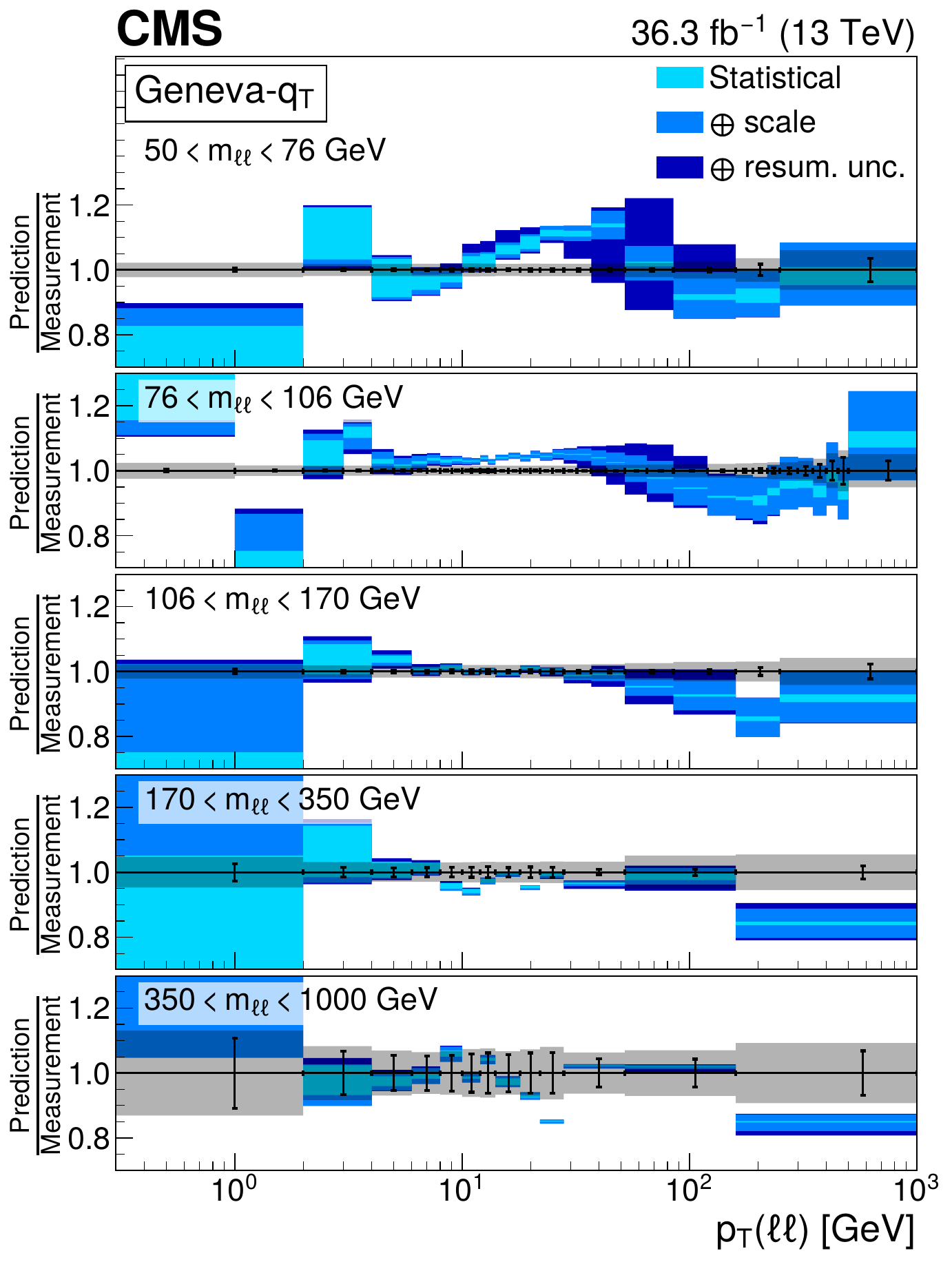}
  \caption{Comparison to resummation based predictions.
           The ratio of \GE-$\tau$ (left) and \GE-\qt (right) predictions to the measured differential cross sections
           in \PTll are presented for various \Mll ranges.
           The error bars correspond to the statistical uncertainty of the measurement and the shaded bands to the total experimental uncertainty.
           The light color bands around the predictions represents the statistical uncertainties and the middle color bands 
           represents the scale uncertainties. The dark outer bands of \GE-\qt prediction represent the resummation uncertainties.
          }
 \label{fig:UnfCombPt0c}
\end{figure*}

\begin{figure*}[htbp!]
    \flushright
    \includegraphics[height=0.47\textwidth]{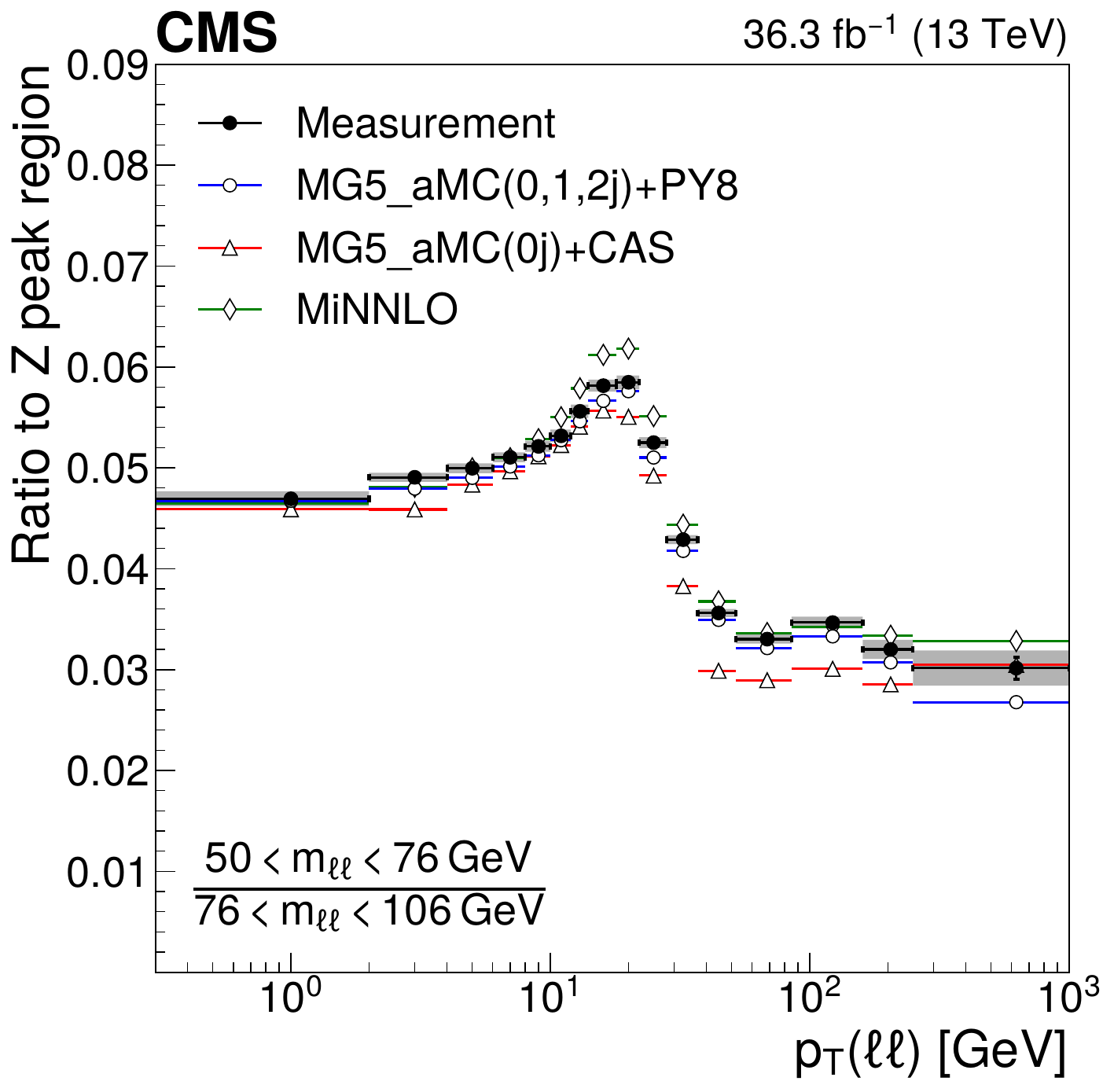}
    \hspace{1.6mm}
    \includegraphics[height=0.47\textwidth]{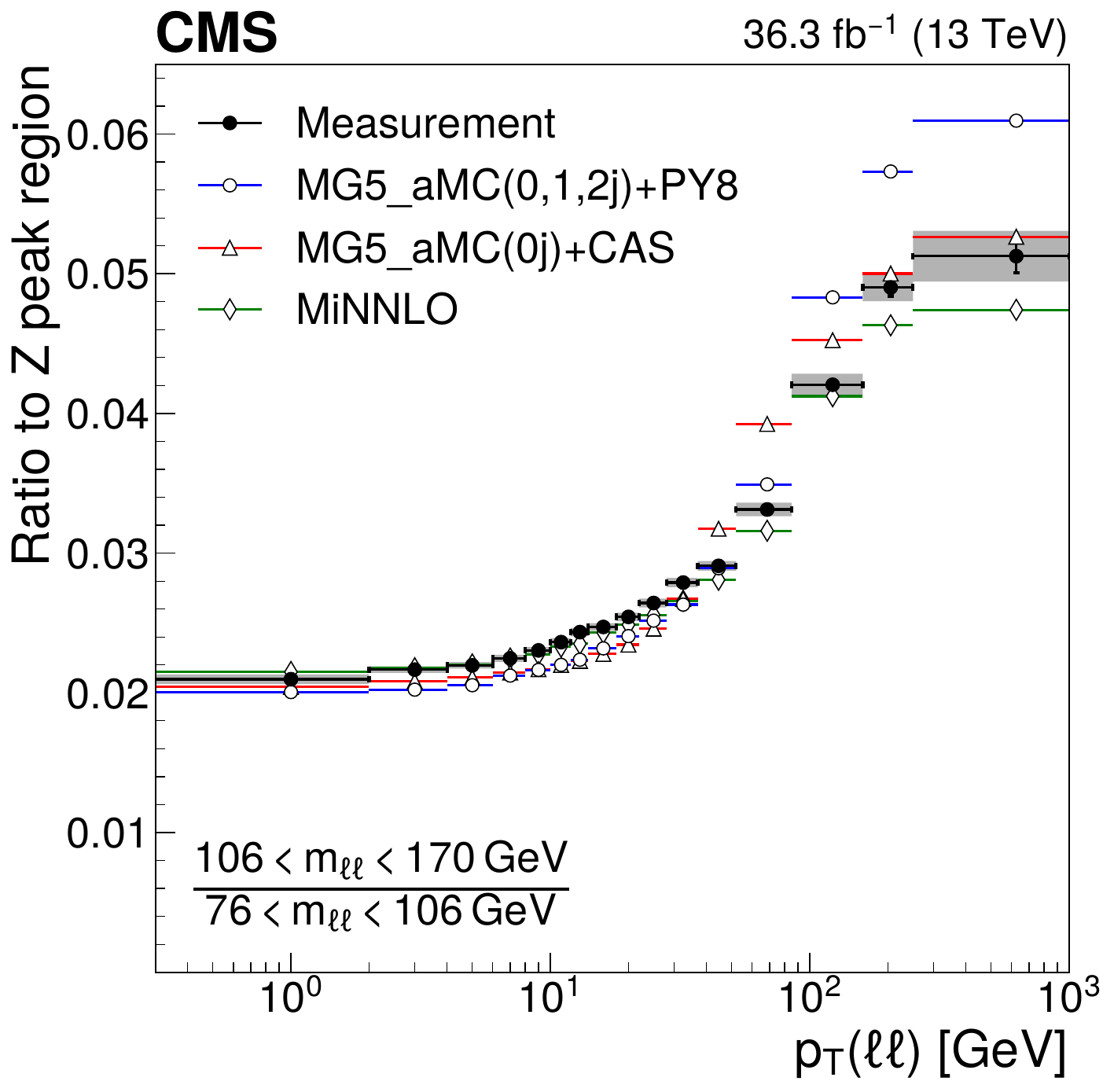}
    \includegraphics[height=0.47\textwidth]{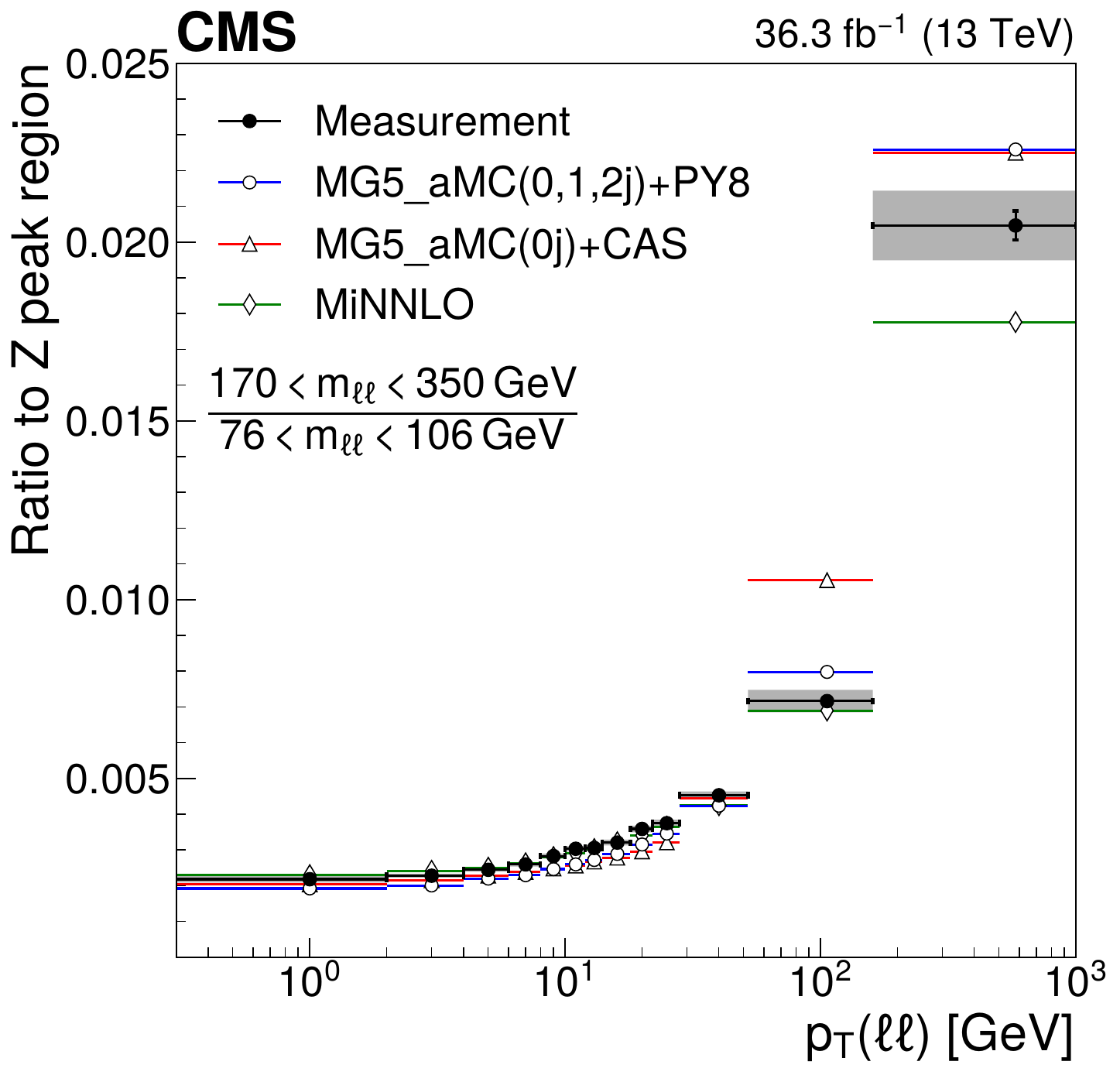}
    \includegraphics[height=0.47\textwidth]{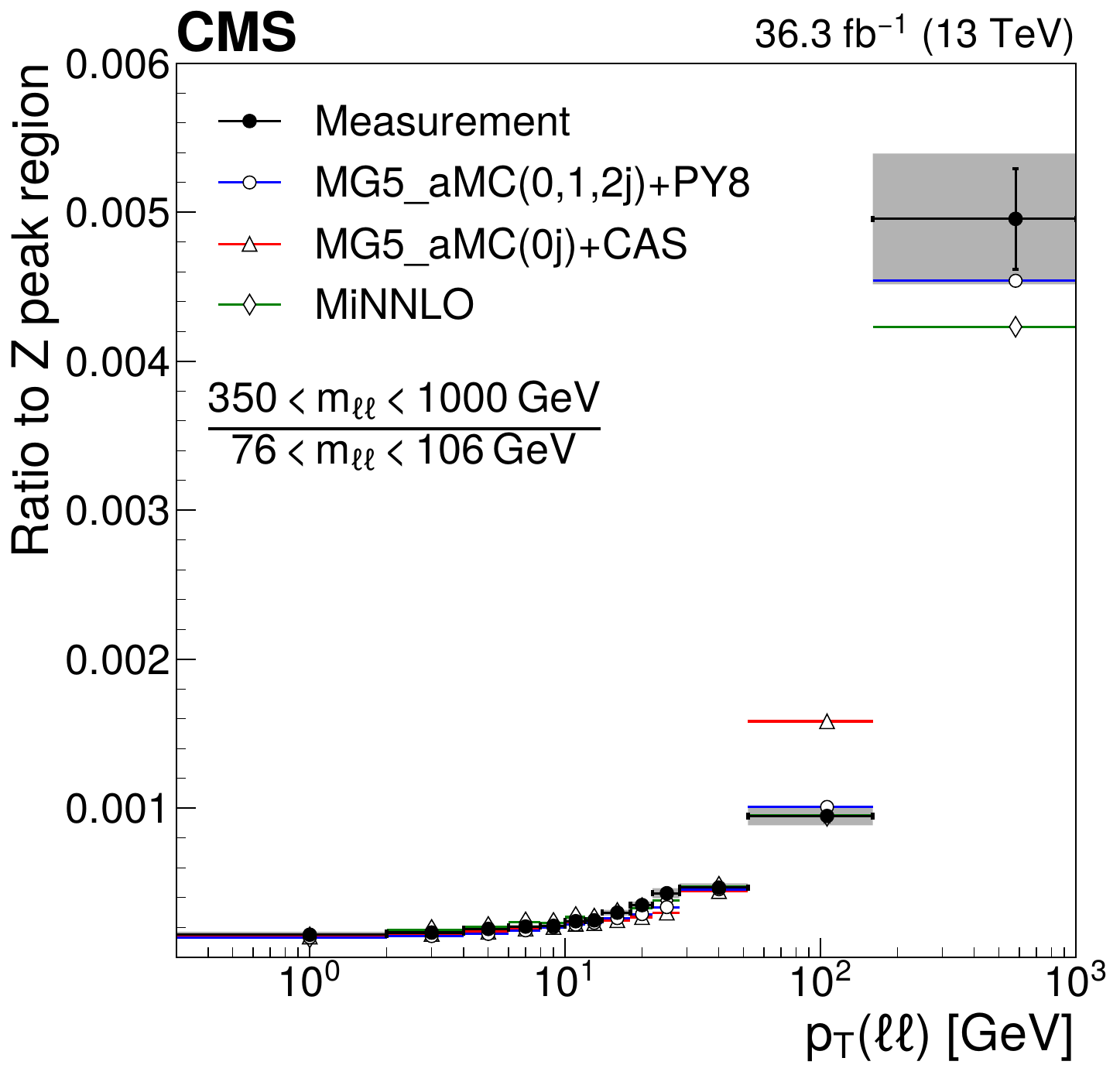}
    \caption{
        Ratios of differential cross sections in \PTll for invariant mass ranges 
        with respect to the peak region \mrangeb:
         \mrangea (upper left), \mrangec (upper right), \mranged (lower left),
        and \mrangee (lower right).
        Details on the presentation of the results are given in Fig.~\ref{fig:UnfCombPt0} caption.
        }
    \label{fig:UnfCombPt0Ratios}
\end{figure*}

\begin{figure*}[htbp!]
  \centering
  \includegraphics[width=0.48\textwidth]{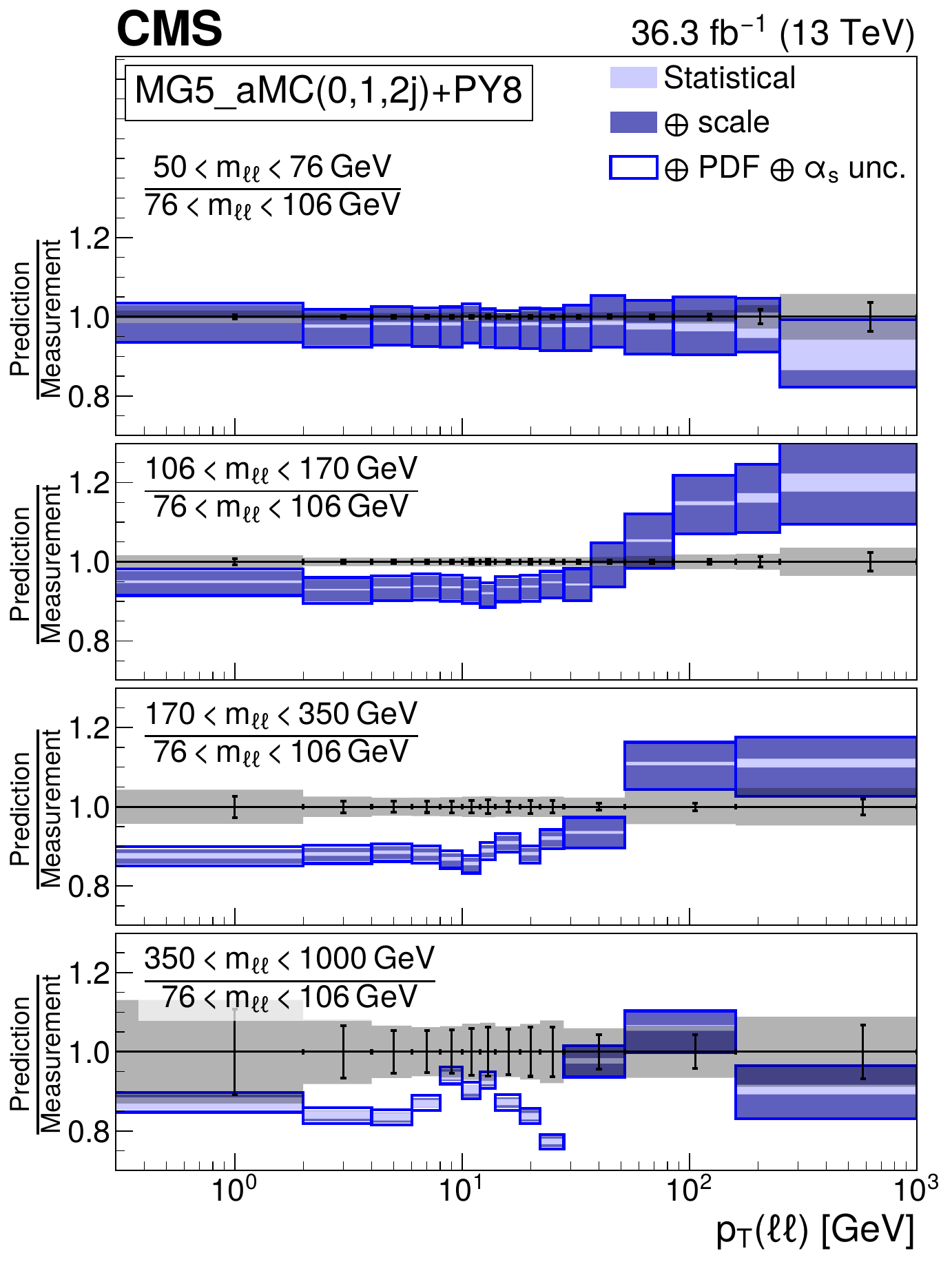}
  \includegraphics[width=0.48\textwidth]{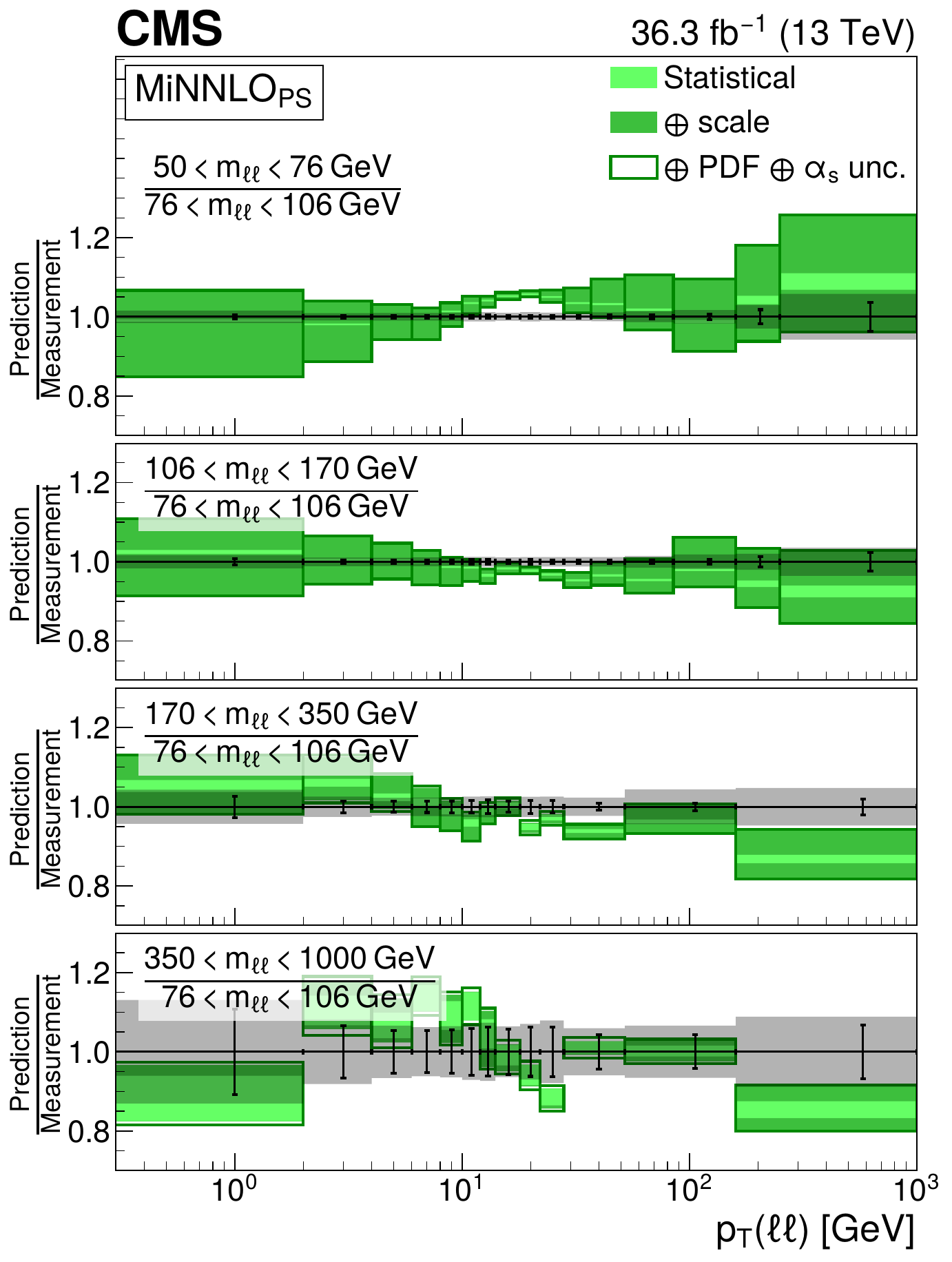}
  \caption{Comparison to Monte Carlo predictions based on a matrix element with parton shower merging. 
   The distributions show the ratio of differential cross sections as a function of \PTll for a given \Mll range to the cross section at the peak region \mrangeb.
   The predictions are \MGaMC (0, 1, and 2 jets at NLO) + \PYTHIA8 (left) and \minnlo (right).
   Details on the presentation of the results are given in Fig.~\ref{fig:UnfCombPt0a} caption.
          }
 \label{fig:UnfCombPt0Ratiosa}
\end{figure*}

\begin{figure*}[htbp!]
  \centering
  \includegraphics[width=0.48\textwidth]{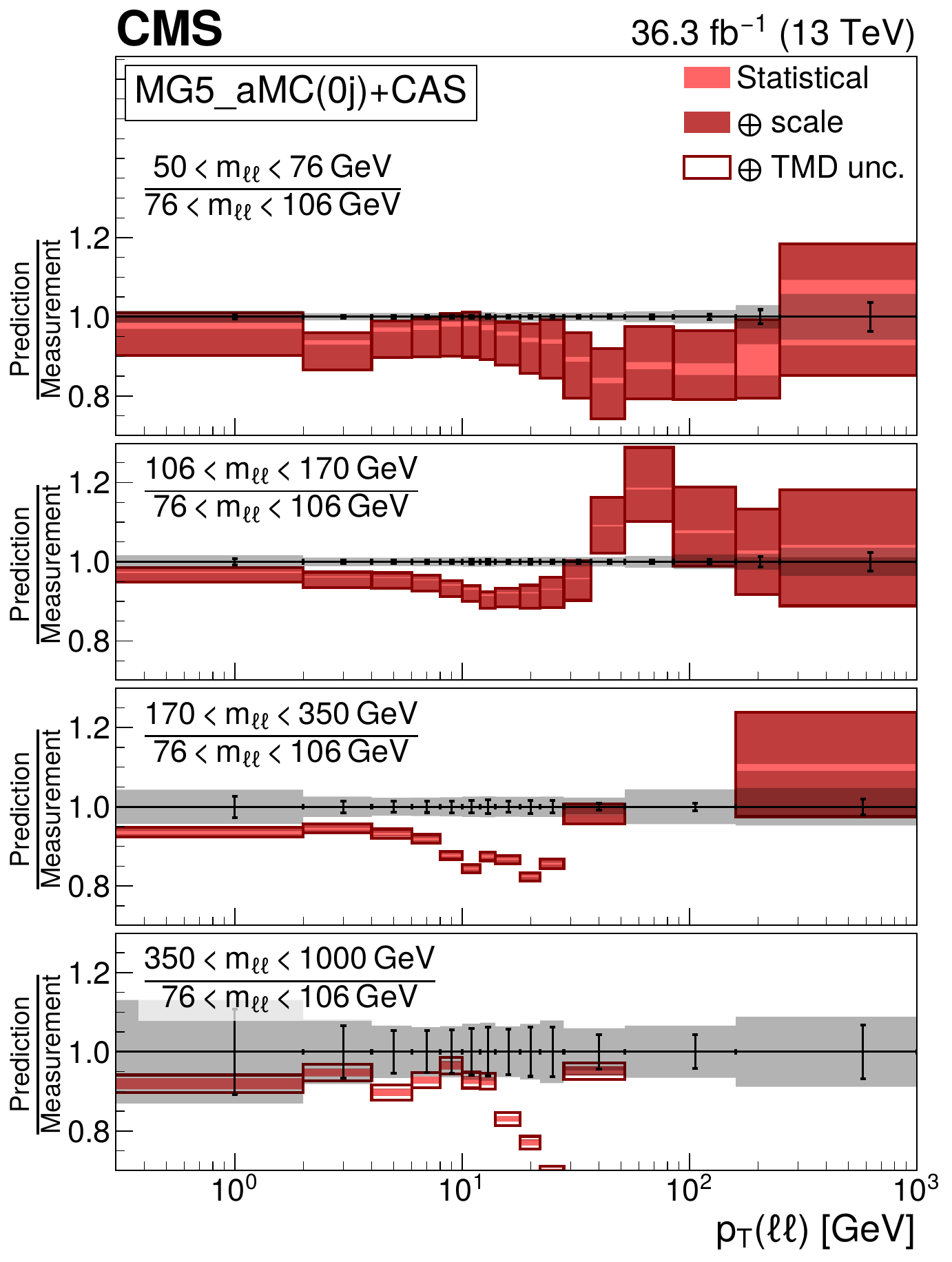}
  \includegraphics[width=0.48\textwidth]{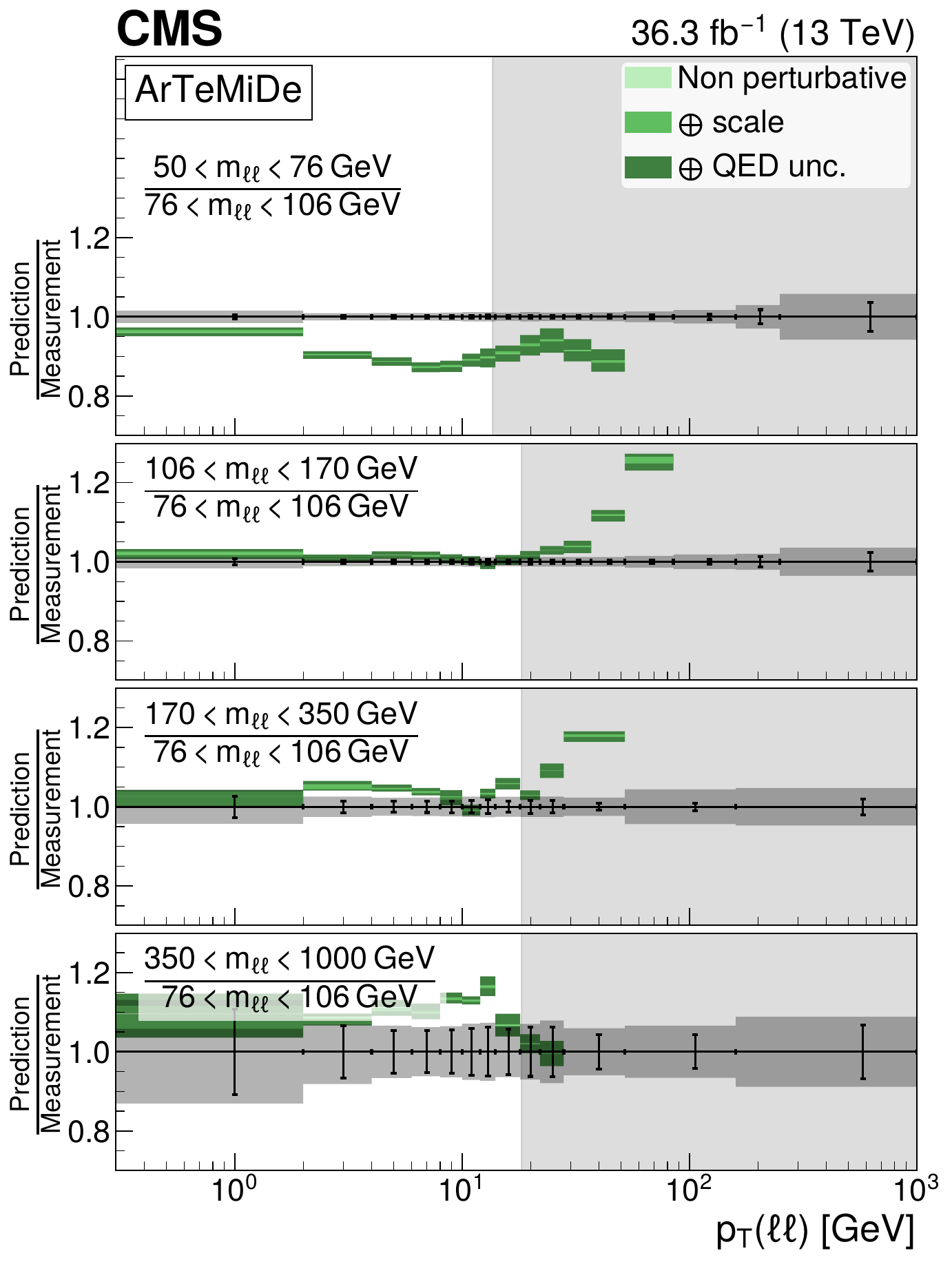}
  \caption{Comparison to TMD based predictions.
      The distributions show the ratio of differential cross sections as a function of \PTll for a given \Mll range to the cross section at the peak region \mrangeb.
   The predictions are \MGaMC (0 jet at NLO) + PB (\cascade) (left) and \artemide (right).
   Details on the presentation of the results are given in Fig.~\ref{fig:UnfCombPt0b} caption.}
 \label{fig:UnfCombPt0Ratiosb}
\end{figure*}

\begin{figure*}[htbp!]
  \centering
  \includegraphics[width=0.48\textwidth]{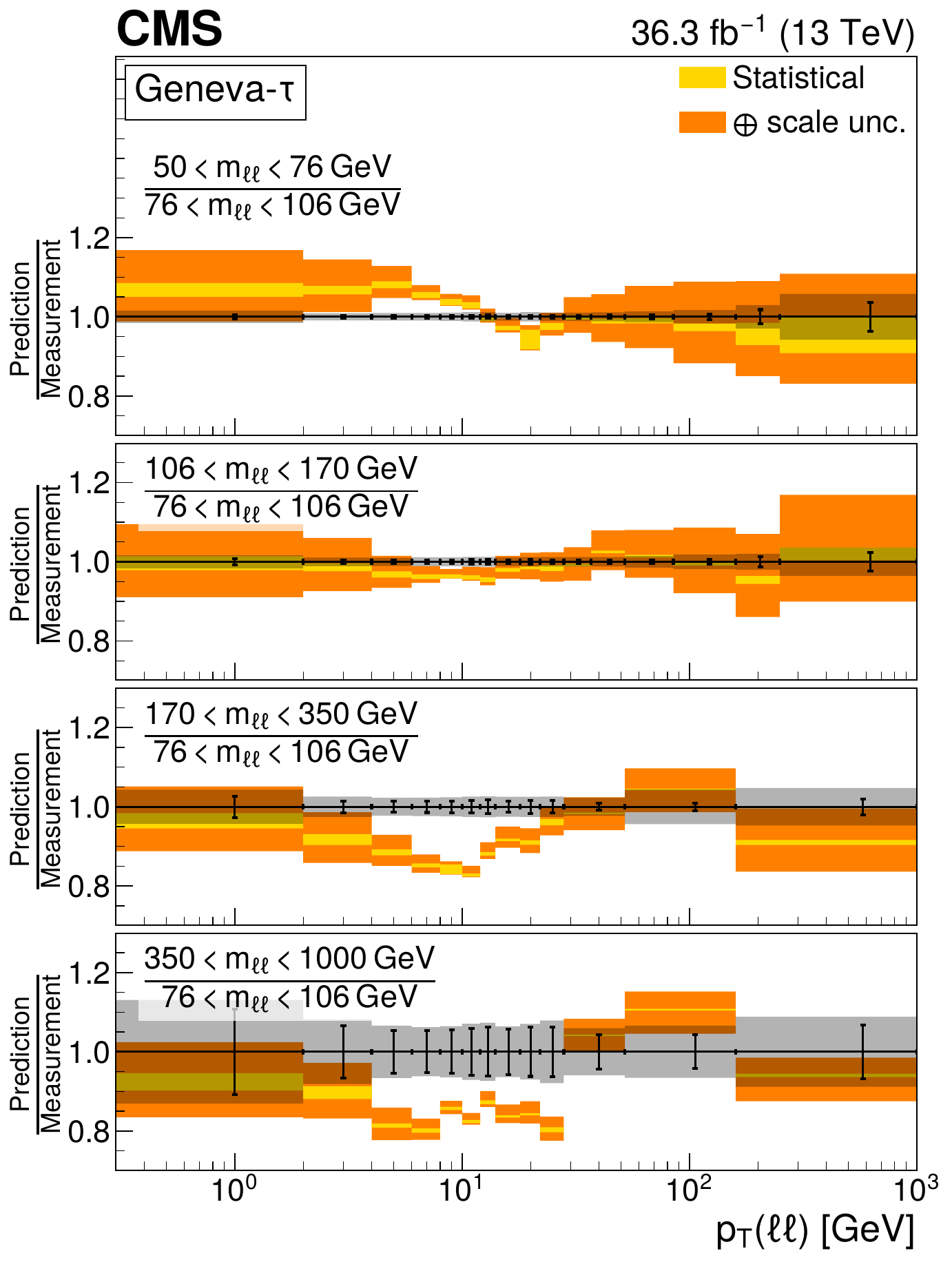}
  \includegraphics[width=0.48\textwidth]{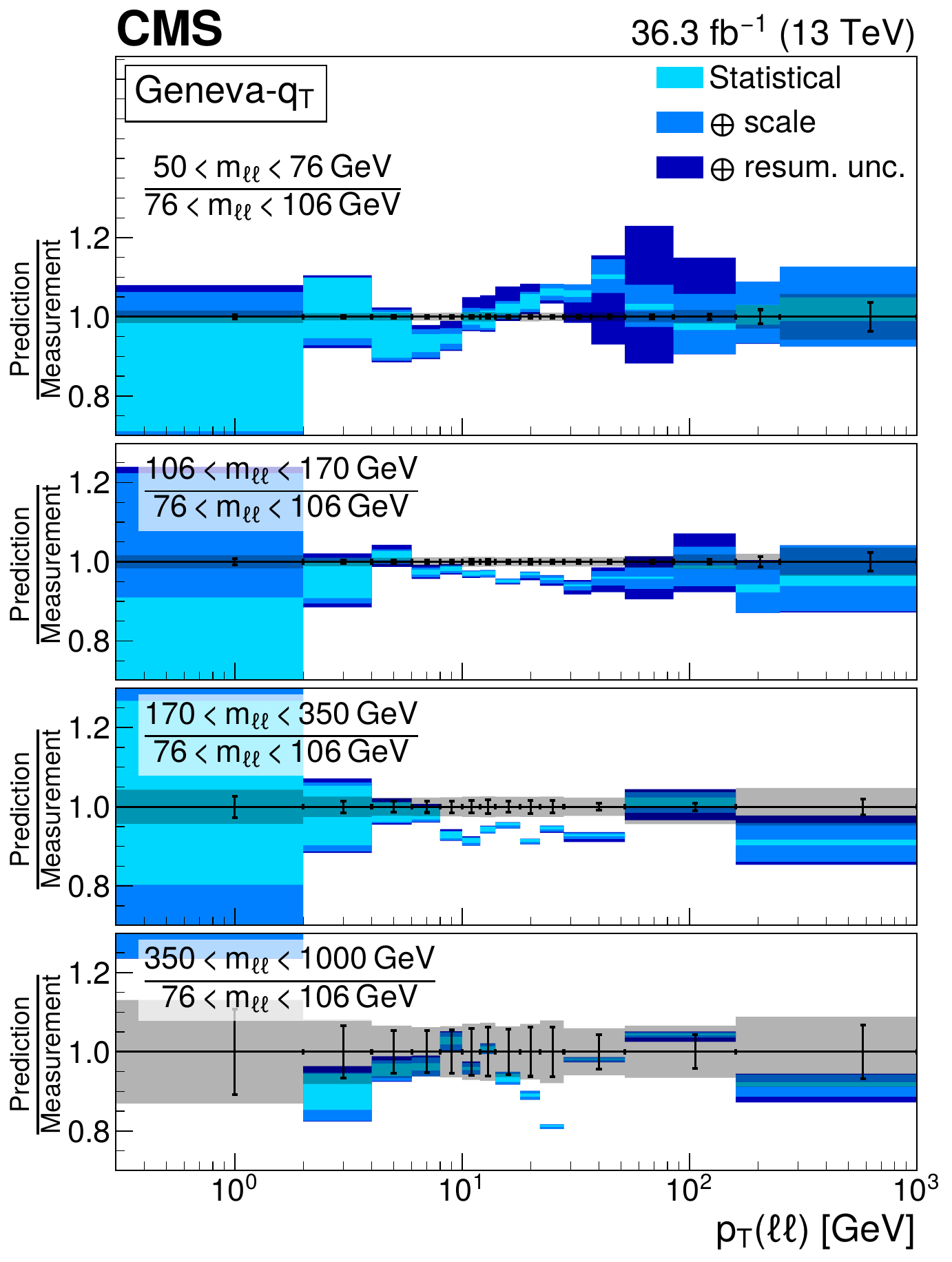}
  \caption{Comparison to resummation based predictions.
      The distributions show the ratio of differential cross sections as a function of \PTll for a given \Mll range to the cross section at the peak region \mrangeb.
   The predictions are \GE-$\tau$ (left) and \GE-\qt (right).
   Details on the presentation of the results are given in Fig.~\ref{fig:UnfCombPt0c} caption.
          }
 \label{fig:UnfCombPt0Ratiosc}
\end{figure*}

The differential cross sections in \PTll are shown in Fig.~\ref{fig:UnfCombPt0} for invariant mass
ranges between 50\GeV and 1\TeV. Because of the lack of precision of the muon transverse momentum measurement
at high \pt, the cross section measurement in the highest mass range is based on the electron channel only.
The ratio of the predictions to the data are presented in Figs.~\ref{fig:UnfCombPt0a},~\ref{fig:UnfCombPt0b} 
and~\ref{fig:UnfCombPt0c}.  
The comparison with different predictions is discussed later in the text.
The ratios of the unfolded distributions for invariant masses outside the \PZ boson peak to the distribution within the \PZ boson
peak (\mrangeb) are shown in Fig.~\ref{fig:UnfCombPt0Ratios}, and the comparisons to predictions in Figs.~\ref{fig:UnfCombPt0Ratiosa},~\ref{fig:UnfCombPt0Ratiosb} and~\ref{fig:UnfCombPt0Ratiosc}.

\begin{figure*}[htbp!]
  \includegraphics[height=0.48\textwidth]{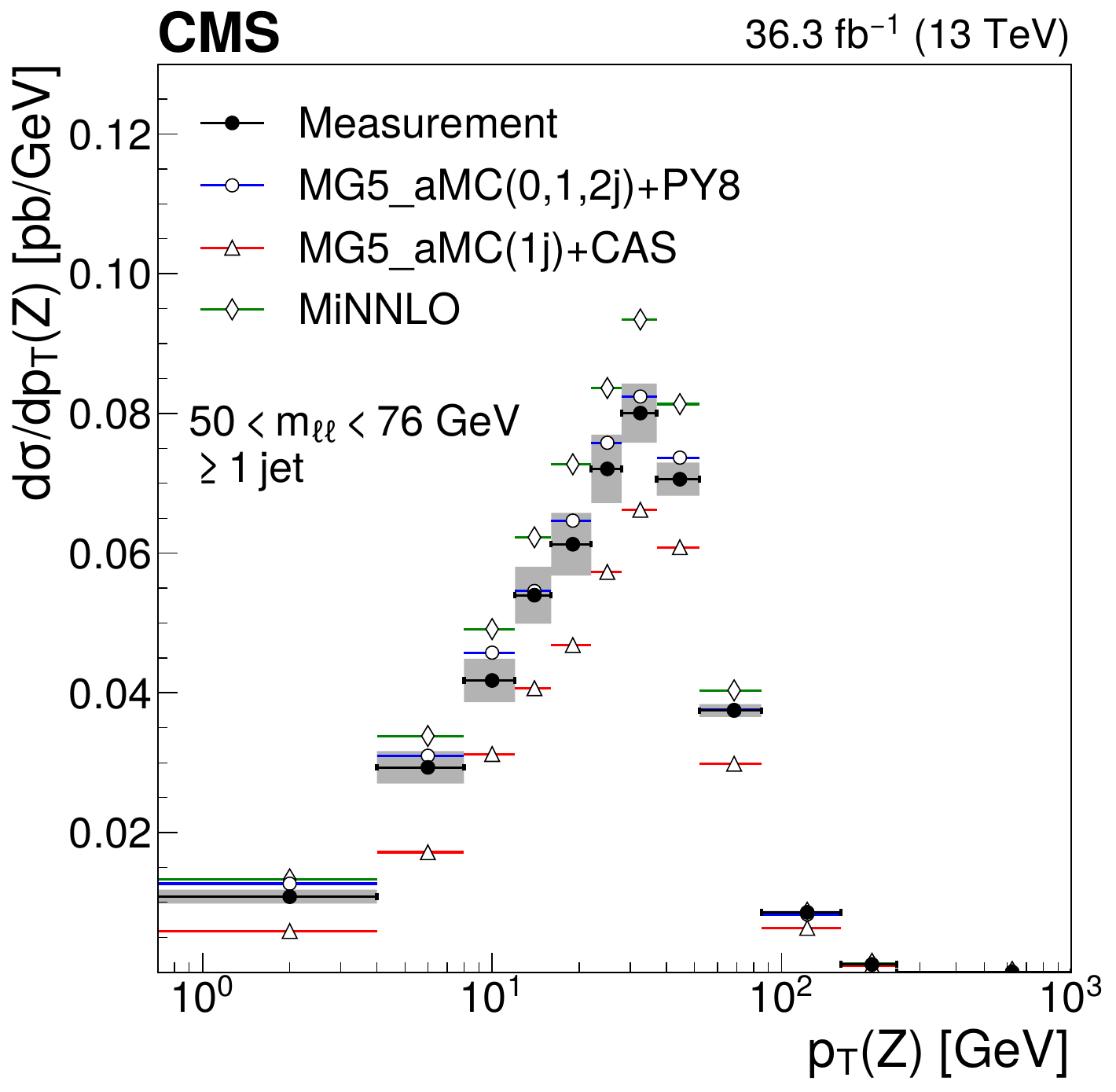}
  \hspace{3.4mm}
  \includegraphics[height=0.48\textwidth]{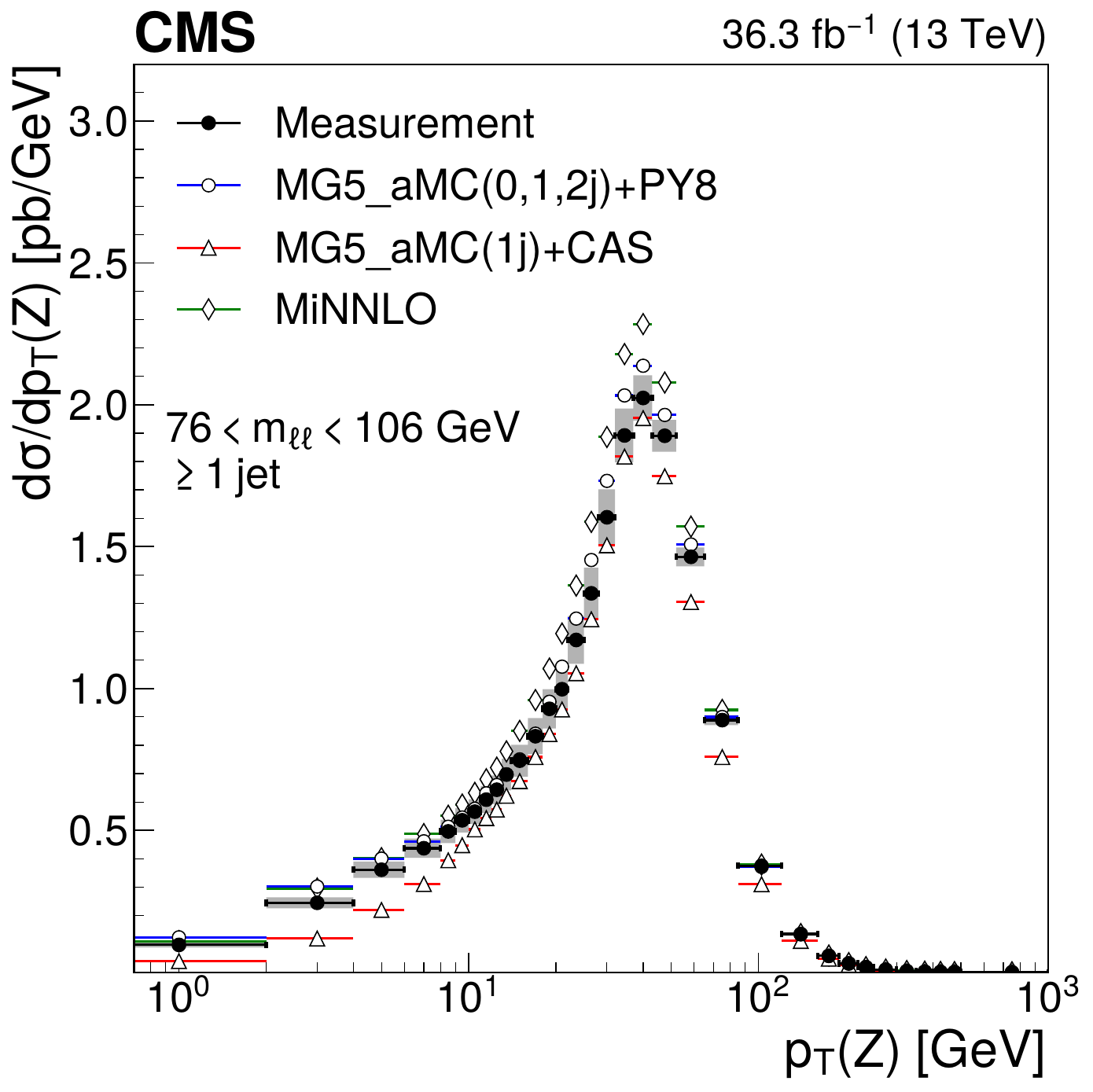}
  \includegraphics[height=0.48\textwidth]{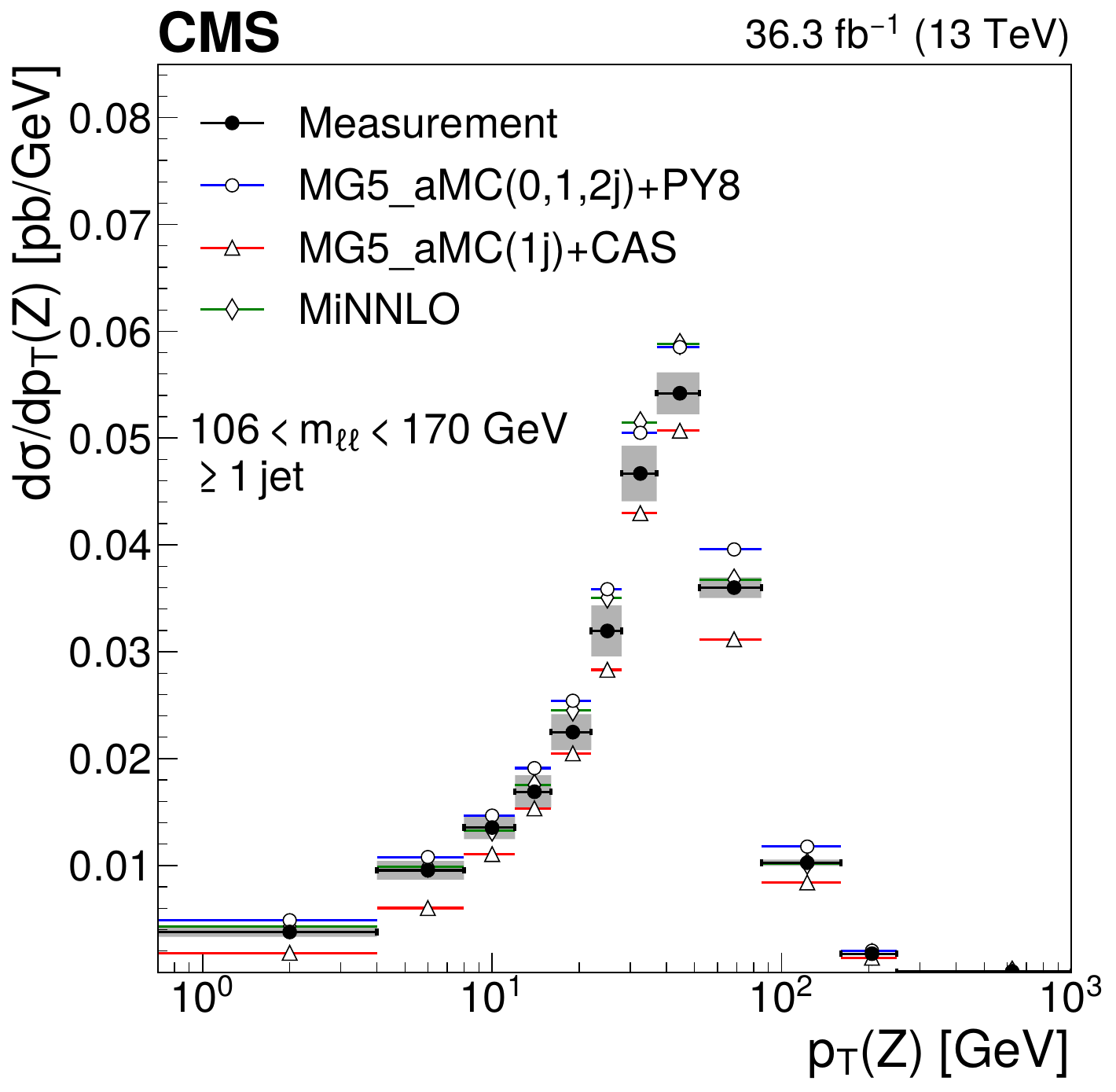}
  \includegraphics[height=0.48\textwidth]{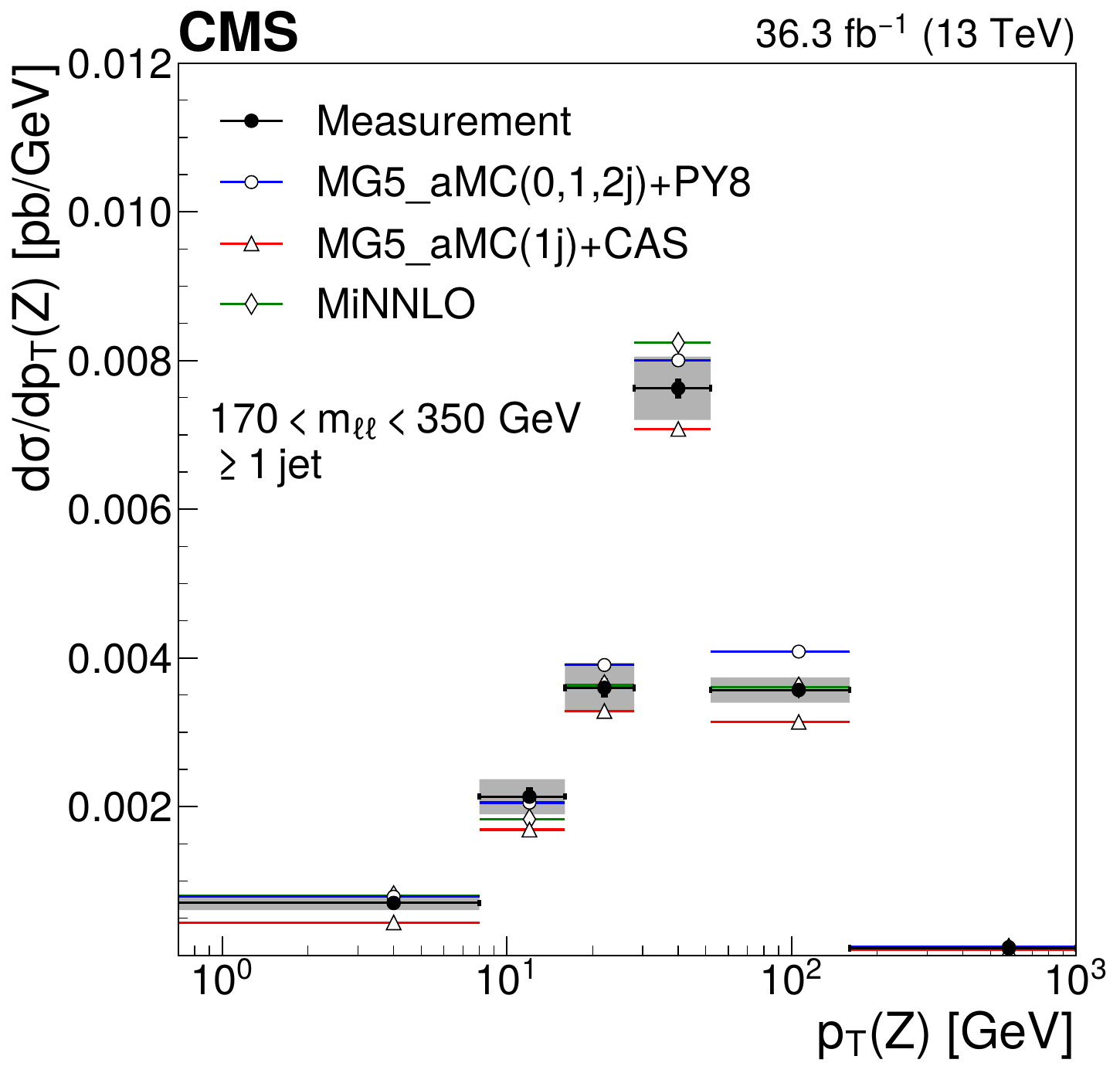}
  \caption{Differential cross sections in \PTll for one or more jets in various invariant mass ranges:
          \mrangea (upper left), \mrangeb (upper right), \mrangec
          (lower left), and \mranged (lower right).
          The measurement is compared with \MGaMC (0, 1, and 2 jets at NLO) + \PYTHIA8 (blue dots), \minnlo (green diamonds) and \MGaMC (1 jet at NLO)+ PB (\cascade) (red triangles). Details on the presentation of the results are given in Fig.~\ref{fig:UnfCombPt0} caption.
         }
  \label{fig:UnfPt1}
\end{figure*}

\begin{figure*}[htbp!]
 \centering
 \includegraphics[width=0.48\textwidth]{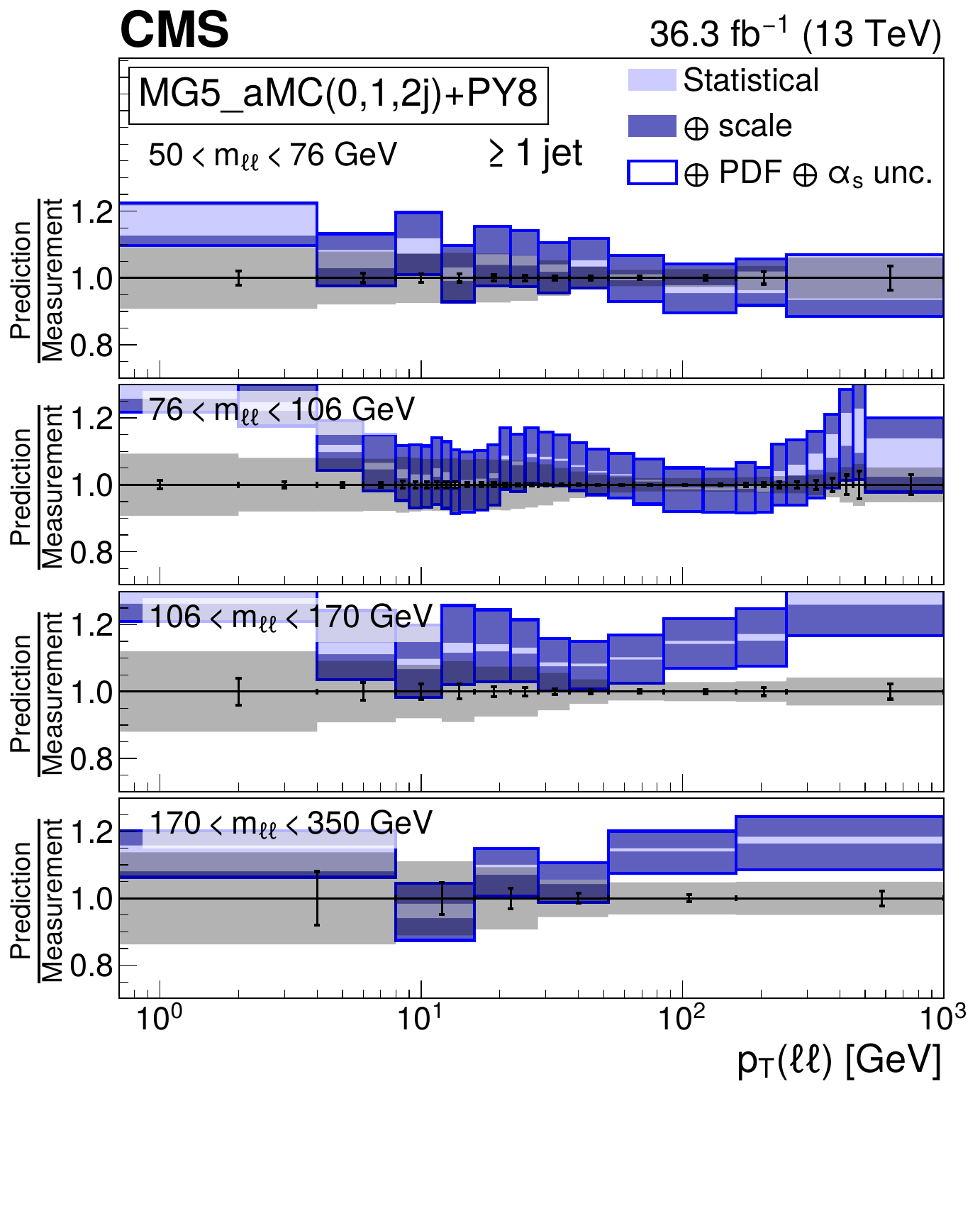}
 \includegraphics[width=0.48\textwidth]{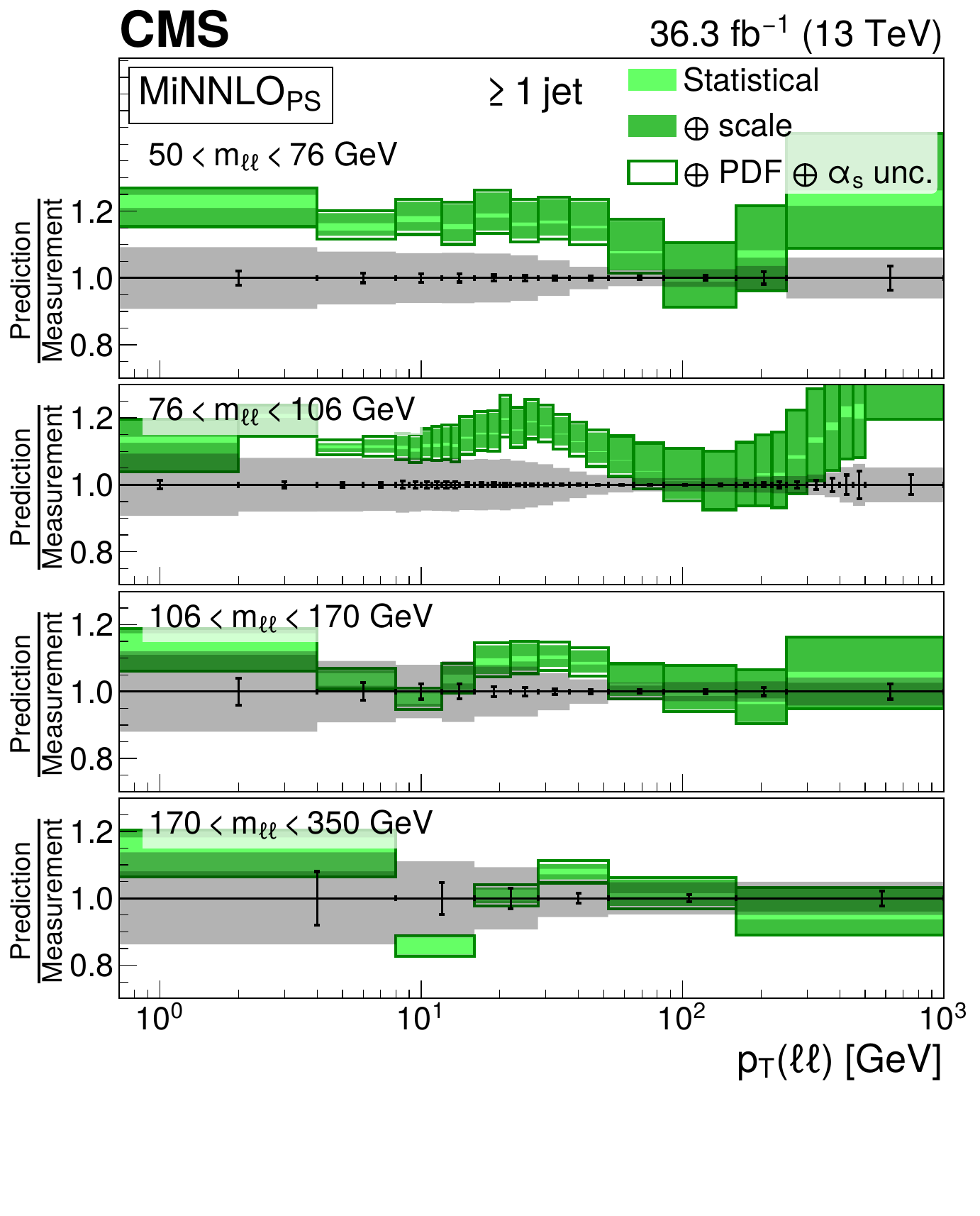}
 \includegraphics[width=0.48\textwidth]{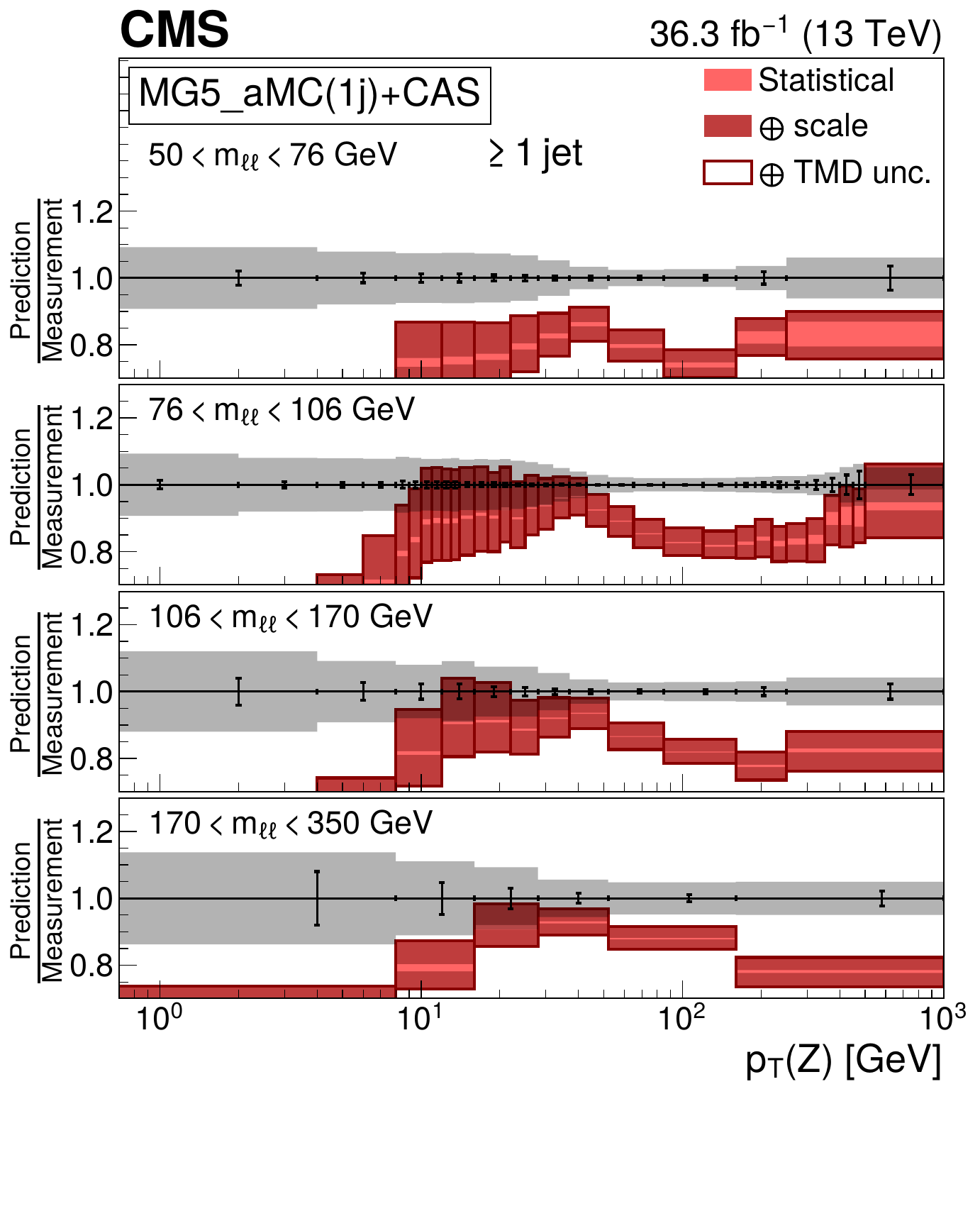}
 \caption{Comparison of the differential cross sections in \PTll to predictions in various 
  invariant mass ranges for the one or more jets case.
  The measurement is compared with \MGaMC (0, 1, and 2 jets at NLO) + \PYTHIA8 (upper left), \minnlo (upper right) and \MGaMC (1 jet at NLO) + PB (\cascade) 
  (lower). 
  Details on the presentation of the results are given in Fig.~\ref{fig:UnfCombPt0a} caption.
         }
 \label{fig:UnfCombPt1a}
\end{figure*}

\begin{figure*}[htbp!]
  \centering
  \includegraphics[width=0.48\textwidth]{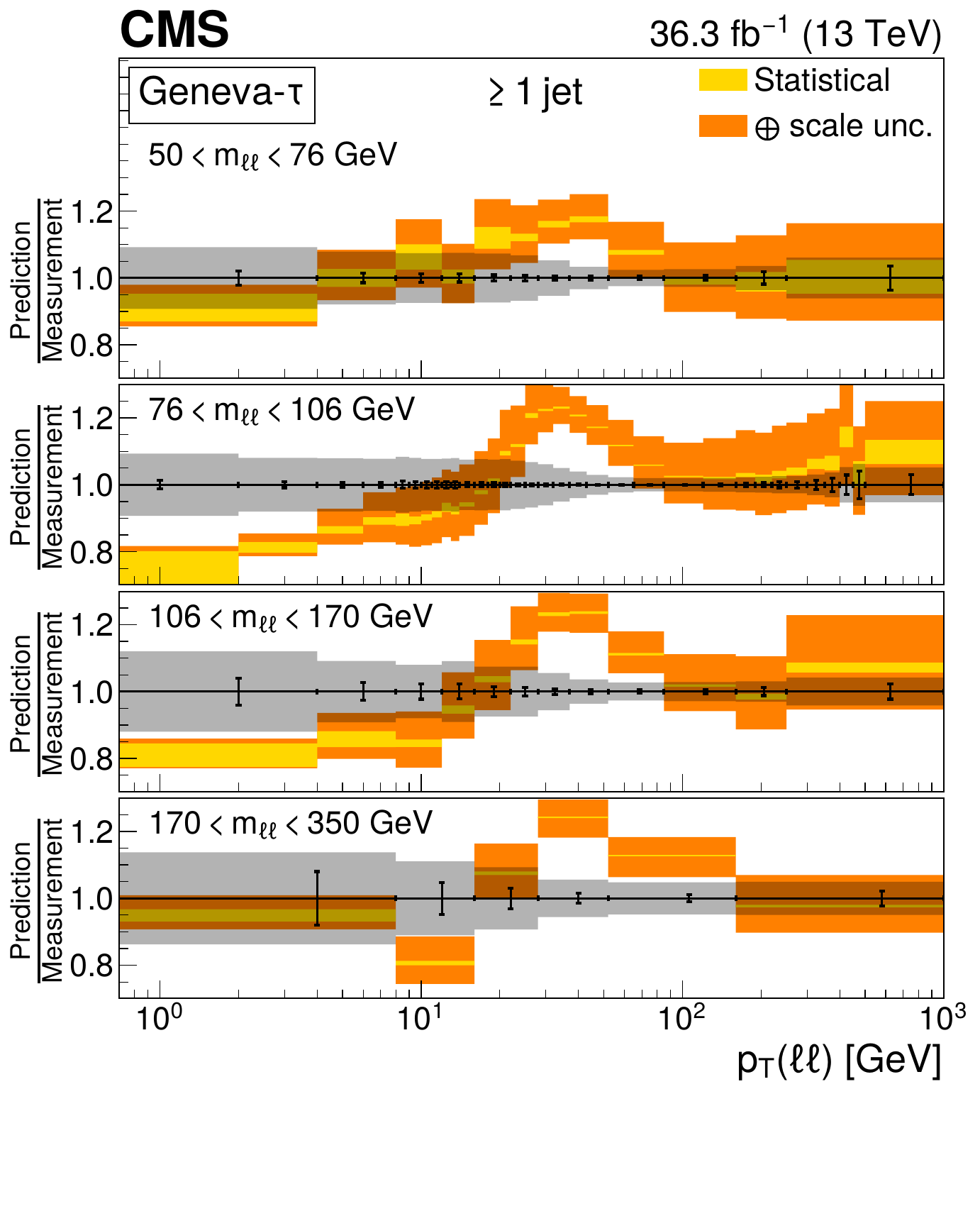}
  \includegraphics[width=0.48\textwidth]{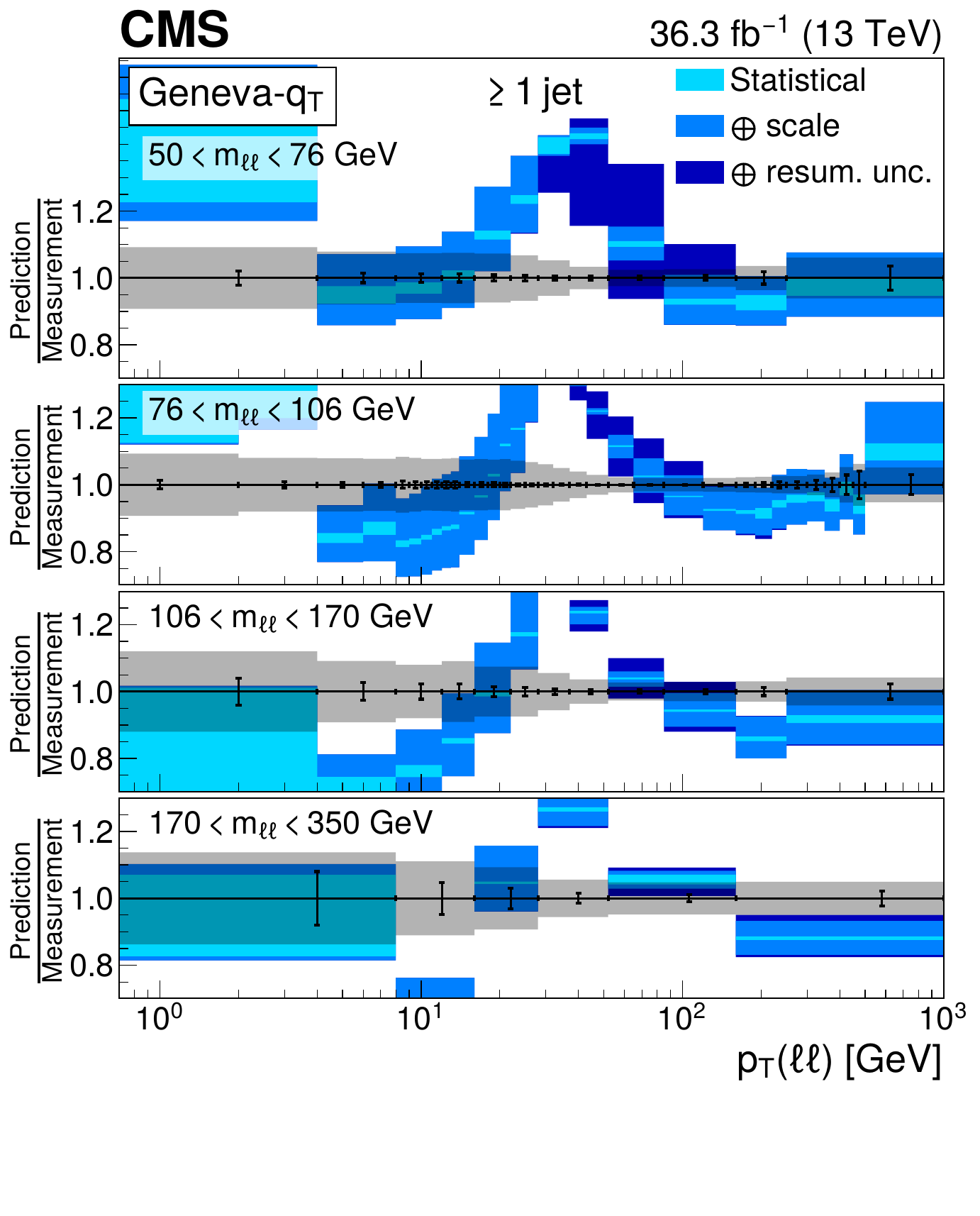}
  \caption{Comparison of the differential cross sections in \PTll to predictions in various 
   invariant mass ranges for the one or more jets case.
   The measurement is compared with \GE-$\tau$ (left) and \GE-\qt (right) predictions.
   Details on the presentation of the results are given in Fig.~\ref{fig:UnfCombPt0c} caption.
          }
 \label{fig:UnfCombPt1b}
\end{figure*}

\begin{figure*}[htbp!]
 \centering
 \includegraphics[height=0.48\textwidth]{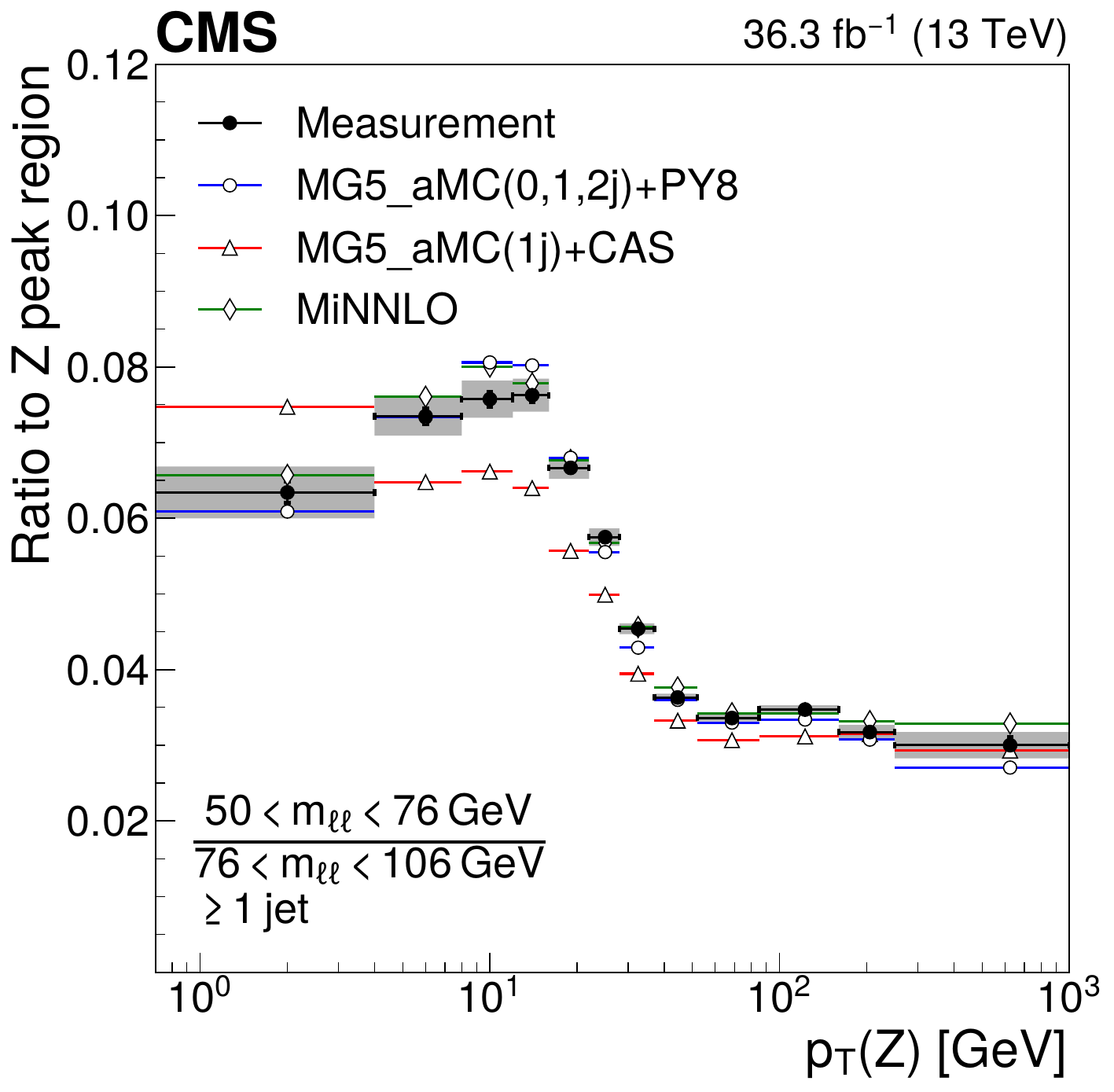}
 \includegraphics[height=0.48\textwidth]{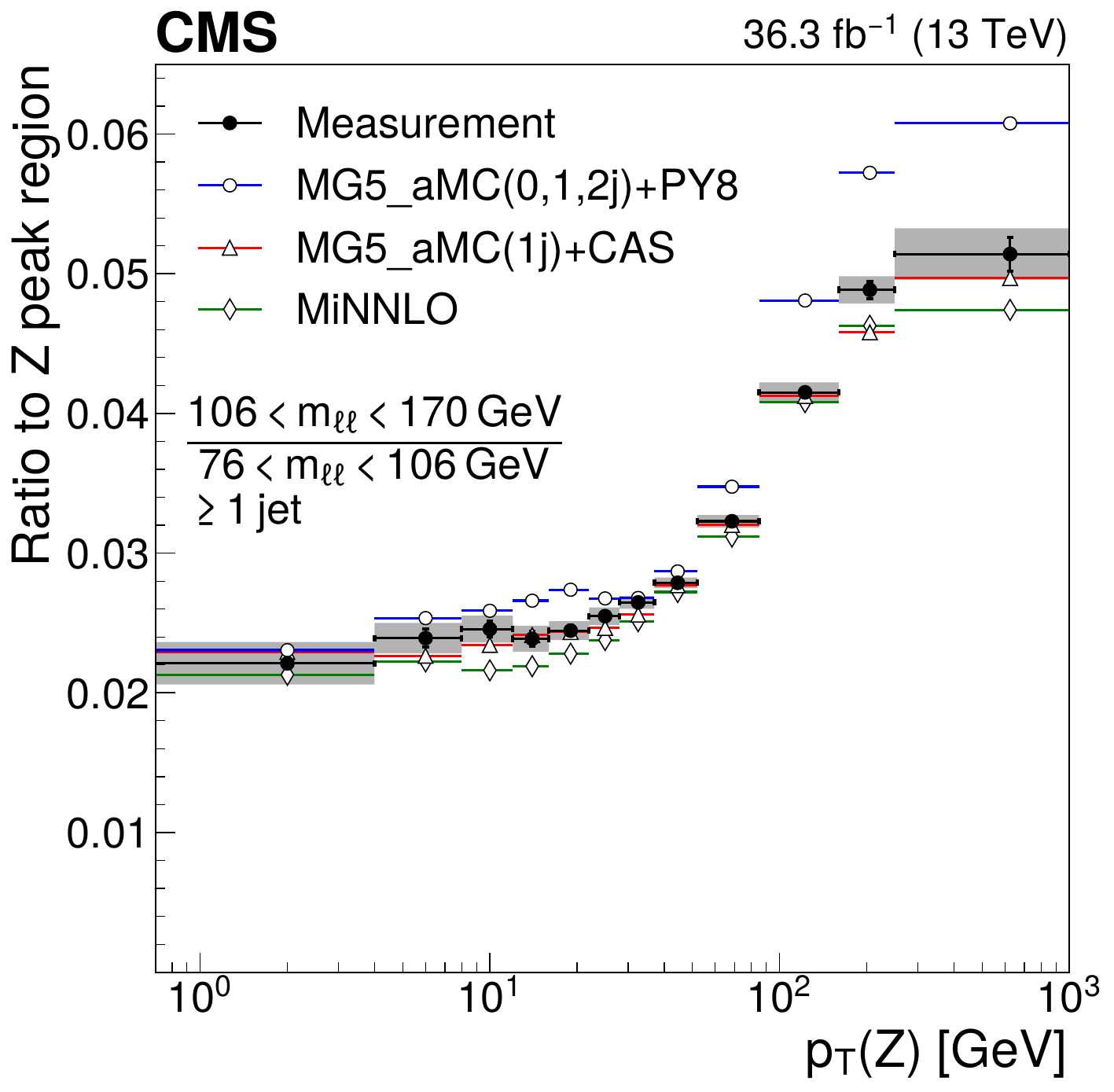}
 \includegraphics[height=0.48\textwidth]{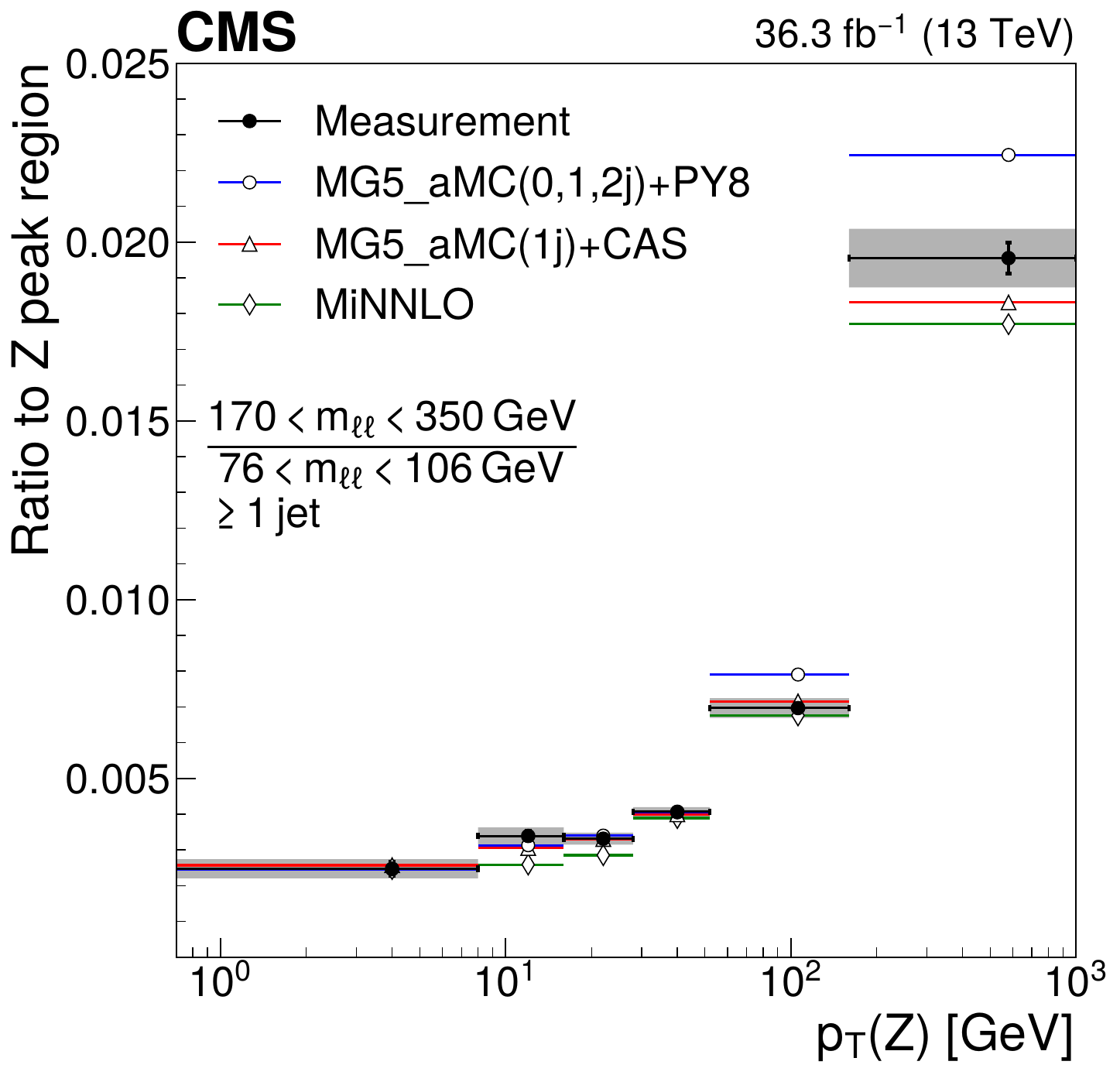}
 \caption{
  Ratios of differential cross sections in \PTll for one or more jets in various invariant 
  mass ranges with respect to the peak region \mrangeb:
  \mrangea (upper left), \mrangec (upper right), and \mranged (lower).
  The measurement is compared with \MGaMC (0, 1, and 2 jets at NLO) + \PYTHIA8 (blue dots), \minnlo (green diamonds) and \MGaMC (1 jet at NLO)+ PB (\cascade) (red triangles). Details on the presentation of the results are given in Fig.~\ref{fig:UnfCombPt0} caption.
        }
 \label{fig:UnfCombPt1Ratios}
\end{figure*}

\begin{figure*}[htbp!]
 \centering
 \includegraphics[width=0.48\textwidth]{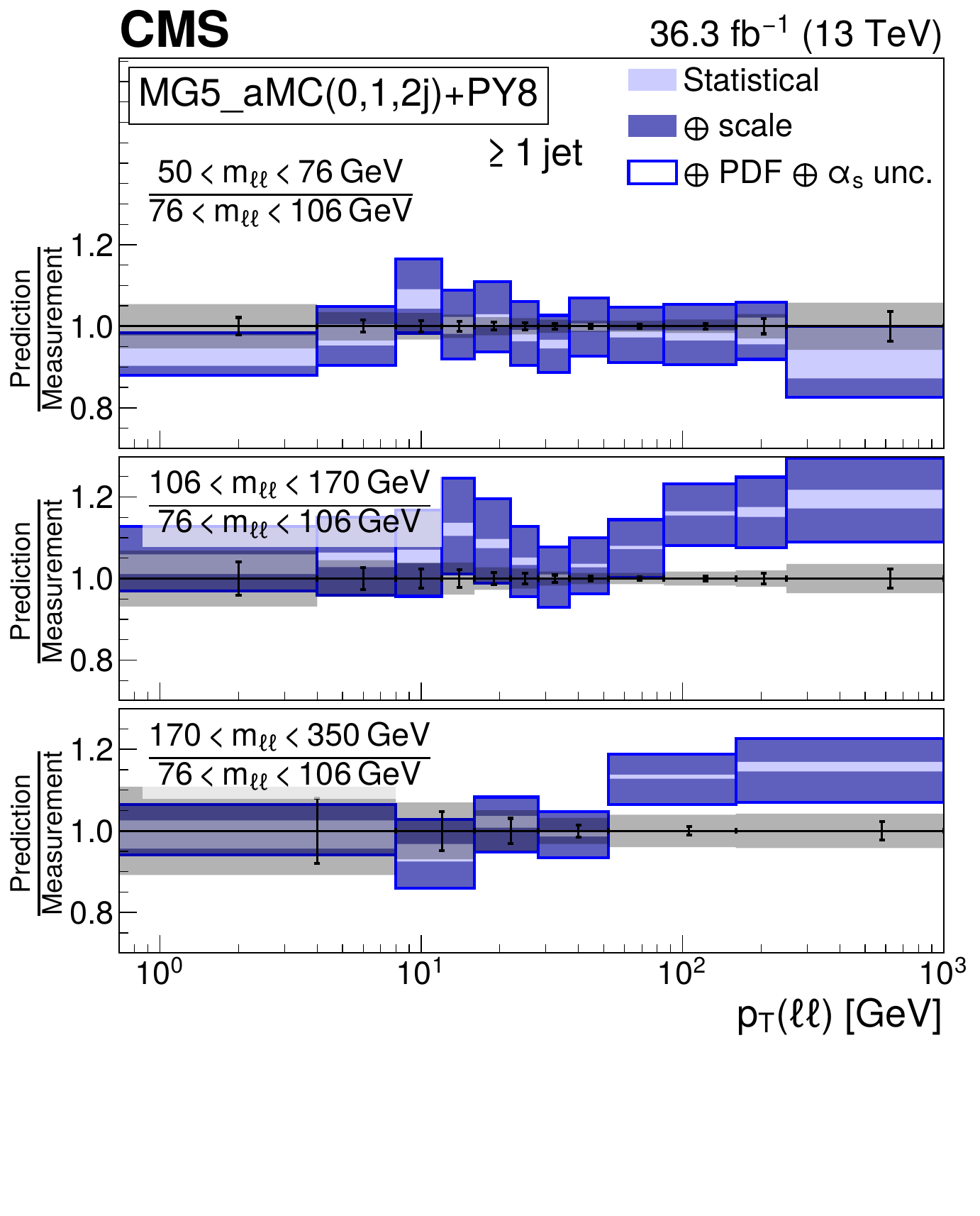}
 \includegraphics[width=0.48\textwidth]{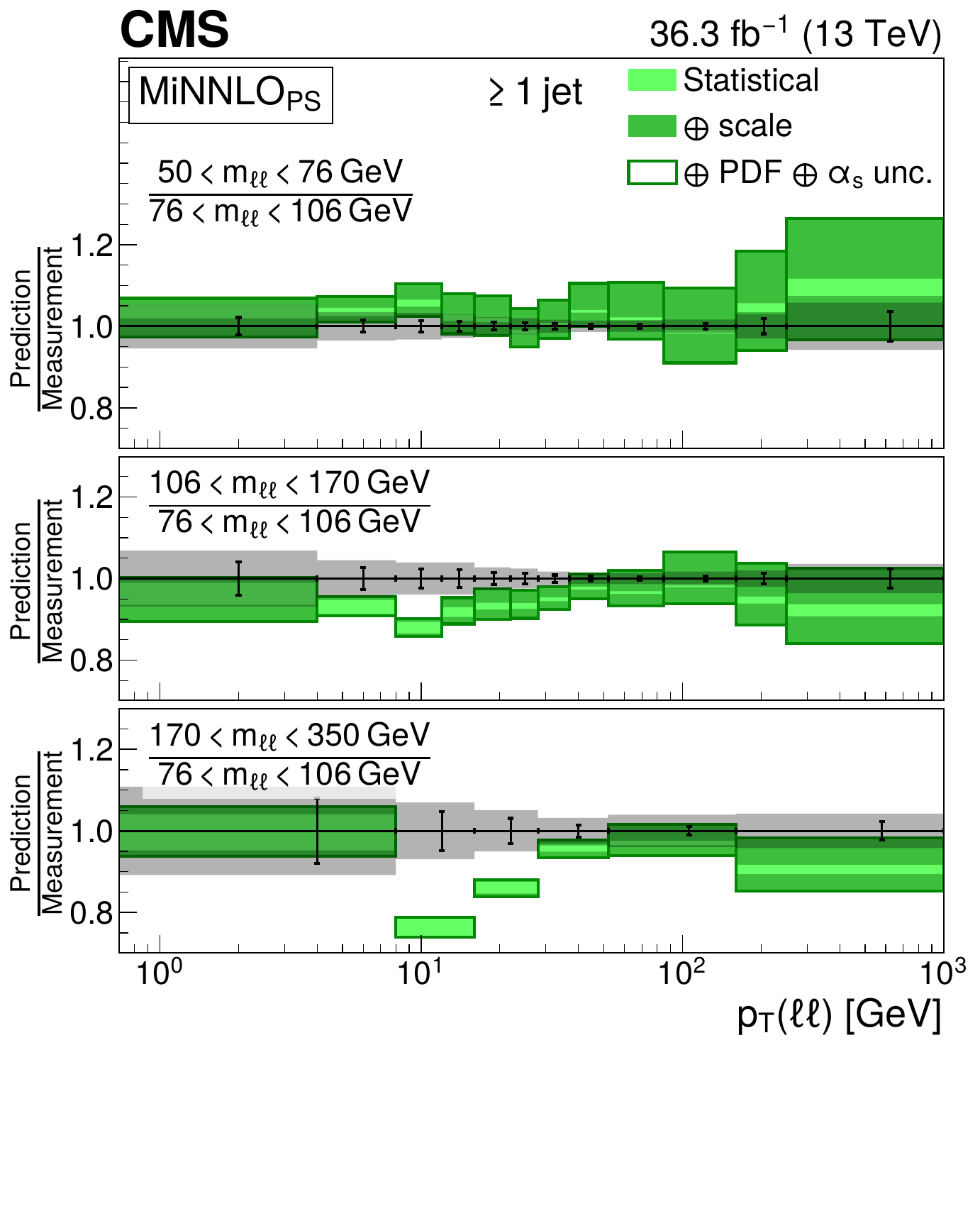}
 \includegraphics[width=0.48\textwidth]{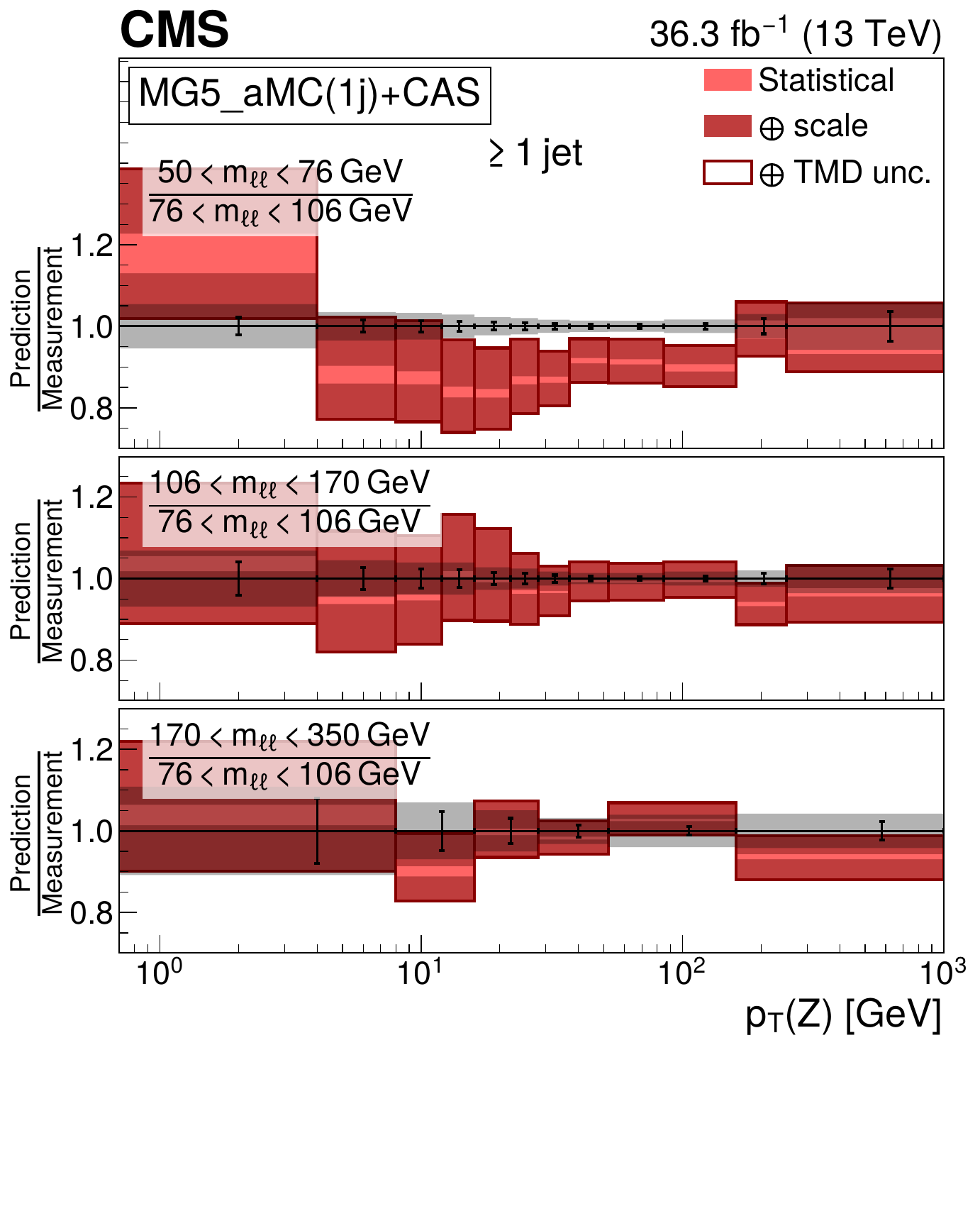}
 \caption{Comparison of the ratios of differential cross sections in \PTll for one or more jets in 
  various invariant mass ranges with respect to the peak region \mrangeb.
  The measured ratio is compared with \MGaMC (0, 1, and 2 jets at NLO) + \PYTHIA8 (upper left), \minnlo (upper right) 
  and \MGaMC (1 jet at NLO) + PB (\cascade) (lower). 
  Details on the presentation of the results are given in Fig.~\ref{fig:UnfCombPt0a} caption.
 \label{fig:UnfCombPt1Ratiosa}
        }
\end{figure*}

\begin{figure*}[htbp!]
 \centering
 \includegraphics[width=0.48\textwidth]{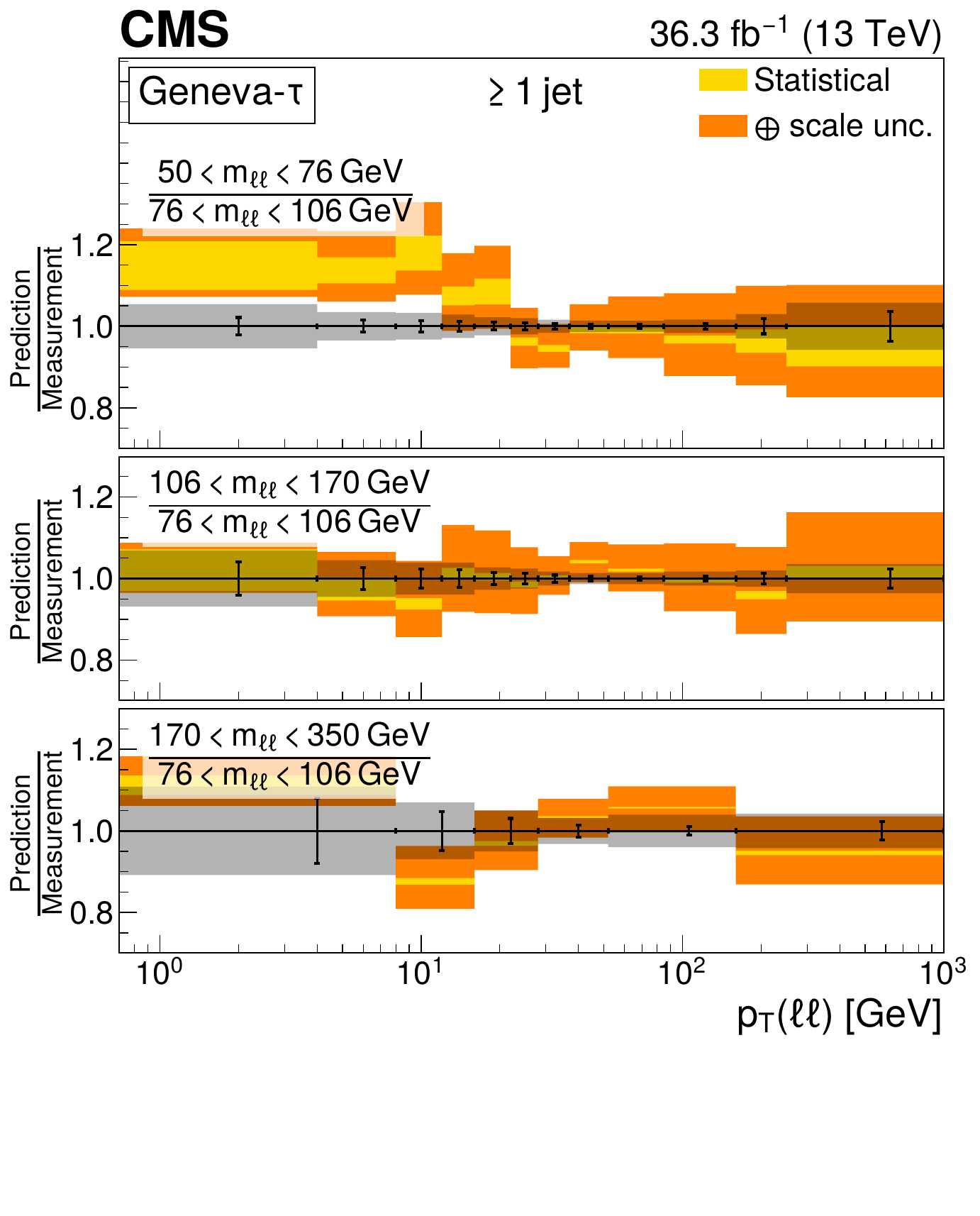}
 \includegraphics[width=0.48\textwidth]{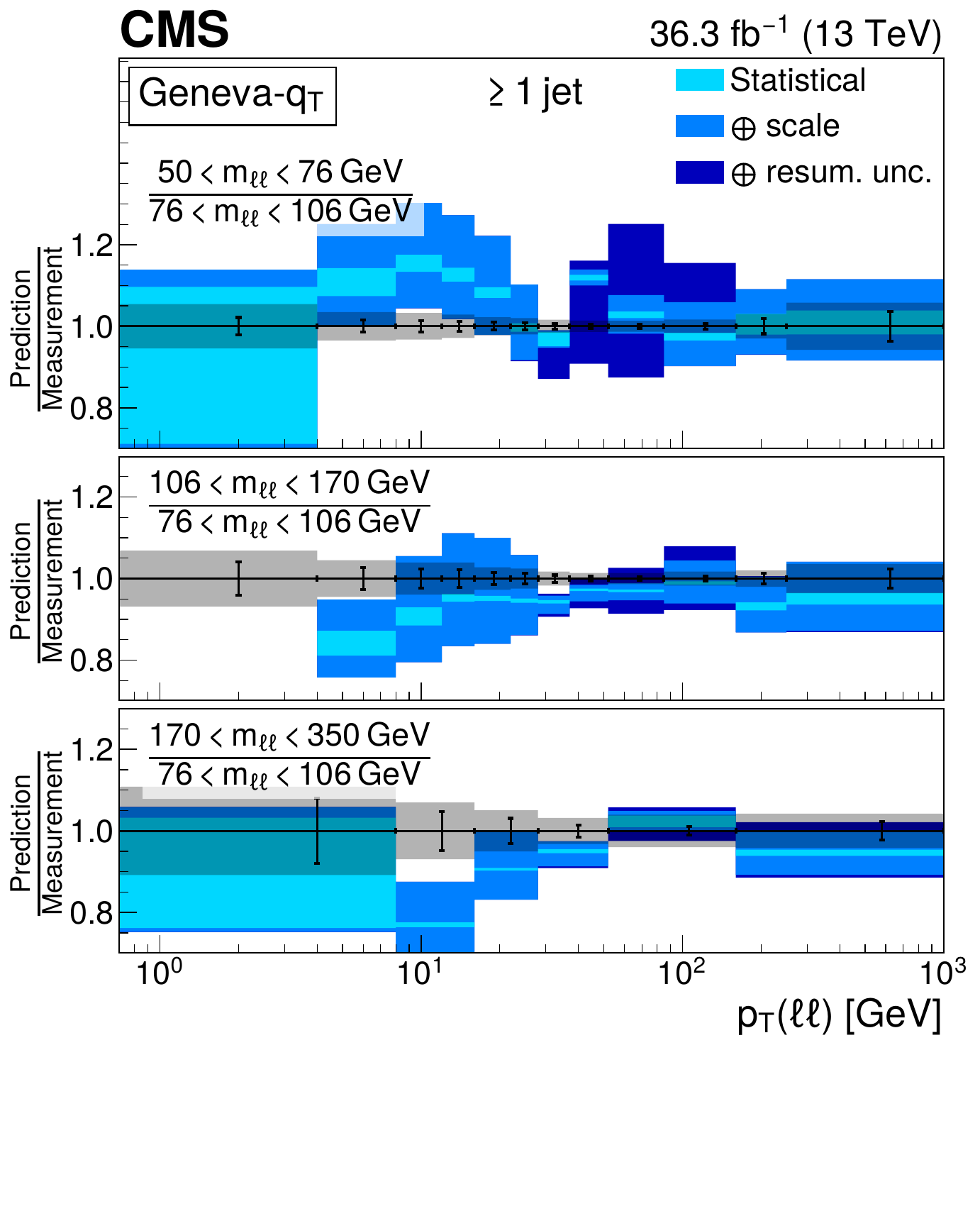}
 \caption{Comparison of the ratios of differential cross sections in \PTll for one or more jets in 
  various invariant mass ranges with respect to the peak region \mrangeb.
  The measured ratio is compared with \GE-$\tau$ (left) and \GE-\qt (right) predictions.
  Details on the presentation of the results are given in Fig.~\ref{fig:UnfCombPt0c} caption.
         }
 \label{fig:UnfCombPt1Ratiosb}
\end{figure*}

The measured cross sections are presented in Fig.~\ref{fig:UnfPt1} as a function of \PTll for at least one jet, for the same mass ranges except the 
highest. Ratios of the predictions to the data are presented 
in Figs.~\ref{fig:UnfCombPt1a} and~\ref{fig:UnfCombPt1b}. 
The ratio of these differential cross sections for various mass ranges with respect to the
same distribution in the \PZ boson peak region are shown in Fig.~\ref{fig:UnfCombPt1Ratios}, and the comparisons to predictions in Figs.~\ref{fig:UnfCombPt1Ratiosa} and~\ref{fig:UnfCombPt1Ratiosb}.

The measurements show that the differential cross sections in \PTll
are rising from small \PTll values up to a maximum
between 4~and 6\GeV and then falling towards large \PTll (Fig.~\ref{fig:UnfCombPt0}).
For these cross sections, the variation of the dilepton
invariant mass does not have a visible effect on the peak position (around 5\GeV) or on the rising shape for the values below the peak.
However, the increase of \Mll results in a broader distribution
for \PTll values above the peak. These effects are highlighted by the cross section ratios presented in
Fig.~\ref{fig:UnfCombPt0Ratios}. It has to be noted that the rising ratio for the
lowest \Mll range (Fig.~\ref{fig:UnfCombPt0Ratios} top left) up to a
\PTll value of 20\GeV is due to QED radiative effects on the final-state leptons
(photon radiations at $\Delta R(\Pell,\gamma)>0.1$)
inducing migrations from the \PZ mass peak towards lower masses.
When a jet with a large transverse momentum is required (Fig.~\ref{fig:UnfPt1}), the peak is shifted towards
larger \PTll values corresponding to the jet selection threshold, here 30\GeV regardless of the \Mll.
As in the inclusive case, the distributions become broader for \PTll values larger
than the peak for increasing \Mll.

A description of these measurements based on QCD requires both multi-gluon
resummation and a fixed-order matrix element.
The description of the distributions at small \PTll values requires an approach
taking into account initial-state nonperturbative
and perturbative multi-gluon resummation.
The falling behaviour at large \PTll is sensitive to hard QCD radiation, which is expected to be
well described by matrix element calculations including at least NLO corrections.
The size of the QCD radiation is driven by the available kinematic phase
space and the value of \alpS.
An increase of \Mll extends the phase space for hard radiations, slightly
compensated by the decrease of \alpS with increasing \Mll.
The tail at large \PTll is dominated by jet multiplicities above one. 
For the inclusive cross sections, the resummation effects are concentrated at small \PTll.
The value of the maximum of the distributions is expected to depend weakly on \Mll.
In the presence of a hard jet, multiple gluon emissions also affect the perturbative region located in $\eta$ between the
jet and the vector boson. The corresponding cross section measurements therefore provide additional constraints on the resummation
treatment in the predictions.

The \MGPE prediction describes the data well globally (Fig.~\ref{fig:UnfCombPt0a}), 
although it predicts a too-small cross section for
\PTll values below 30\GeV in the inclusive case.
This disagreement is more pronounced at higher \Mll and reaches
about 20\% for masses above 170\GeV.
The low-\PTll region is sensitive to gluon resummation.
In \MGaMC, the resummation effects are simulated by the parton shower, modelled in \PYTHIA8
depending on parameters tuned on previously published measurements, including
DY cross sections in the \PZ boson mass peak region.
It has to be noted that the low \PTll spectrum is sensitive to the choice of the tuned 
parameters~\cite{CMS:2019csb} and that no related systematic uncertainty is available.
The large \PTll distributions are well described by \MGaMC, which relies on NLO matrix elements 
for 0, 1 and 2 partons in the final state. Nevertheless, \MGaMC predicts cross sections larger than those
observed for the highest \PTll values measured in the mass ranges
\mrangec~for both the inclusive and 1 jet cases. 
Since the theoretical uncertainty is dominant in that region, a better agreement
might be found using higher-order (\eg, NNLO) multiparton predictions.

The \minnlo prediction provides the best global description of the data among the predictions 
presented in this paper. This approach, based on NNLO matrix element and \PYTHIA8 parton shower and MPI, describes well the large \PTll
cross sections (Fig.~\ref{fig:UnfCombPt0a}) and ratios (Fig.~\ref{fig:UnfCombPt1Ratiosa}), except above 400\GeV, for \Mll around the \PZ boson peak. The medium and low 
\PTll cross sections are also well described by \minnlo which relies on parton showers, a harder primordial \kt and Sudakov 
form factors. The same observation can be made in the
one jet case. The inclusion of an NNLO matrix element
reduces significantly the scale uncertainties, in particular for the inclusive cross section 
in for the medium \PTll values where the PDF uncertainty becomes significant with respect to other model
uncertainties. 
It has to be noted that no parton shower tune uncertainty is assigned in the case of \minnlo
as well as in the case of \MGaMC.

\begin{tolerant}{2000}
We see that the \cascade predictions (\MGCAS)  involving TMDs produce a better description 
in the low-\PTll part than \MGPE, which is valid for all \Mll bins.
The predicted cross section for medium \PTll values is 5 to 10\% too low (Fig.~\ref{fig:UnfCombPt0b}).
What is remarkable is that this prediction is based on TMDs obtained from totally independent data, 
from a fit to electron-proton deep inelastic scattering measurements performed at HERA. 
The high \PTll part is not described by the \PZ+0,1 jet matrix element calculations from \MGaMC with  \cascade due to missing higher fixed-order calculations. 
The range of \PTll values well described extends with increasing \Mll. 
For the one jet case (Fig.~\ref{fig:UnfCombPt1a}), the low-\PTll part is mainly dominated by $\PZ + 2$ jet events, and the \cascade predictions are missing  the contributions from the double parton scattering. It thus fails to describe the low \PTll region.   
In the low-\PTll region  of the 1 jet case double parton scattering contributions play a significant role and thus \cascade without it cannot describe this region.   
The  \cascade predictions give an overall good description of the ratio measurements (Fig.~\ref{fig:UnfCombPt1Ratiosa}).
Recently the predictions have been extended by including multi-jet merging~\cite{Martinez:2021chk} for an improved description of the full \PTll spectrum, 
shown in the Appendix~\ref{app:suppMat}. 
\end{tolerant}

Within its range of validity, $\PTll < 0.2 \ \Mll$, the \artemide prediction describes the measurements very well.
For all \Mll, the low-\PTll distributions predicted by \artemide, based on TMDs corresponding to an
N$^3$LL approximation, are in very good agreement with the data, except for the highest masses.
Figure~\ref{fig:UnfCombPt0b} shows the prediction with and without QED FSR corrections.
This underlines the importance of migrations from the \PZ boson peak towards lower
masses, inducing the peak structure in the \PTll ratio distribution of Fig.~\ref{fig:UnfCombPt0Ratios}.
The remarkable agreement of the \artemide prediction with the measurement at the \PZ boson peak  
is expected since the prediction relies on TMDs fitted on previous DY measurements 
at the \PZ boson peak though at lower centre of mass energies. The excellent agreement for higher \Mll 
confirms the validity of the approach and in particular of the TMD factorization when the mass scale 
largely dominates over the transverse momentum.
No prediction is provided by \artemide for the 1 jet case nor for the \phistar cross section dependence.

Comparisons of the inclusive cross section as a function of \PTll with two predictions of \GE are presented
in Fig.~\ref{fig:UnfCombPt0c} for the inclusive cross sections and in Fig.~\ref{fig:UnfCombPt1b} for the one 
jet cross sections.
The original prediction combining NNLL resummation on the 0-jettiness variable $\tau_0$ (\GE-$\tau$) 
and NNLO corrections does not describe the data well for \PTll values below 40\GeV. 
This too hard \PTll spectrum might be related to the choice of \alpS, 
as discussed in Ref.~\cite{Alioli:2015toa}.
For the high \PTll region, which is dominated by the fixed-order effects, the inclusion of NNLO 
corrections provides a good description of the measured cross section.
The more recent \GE prediction (\GE-\qt), using a \qt resummation at N$^3$LL, provides a much better 
description of the measured
inclusive cross sections, describing very well the data in the full \PTll range except for middle \PTll
values in the lowest mass bin. Here, as in \minnlo case, the inclusion of NNLO corrections provides a
significant reduction of the scale uncertainties, leading to very small theory uncertainties in the middle \PTll
range.
The two \GE predictions compared with the measured one jet cross sections are similar because both use 1-jettiness in this part of the phase space. 
This could explain that \GE predicts a too hard \PTll spectrum, similarly to the 0-jettiness inclusive case.



\subsection{\texorpdfstring{\phistar}{phi*(eta)} results}

\begin{figure*}[htbp!]
 \centering
 \includegraphics[height=0.40\textwidth]{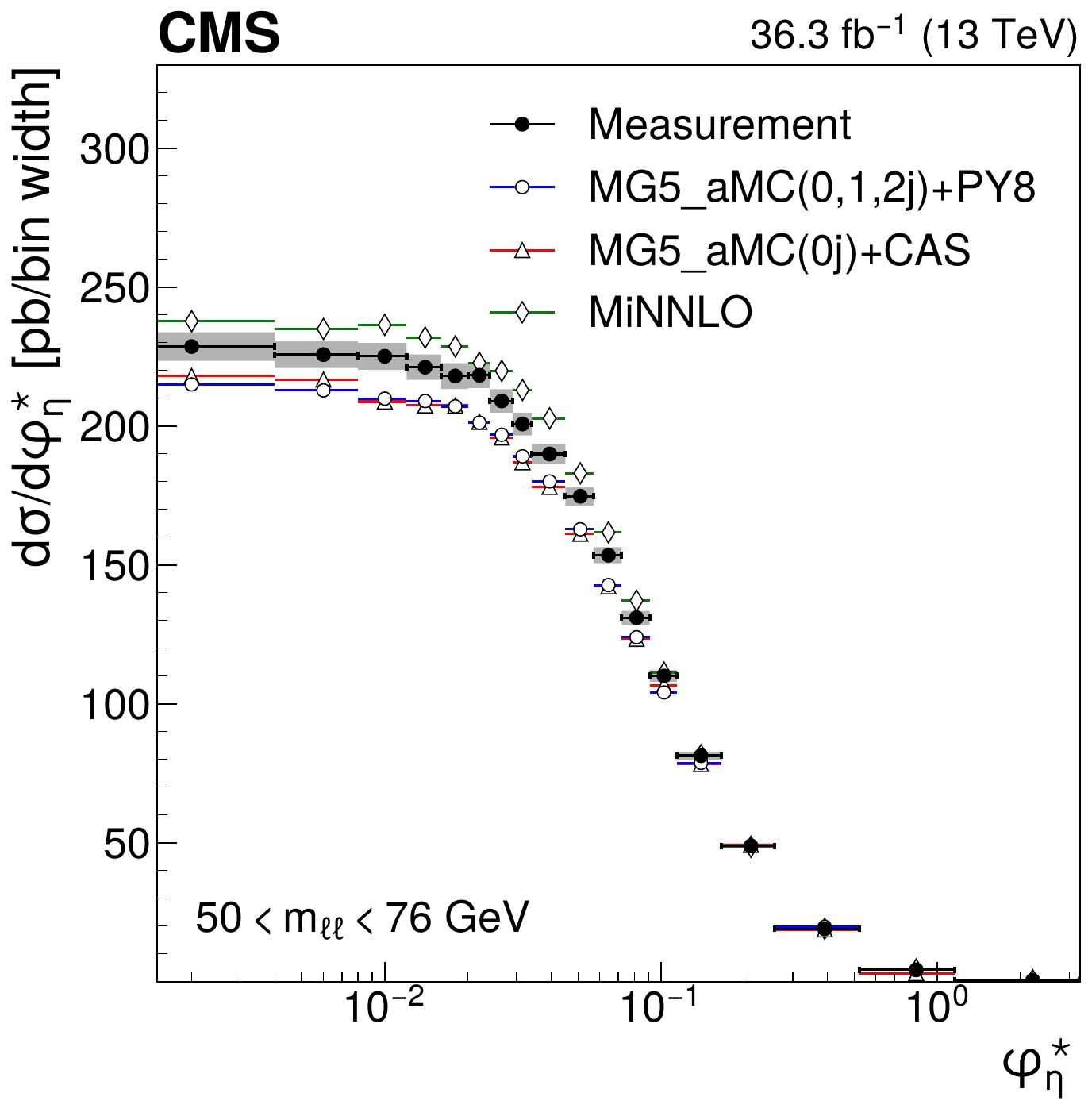}
 \includegraphics[height=0.40\textwidth]{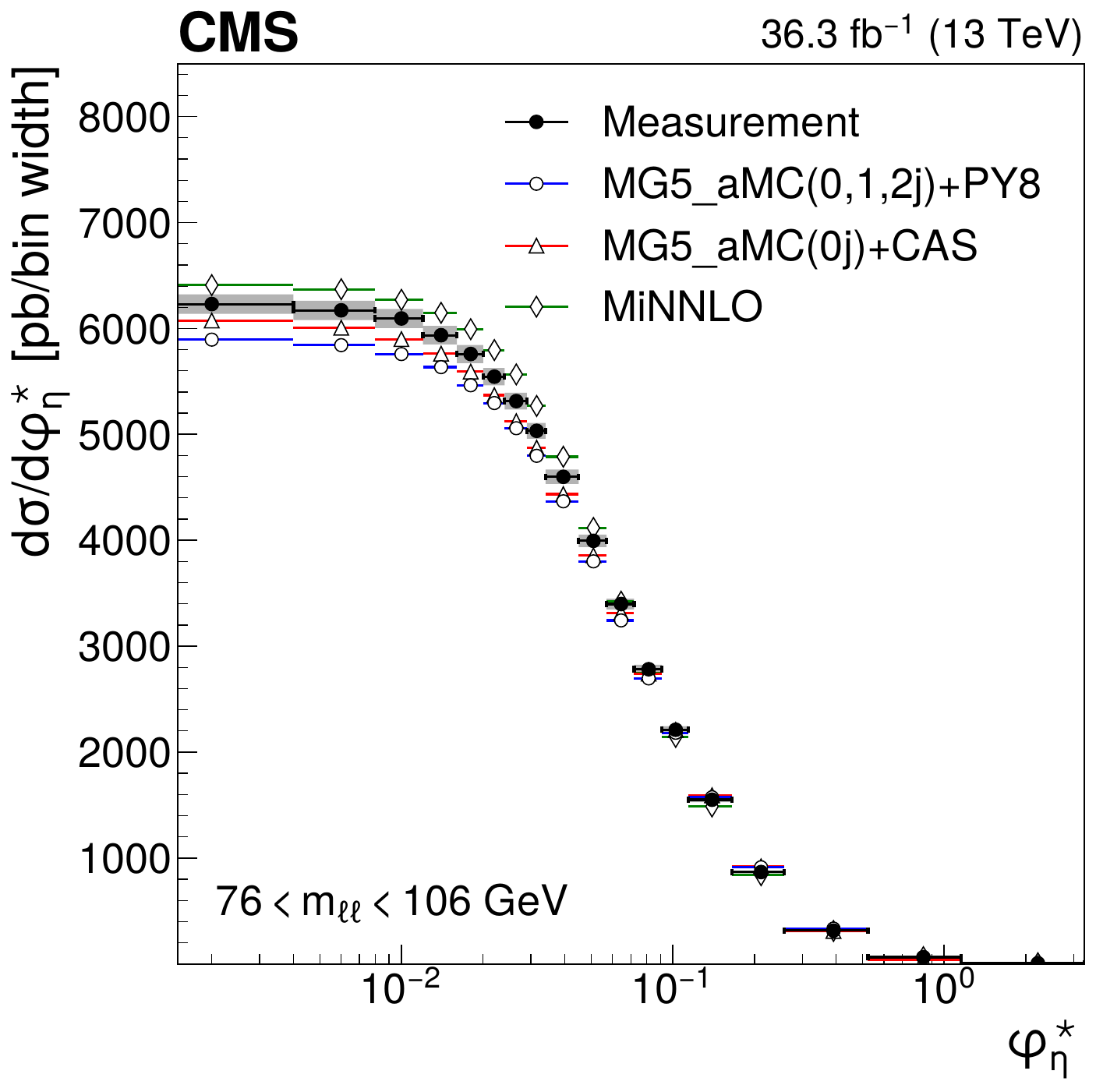}
 \includegraphics[height=0.40\textwidth]{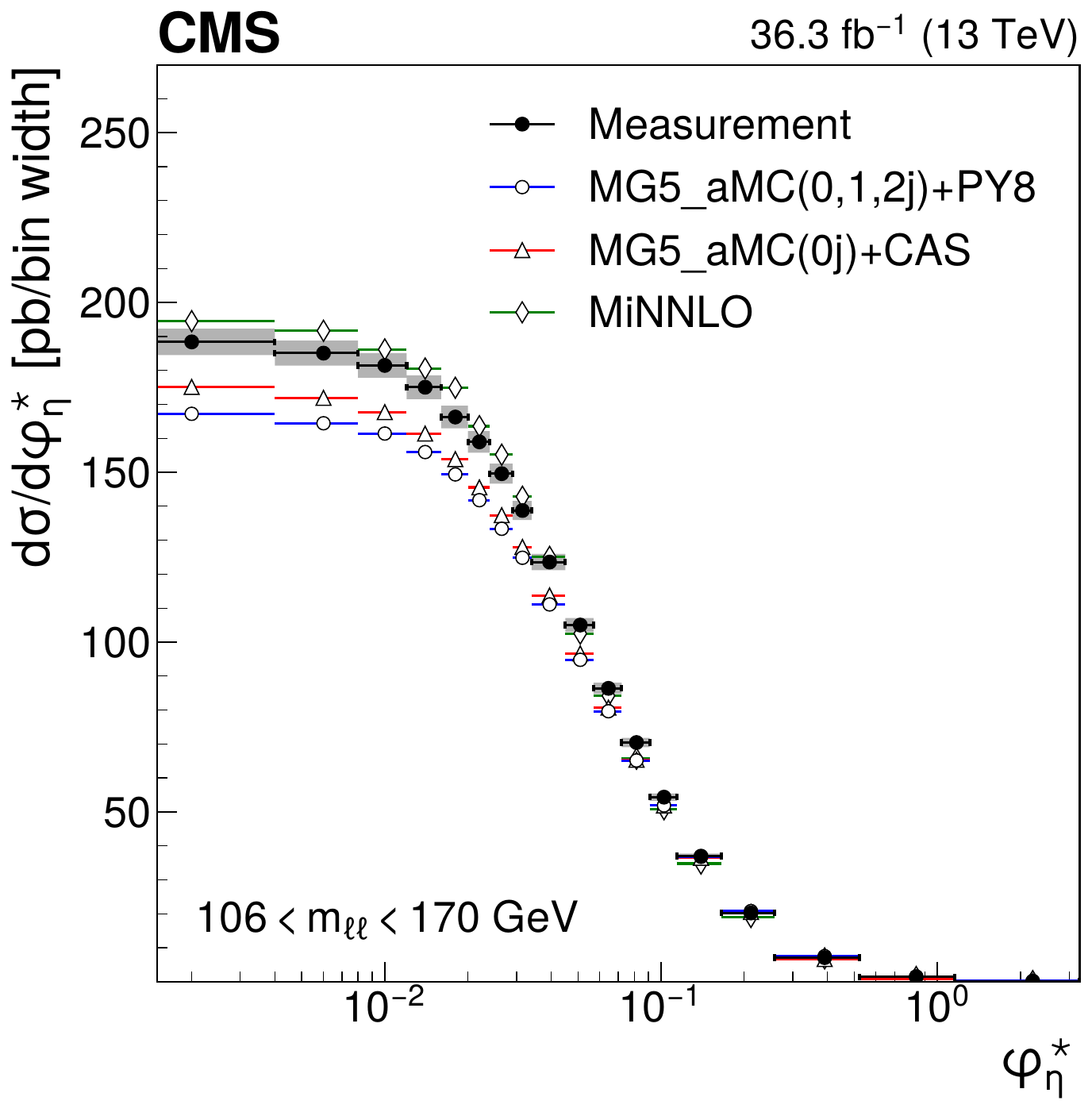}
 \hspace{2.7mm}
 \includegraphics[height=0.40\textwidth]{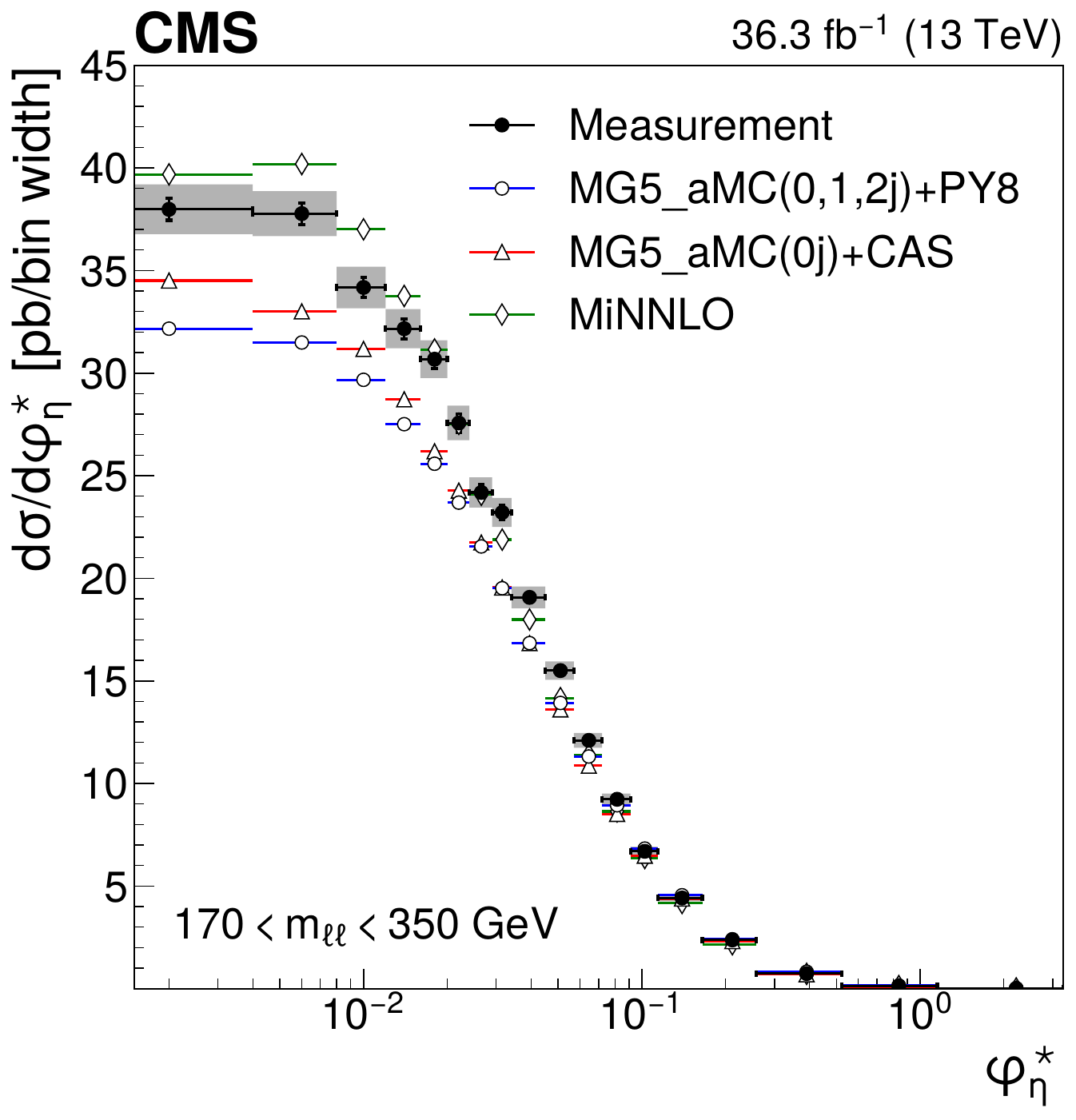}
 \includegraphics[height=0.40\textwidth]{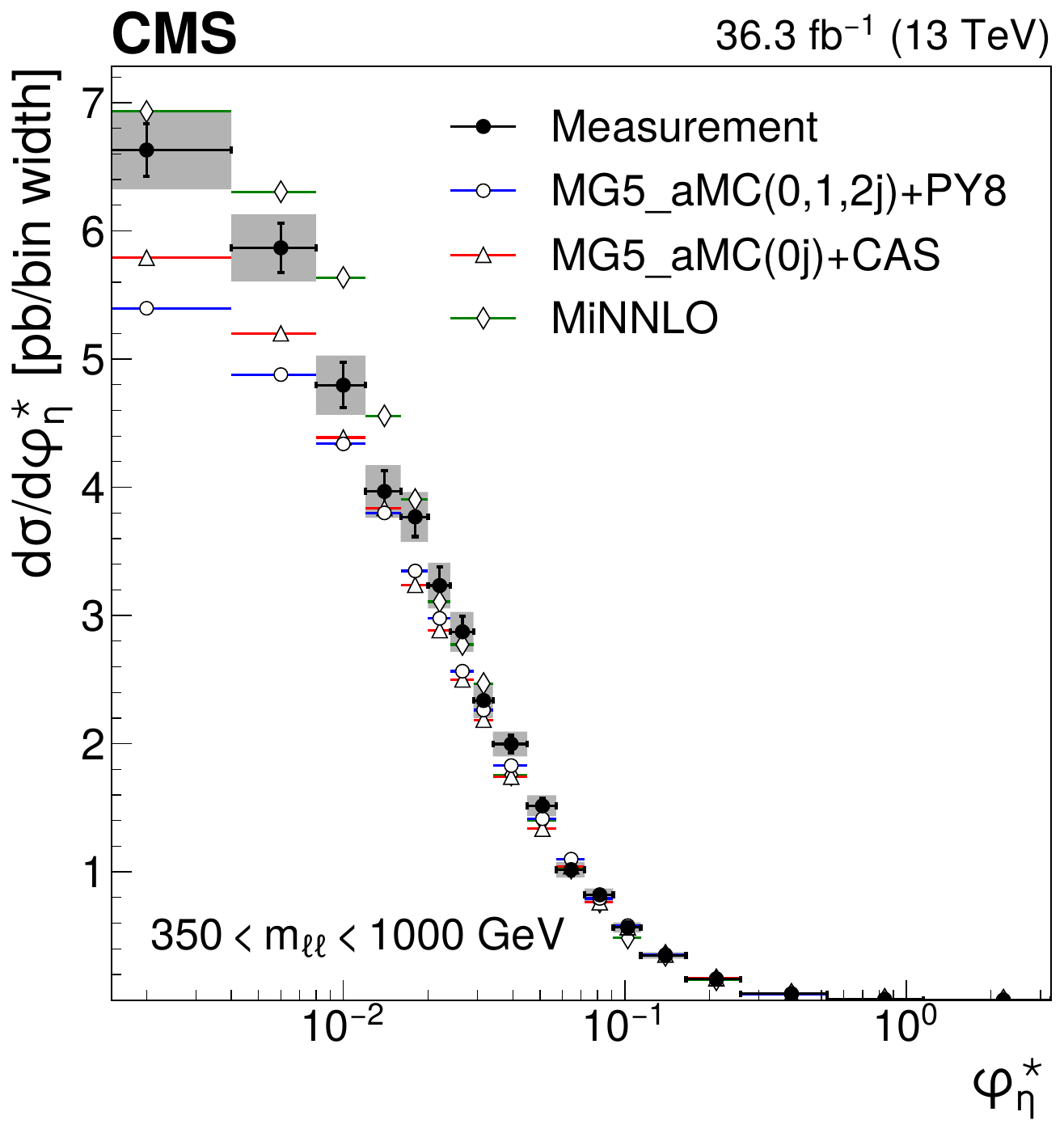}
 \caption{Differential cross sections in $\phistar(\Pell\Pell)$ in various invariant mass ranges: 
  \mrangea (upper left), \mrangeb (upper right), \mrangec 
  (middle left), \mranged (middle right), and \mrangee (lower).
  The measurement is compared with \MGaMC (0, 1, and 2 jets at NLO) + \PYTHIA8 (blue dots), \minnlo (green diamonds) and \MGaMC (0 jet at NLO)+ PB (\cascade) (red triangles). Details on the presentation of the results are given in Fig.~\ref{fig:UnfCombPt0} caption.
         }
 \label{fig:UnfCombPhi}
\end{figure*}

\begin{figure*}[htbp!]
 \centering
 \includegraphics[width=0.48\textwidth]{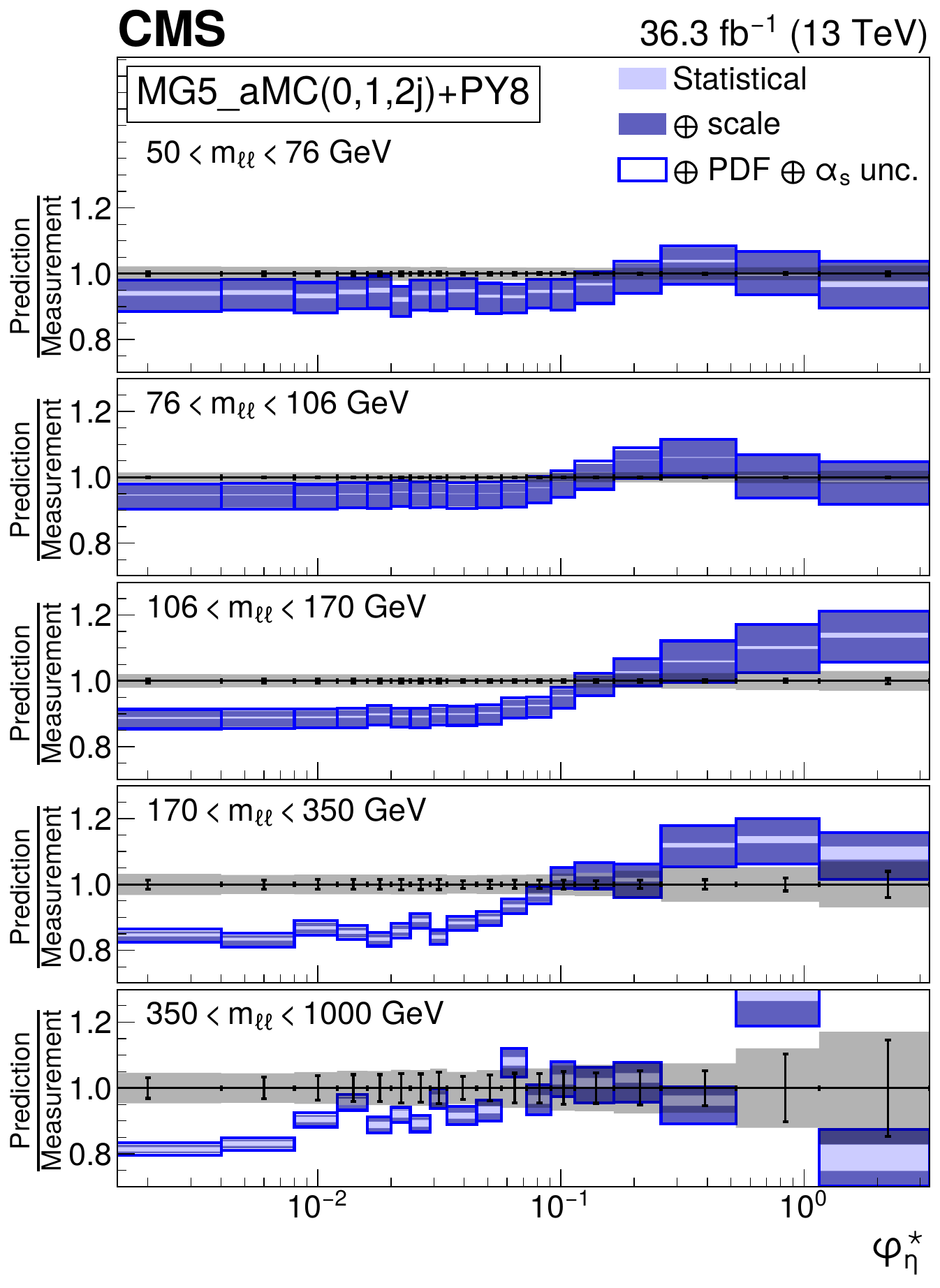}
 \includegraphics[width=0.48\textwidth]{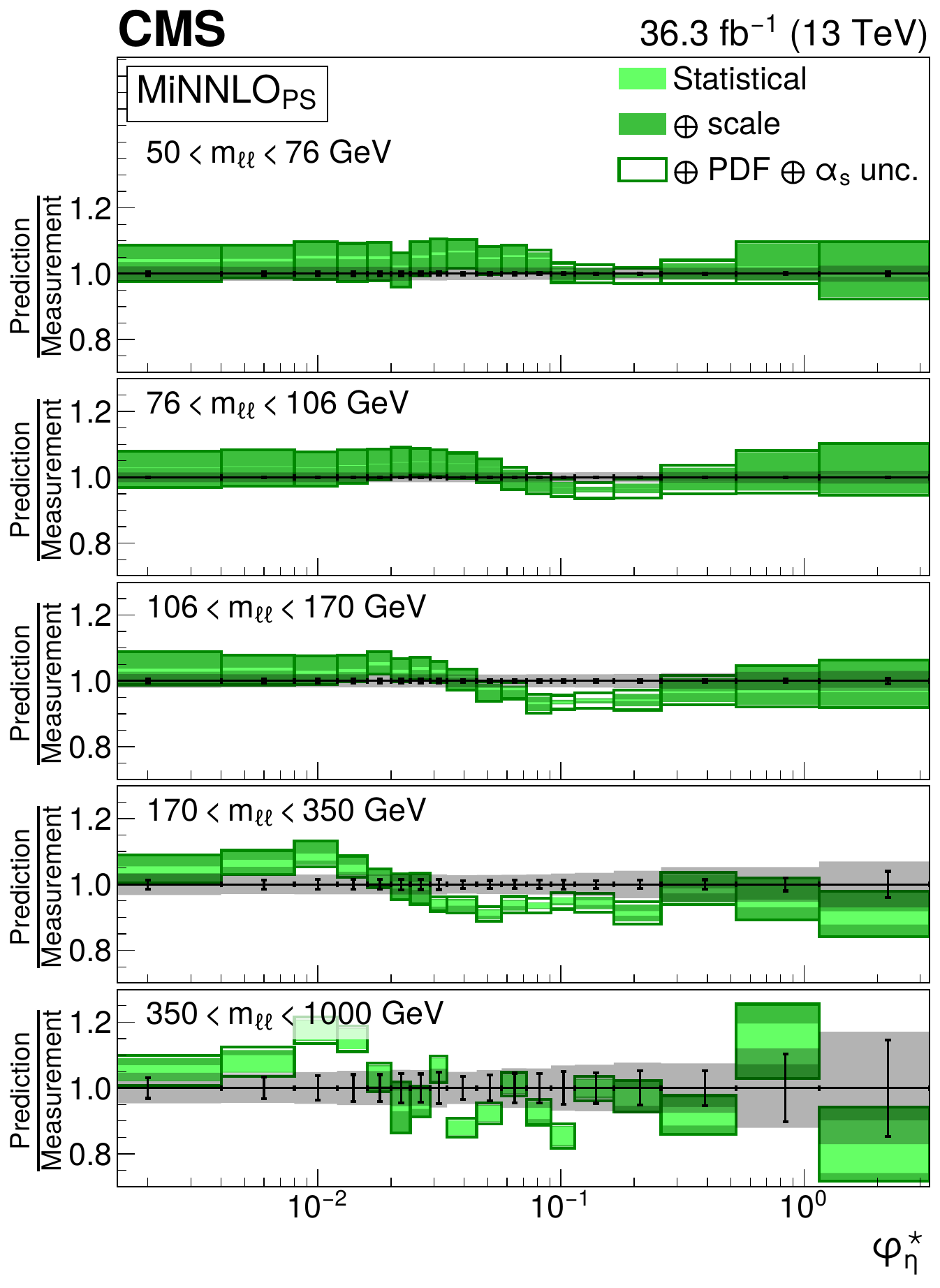}
 \includegraphics[width=0.48\textwidth]{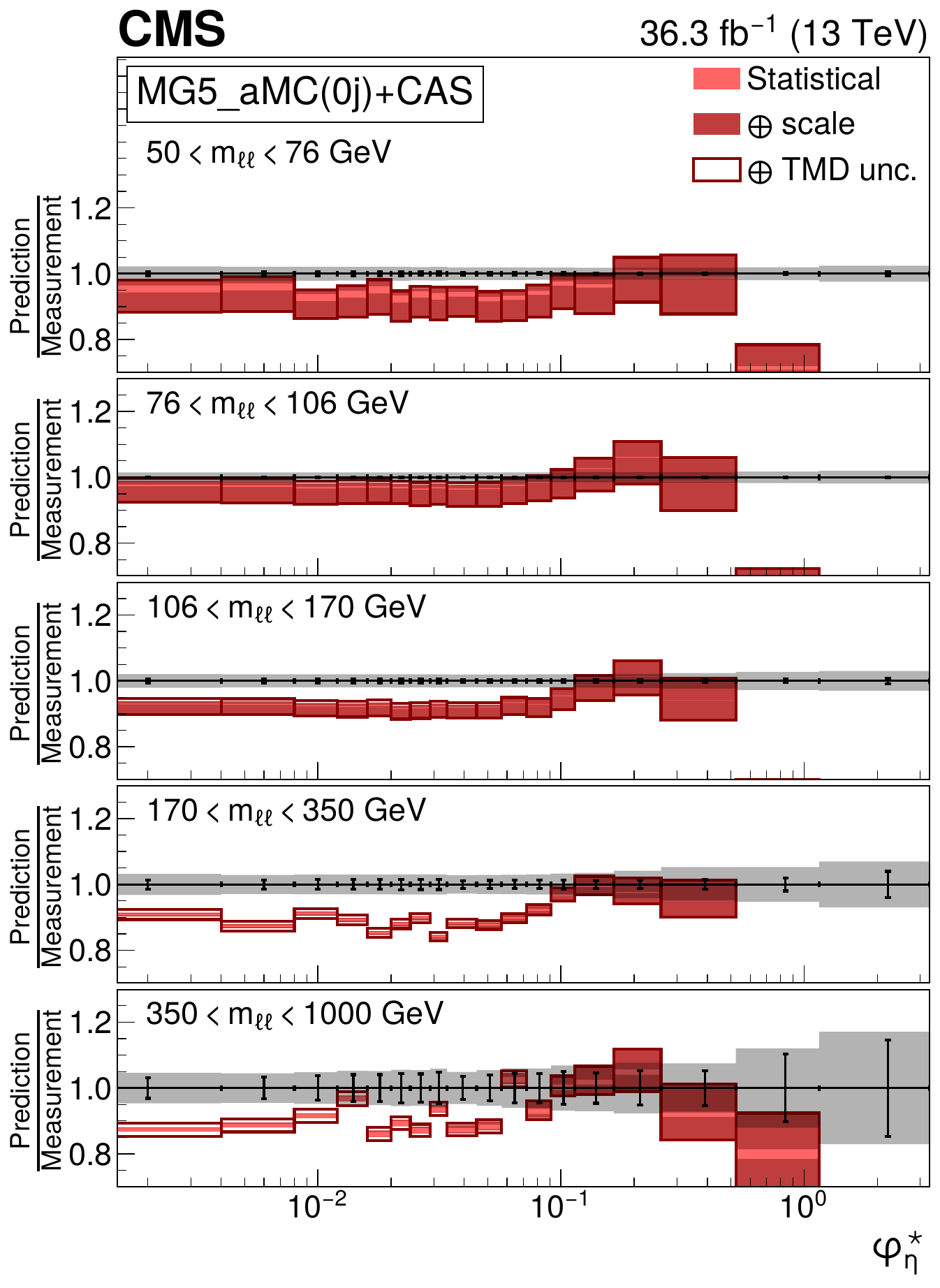}
 \caption{Comparison of the differential cross sections in $\phistar(\Pell\Pell)$ to predictions 
  in various \Mll ranges.
 The measurement is compared with \MGaMC (0, 1, and 2 jets at NLO) + \PYTHIA8 (upper left), \minnlo (upper right) and \MGaMC + PB (\cascade)
 (lower).
  Details on the presentation of the results are given in Fig.~\ref{fig:UnfCombPt0a} caption.
         }
 \label{fig:UnfCombPhia}
\end{figure*}

\begin{figure*}[htbp!]
  \centering
 \includegraphics[width=0.48\textwidth]{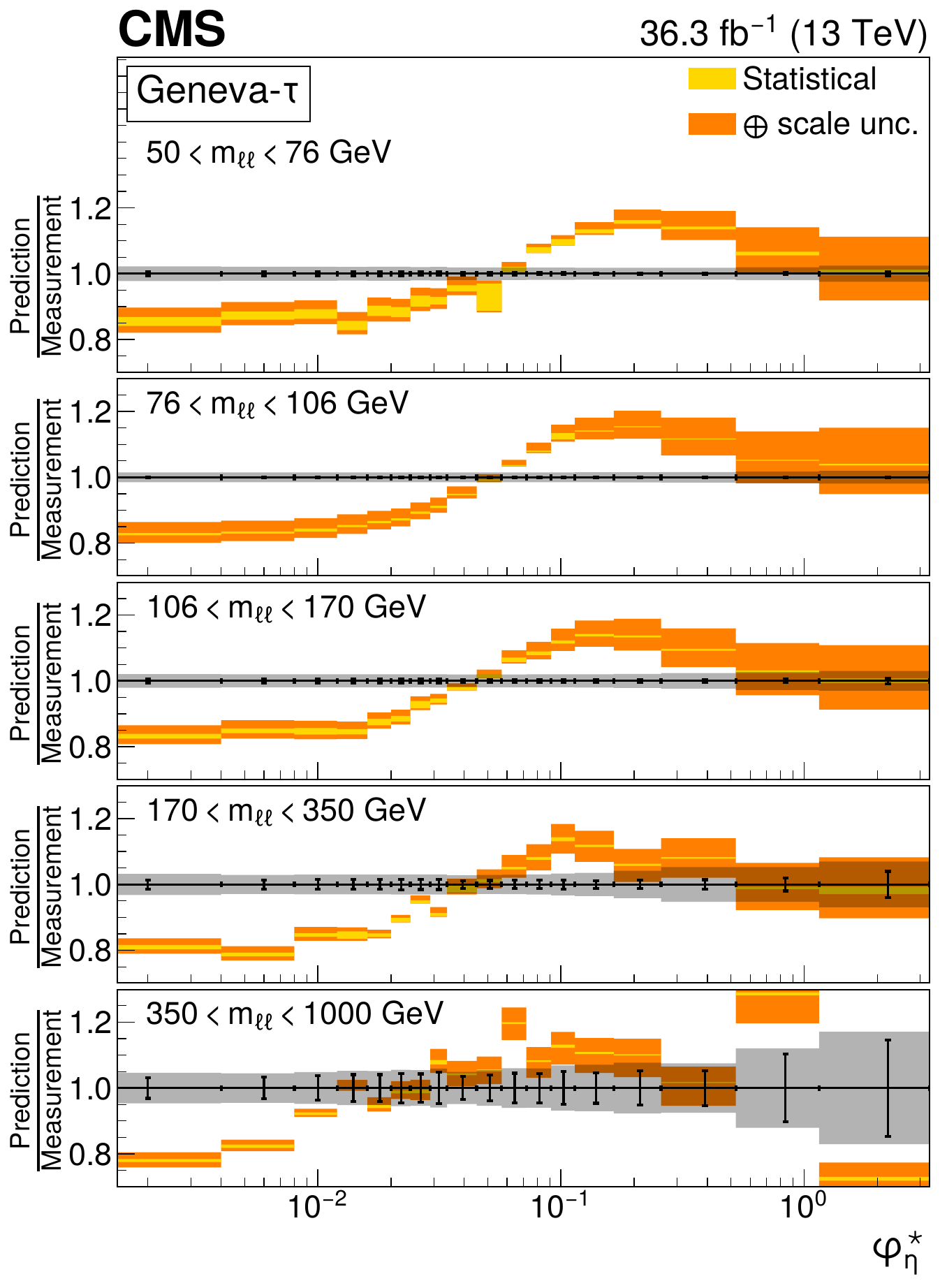}
 \includegraphics[width=0.48\textwidth]{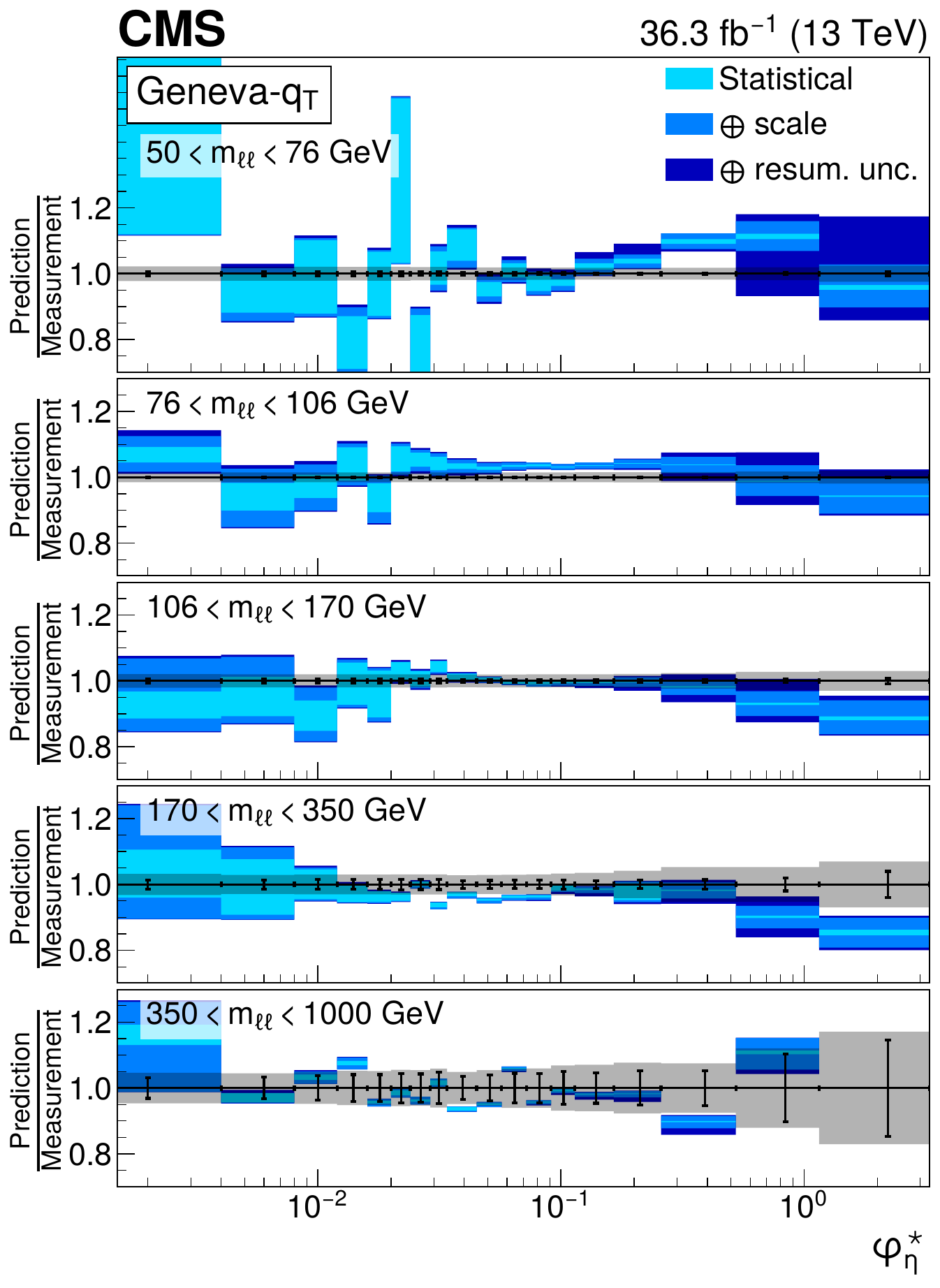}
 \caption{Comparison of the differential cross sections in $\phistar(\Pell\Pell)$ to predictions 
  in various \Mll ranges.
  The measurement is compared with \GE-$\tau$ (left) and \GE-\qt (right) predictions.
  Details on the presentation of the results are given in Fig.~\ref{fig:UnfCombPt0c} caption.
         }
 \label{fig:UnfCombPhib}
\end{figure*}

\begin{figure*}[htbp!]
 \flushright
 \includegraphics[height=0.49\textwidth]{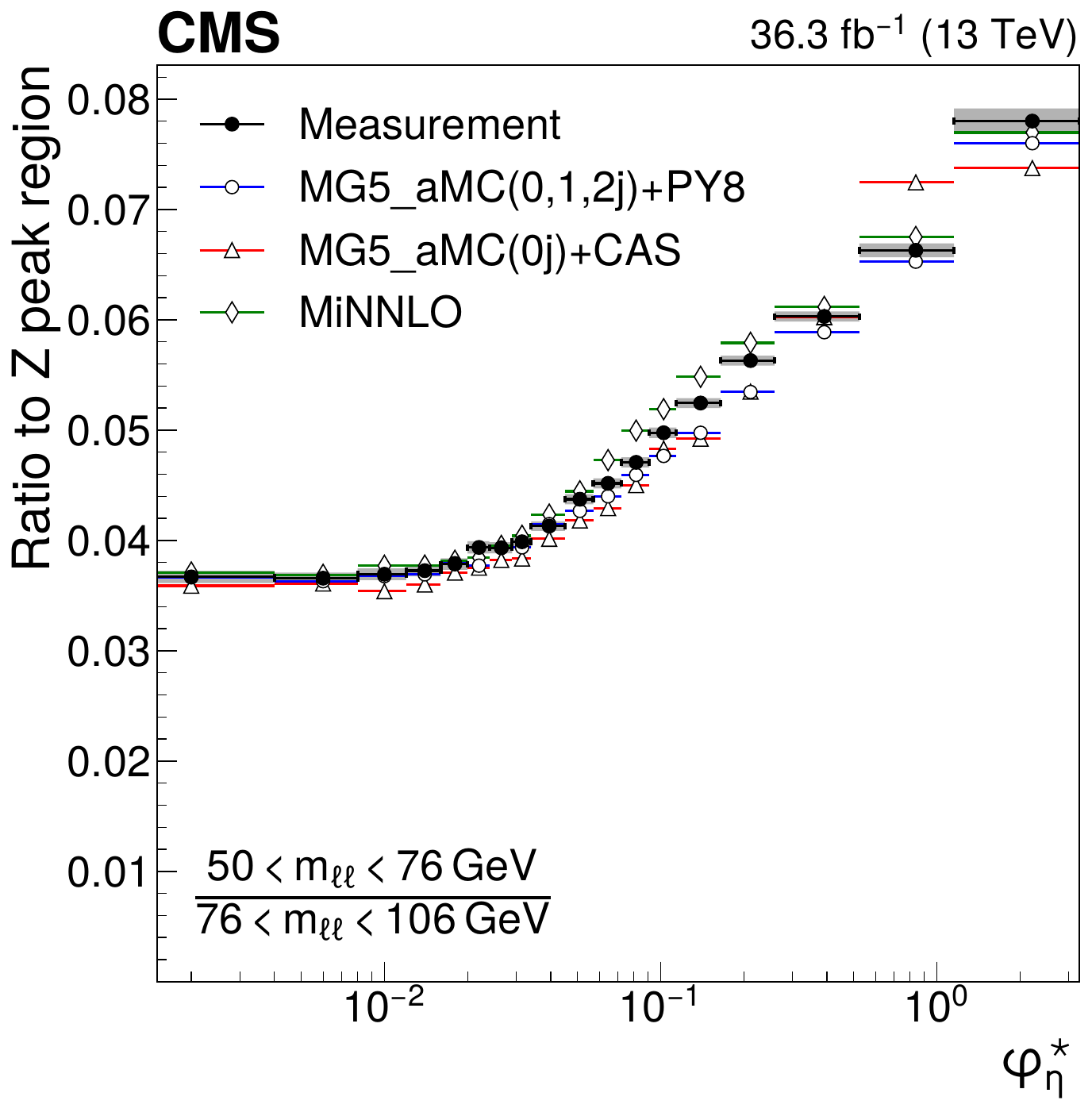}
 \hspace{1.8mm}
 \includegraphics[height=0.49\textwidth]{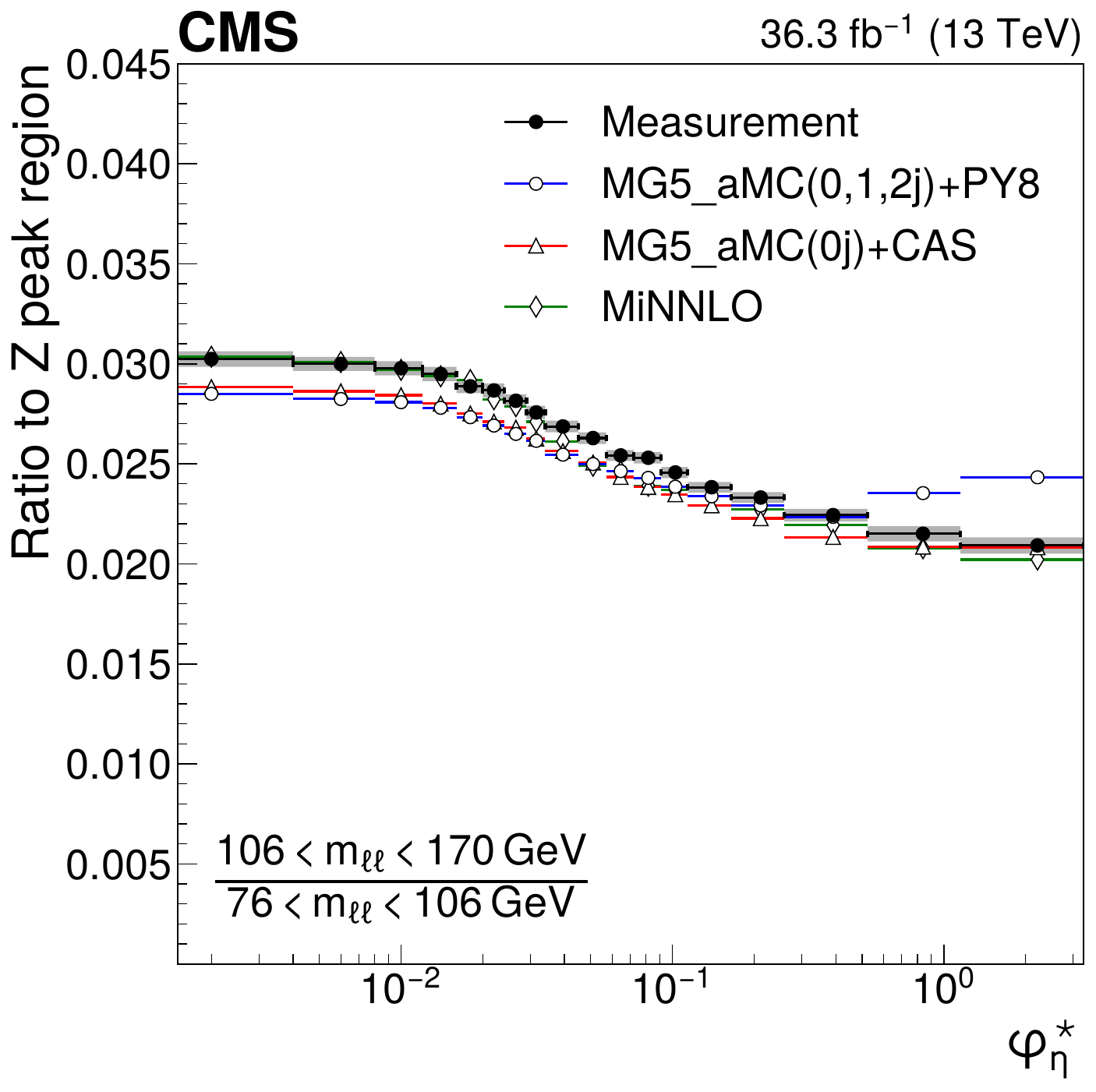}
 \includegraphics[height=0.49\textwidth]{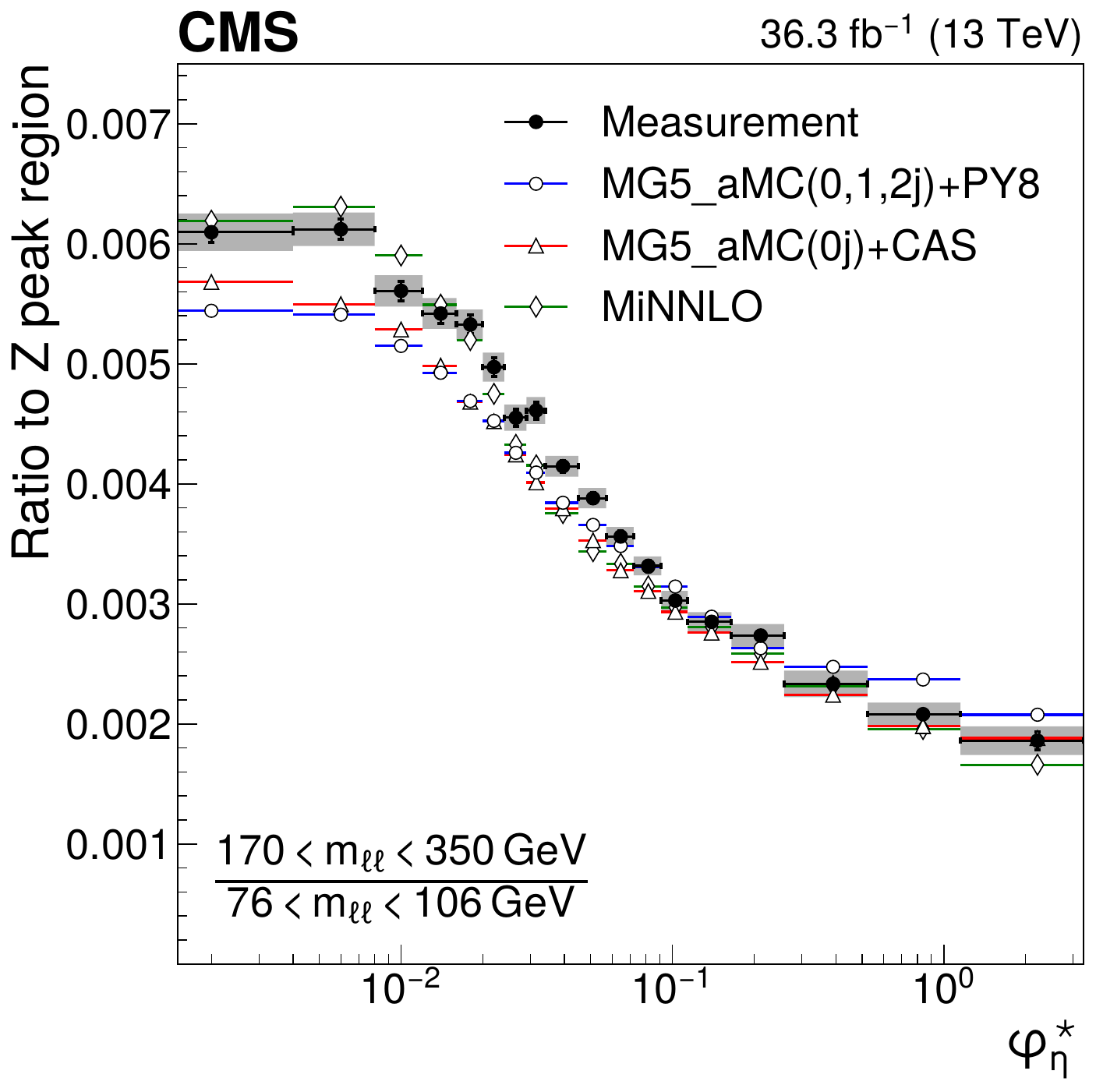}
 \includegraphics[height=0.49\textwidth]{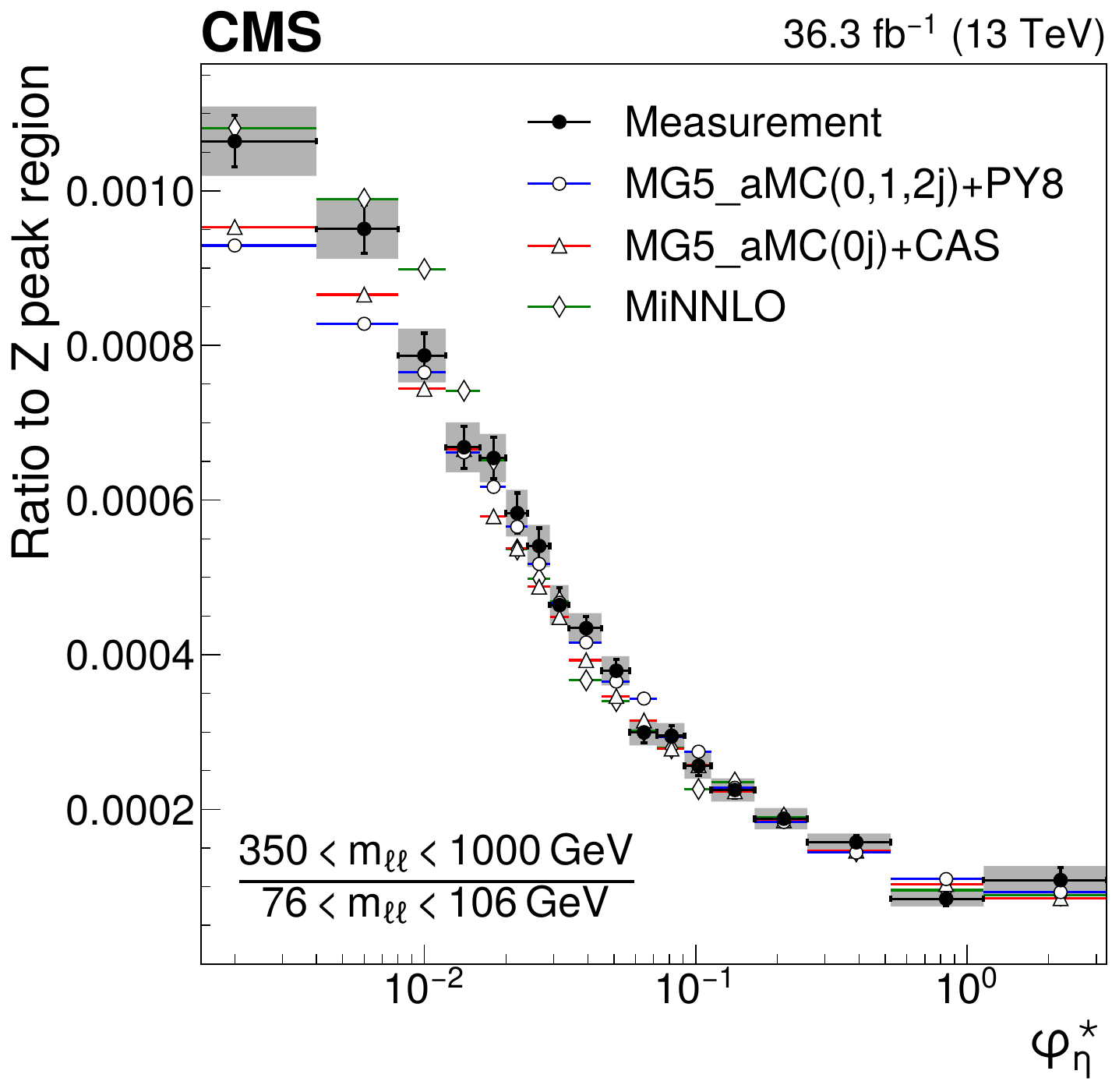}
 \caption{
  Ratios of differential cross sections in $\phistar(\Pell\Pell)$ for invariant mass ranges
  with respect to the peak region \mrangeb:
  \mrangea (upper left), \mrangec (upper right), \mranged (lower left),
  and \mrangee (lower right).
 The measurement is compared with \MGaMC (0, 1, and 2 jets at NLO) + \PYTHIA8 (blue dots), \minnlo (green diamonds) and \MGaMC (0 jet at NLO)+ PB (\cascade) (red triangles).
 Details on the presentation of the results are given in Fig.~\ref{fig:UnfCombPt0} caption.
         }
 \label{fig:UnfCombPhiRatios}
\end{figure*}

\begin{figure*}[htbp!]
 \centering
 \includegraphics[width=0.48\textwidth]{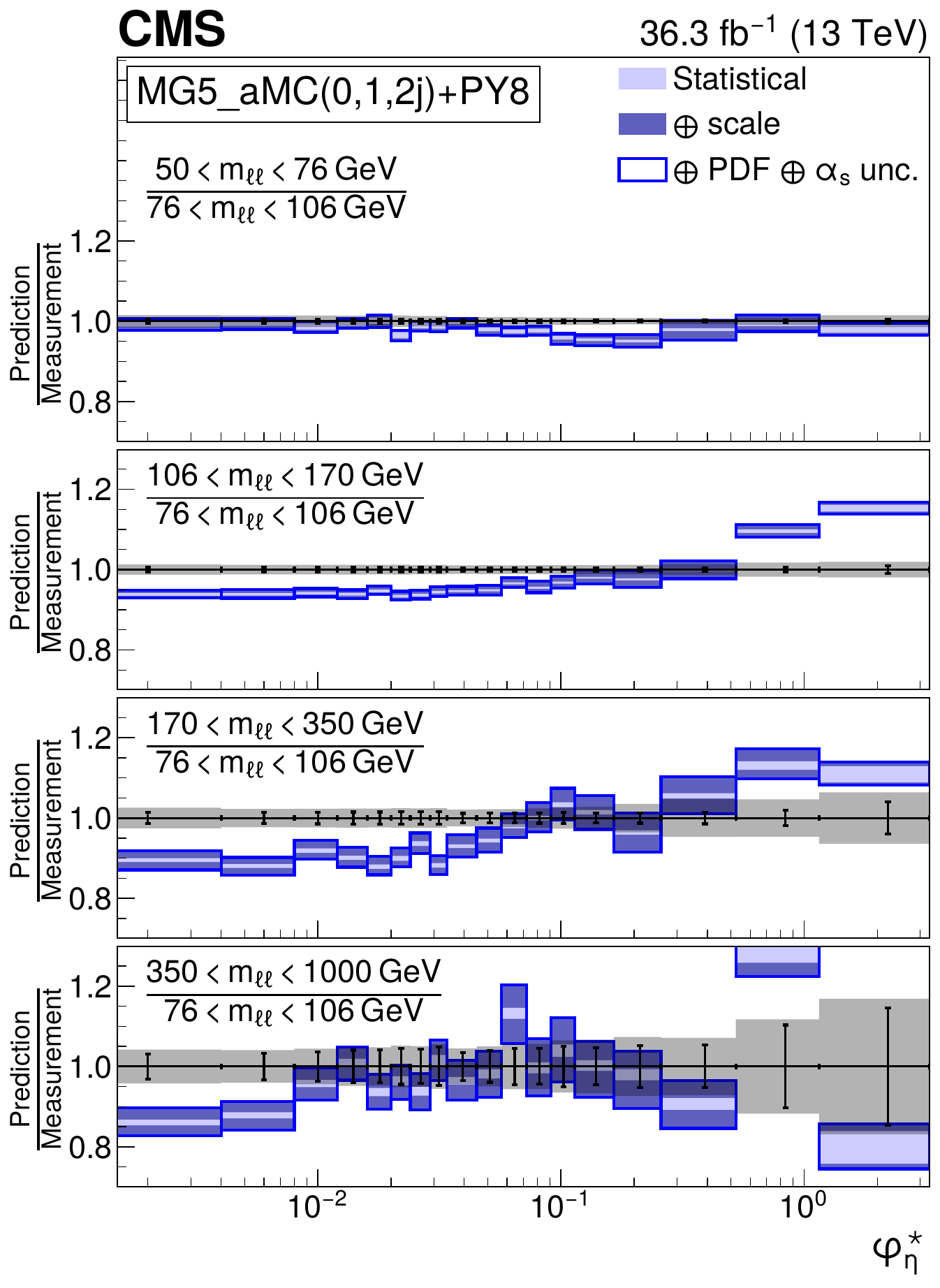}
 \includegraphics[width=0.48\textwidth]{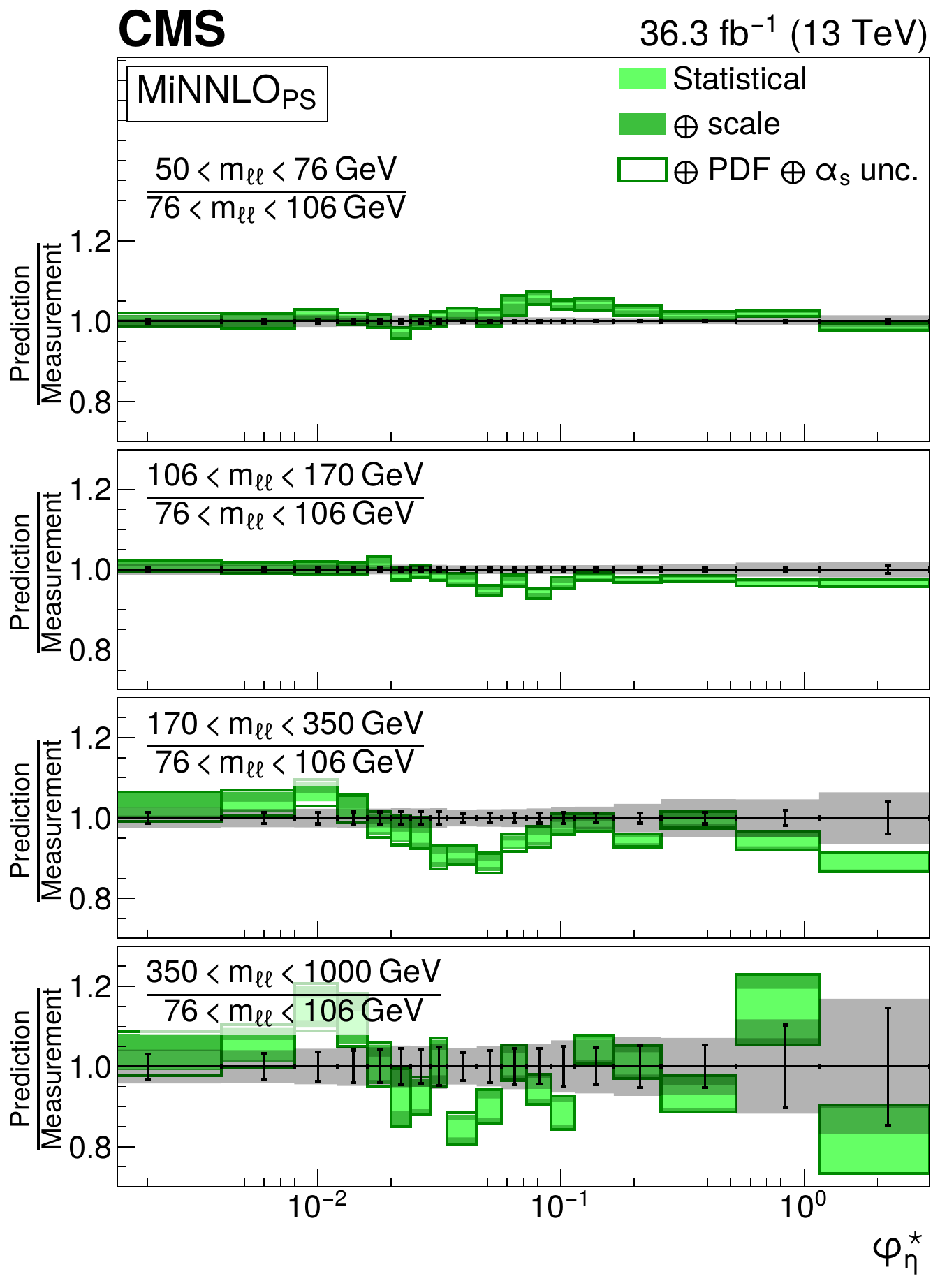}
 \includegraphics[width=0.48\textwidth]{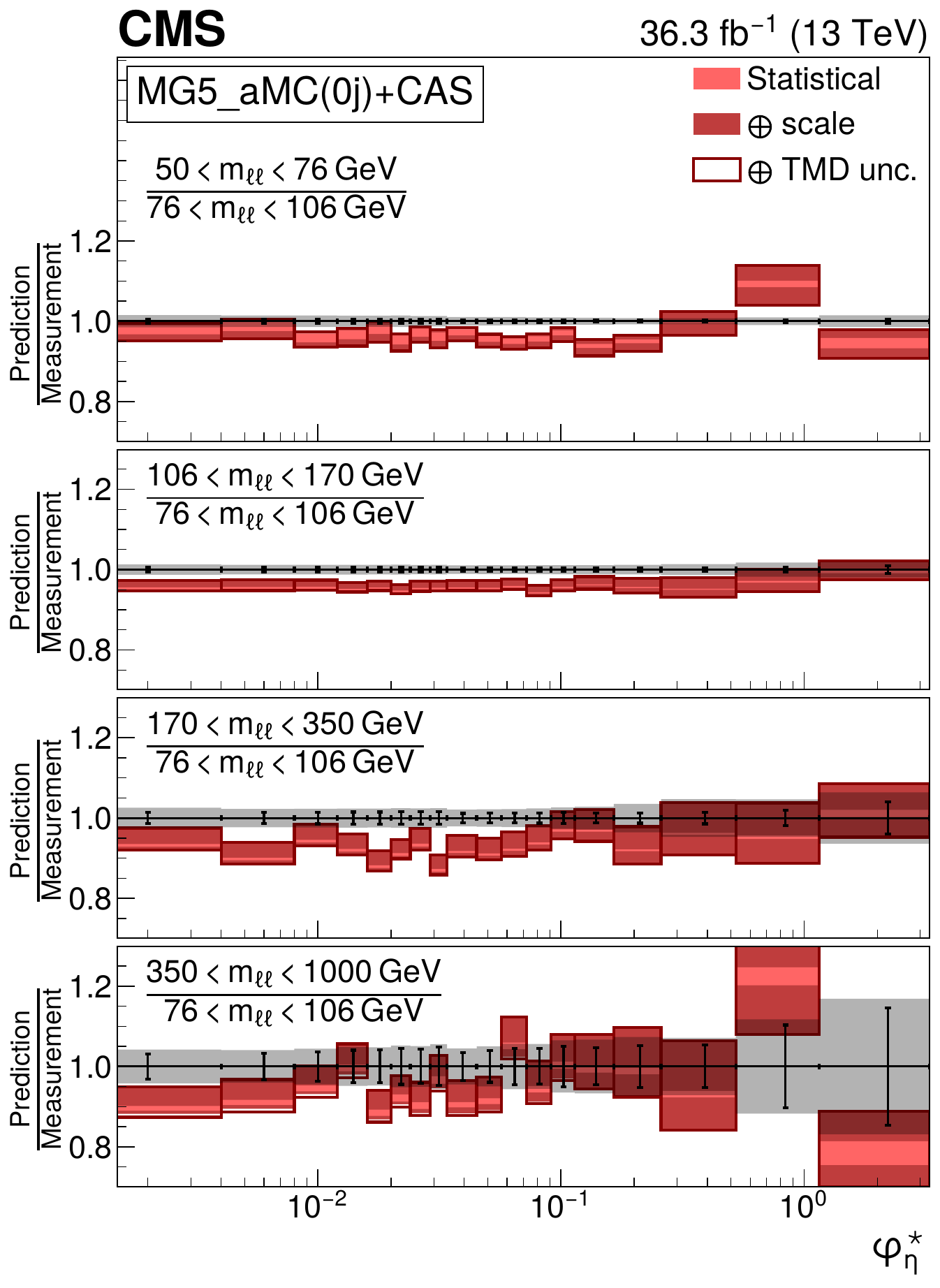}
 \caption{
  Ratios of differential cross sections in $\phistar(\Pell\Pell)$ for invariant \Mll
  with respect to the peak region \mrangeb. Compared to model predictions
  from \MGaMC (0, 1, and 2 jets at NLO) + \PYTHIA8 (upper left), \minnlo (upper right)
  and \MGaMC (0 jet at NLO) + PB (\cascade) (lower).
  Details on the presentation of the results are given in Fig.~\ref{fig:UnfCombPt0a} caption.
         }
 \label{fig:UnfCombPhiRatiosa}
\end{figure*}

\begin{figure*}[htbp!]
 \centering
 \includegraphics[width=0.48\textwidth]{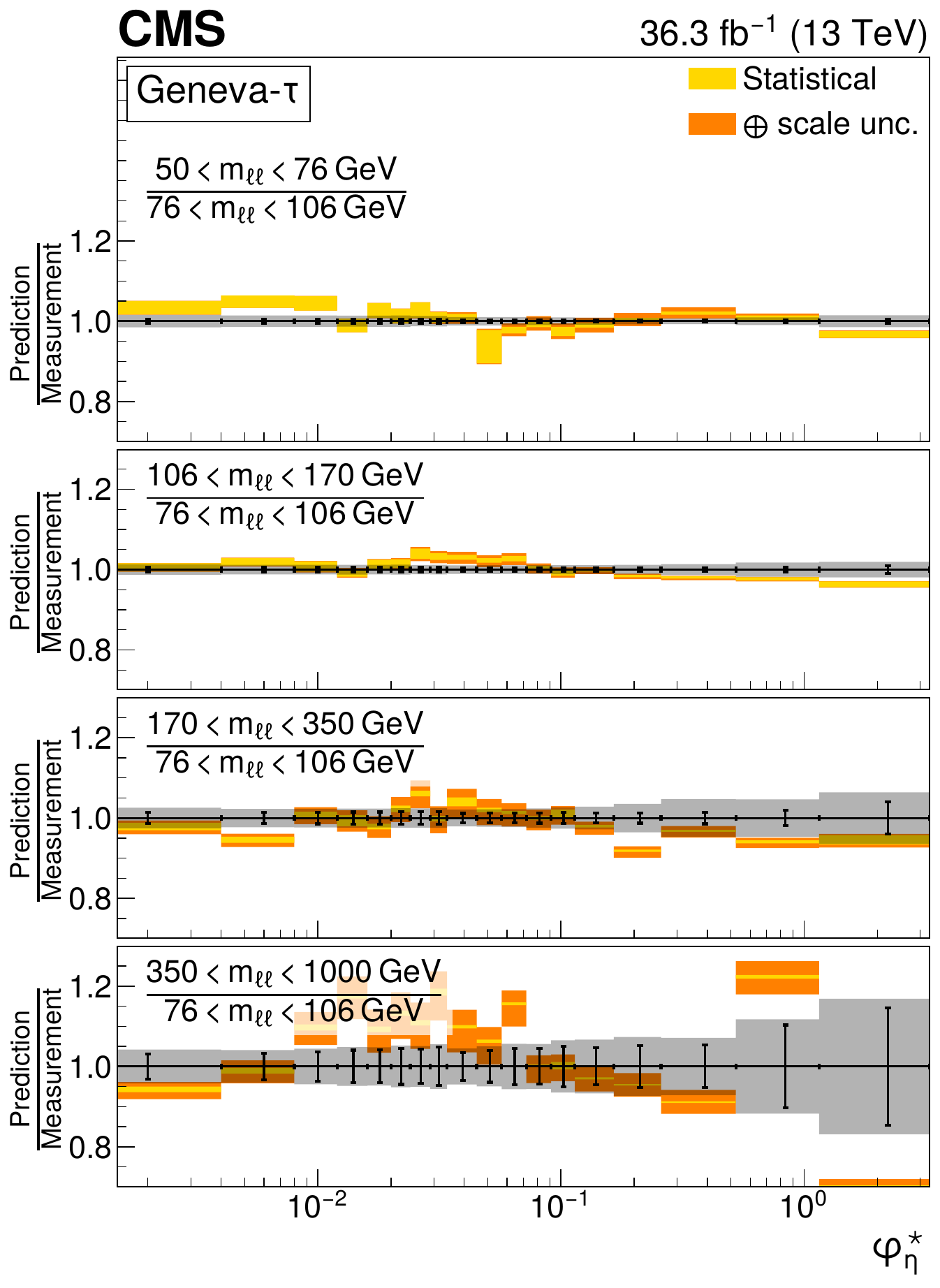}
 \includegraphics[width=0.48\textwidth]{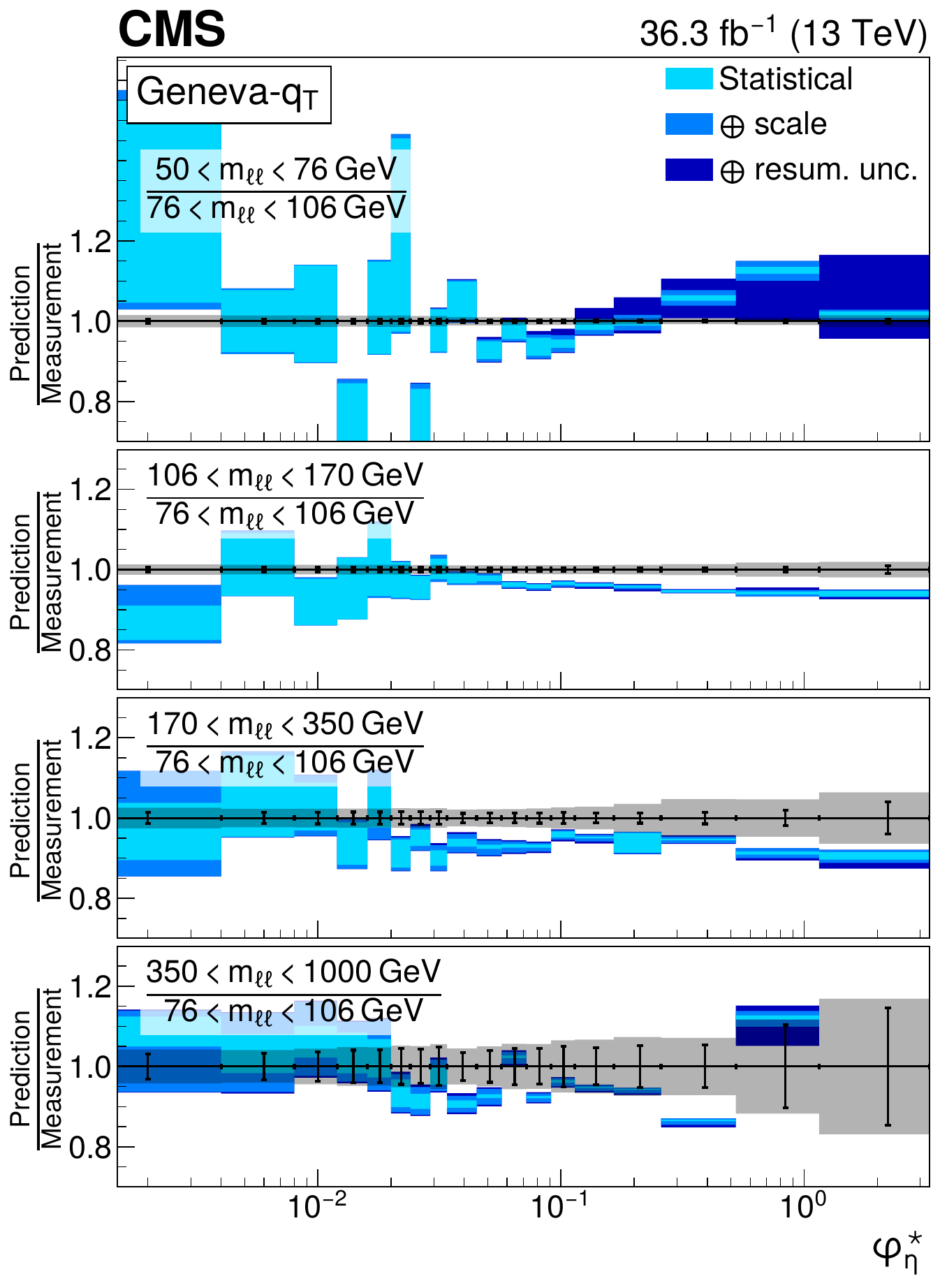}
 \caption{
  Ratios of differential cross sections in $\phistar(\Pell\Pell)$ for invariant \Mll
  with respect to the peak region \mrangeb. Compared to model predictions
  from \GE-$\tau$ (left) and \GE-\qt (right).
  Details on the presentation of the results are given in Fig.~\ref{fig:UnfCombPt0c} caption.
   }
 \label{fig:UnfCombPhiRatiosb}
\end{figure*}

The \phistar variable is highly correlated with \PTll and it offers a
complementary access to the underlying QCD dynamics.
Being based only on angle measurements of the final-state charged leptons,
the \phistar variable can be measured with greater accuracy which allows us to
include the muon channel for all \Mll ranges.
Figure~\ref{fig:UnfCombPhi} presents the inclusive differential cross sections in \phistar for the same invariant mass
ranges as above and comparisons to models. More complete comparisons to model predictions are
presented as ratios of the prediction divided by the measurement in Figs.~\ref{fig:UnfCombPhia}
and~\ref{fig:UnfCombPhib}. The results are discussed below. 
The ratio of these differential cross sections for various \Mll ranges are computed with respect to the 
same distribution in the \PZ peak region.
They are shown in Fig.~\ref{fig:UnfCombPhiRatios} and further compared with models in 
Figs.~\ref{fig:UnfCombPhiRatiosa} and~\ref{fig:UnfCombPhiRatiosb}.

The \phistar distributions are monotonic functions, in particular they do not 
present a peak structure as measured in the \PTll distributions. At small values, the \phistar distributions
contain a plateau whose length decreases with increasing \Mll, and more generally the \phistar distributions
fall more rapidly with increasing \Mll as clearly shown in Fig.~\ref{fig:UnfCombPhi}.
Because the lepton direction is much less affected by QED FSR than the energy,
the effect of migrations from the \PZ boson mass bin towards lower masses
is relatively invisible in the \phistar shape as highlighted 
by the ratio distribution in Fig.~\ref{fig:UnfCombPhiRatios} (upper left).

Since \phistar is highly correlated with \PTll, the comparison of the \phistar distributions to theoretical predictions
leads to the same basic observations and remarks as related above. 
The \MGPE prediction describes the measured \phistar distributions well globally and predicts a too
small cross section in the region sensitive to gluon resummation, \ie,
$\phistar\lesssim0.1$ on the \PZ boson mass peak, as shown in Fig.~\ref{fig:UnfCombPhia}.
The increase of this disagreement for higher \Mll is also observed, clearly visible in
the ratio distributions of Fig.~\ref{fig:UnfCombPhiRatiosa}.

As for the \PTll distributions, the \minnlo prediction provides the best global description of the data
(Fig.~\ref{fig:UnfCombPhia}).
In contrast to the disagreement for \PTll above 400\GeV for \Mll around the \PZ peak that appeared
both in the inclusive case (Fig.~\ref{fig:UnfCombPt0a}) and in the one jet case 
(Fig.~\ref{fig:UnfCombPt1Ratiosa}) the large \phistar values are well described by \minnlo.
The inclusion of NNLO corrections reduces scale uncertainties making the PDF uncertainty
dominant for medium \phistar values in the central \Mll bins. The PDF uncertainty is significantly reduced in
the ratio distributions (Fig.~\ref{fig:UnfCombPhiRatiosa}) leading to remarkable prediction precision 
of the level of 1.5\% in several bins.

The \MGCAS prediction describes well the measured shapes for $\phistar \lesssim 0.1$ in all
\Mll bins (Fig.~\ref{fig:UnfCombPhia}). This contrasts with the description of the \PTll
dependence by the same prediction (Fig.~\ref{fig:UnfCombPt0b}), owing to the washing out of
the details of the \PTll distribution in the \phistar distribution.
The normalisation of the prediction is good for the \PZ boson mass peak region but underestimates 
more and more the cross section with increasing \Mll, in a way relatively close to \MGaMC predictions.
The ratio distributions (Fig.~\ref{fig:UnfCombPhiRatiosa}) also illustrate this, but a compensation effect 
leads to predictions in agreement over the full \phistar range.

The measured cross sections as a function of \phistar are compared with \GE predictions in
Fig.~\ref{fig:UnfCombPhib}.
Similar to previous discussions of the \PTll distributions, \GE-\qt improves significantly the
description of the data with respect to \GE-$\tau$. The discrepancy of  \GE-\qt  for low \PTll
values in the two lowest \Mll bins is smoothed here leading to a global agreement everywhere. 
The cross section ratio distributions of the different \Mll bins over the \PZ boson mass peak bin, 
as a function of \phistar are shown in Fig.~\ref{fig:UnfCombPhiRatiosb}. Here both \GE predictions
provide a good description of the measurements. This indicates that, although the precise shape in
\phistar is not well reproduced by \GE-$\tau$, the scale dependence is well described 
over the large range covered by the present measurement.

The differential cross section measurements are presented in the HEPData entry~\cite{hepdata}.



\clearpage

\section{Summary}
\label{sec:conclusion}

Measurements of differential Drell--Yan cross sections in
proton-proton collisions at $\sqrt{s} = 13\TeV$ in the dielectron and
dimuon final states are presented, using data collected  with the CMS
detector, corresponding to an integrated luminosity of 36.3\fbinv. The
measurements are corrected for detector effects and the two leptonic
channels are combined. Differential cross sections in the dilepton transverse
momentum, \PTll, and in the lepton angular variable \phistar are measured for
different values of the dilepton mass, \Mll, between 50\GeV and 1\TeV.
To highlight the evolution with the dilepton mass scale, ratios of these distributions 
for various masses are presented. 
In addition, dilepton transverse momentum distributions are shown in the presence of at
least one jet within the detector acceptance. 

The rising behaviour of the Drell--Yan inclusive cross section at small
\PTll is attributed to soft QCD radiations, whereas the tail at
large \PTll is only expected to be well described by models relying on higher-order matrix element 
calculations. Therefore, this variable provides a good
sensitivity to initial-state QCD radiations and can be compared with
different predictions relying on matrix element calculations at different orders and 
using different methods to resum the initial-state soft QCD radiations.
 The measurements show that the peak in the  \PTll distribution, located around 5\GeV, is not significantly modified
by changing the \Mll value in the covered range. However, for higher values of \Mll above the peak,  
the \PTll distributions fall less steeply.

The \phistar variable, highly correlated with \PTll, offers a complementary access 
to the underlying QCD dynamics.
Since it is based only on angle measurements of the final-state charged leptons,
it offers, a priori, measurements with greater accuracy. 
However, these measurements demonstrate that the \phistar distributions discriminate between the models less
than the \PTll distributions, since they wash out the peak structure of the \PTll
distributions, which reflect the initial-state QCD radiation effects in a more detailed way.

\begin{tolerant}{5000}
This publication presents comparisons of the measurements to six predictions
using different treatments of soft initial-state QCD radiations.
Two of them, \MGPE and \minnlo, are based on a matrix element calculation merged with parton
showers. Two others, \artemide and \cascade use transverse momentum dependent parton 
distributions (TMD). Finally, \GE combines a higher-order resummation with a Drell--Yan 
calculation at next-to-next-to-leading order (NNLO), in two different ways. One carries out the resummation 
at next-to-next-to-leading logarithm in the 0-jettiness variable $\tau_0$, the other at \NqLL in the \qt variable.
\end{tolerant}

\begin{tolerant}{5000}
The comparison of the measurement with the \MGPE Monte Carlo predictions using 
matrix element calculations including $\PZ + 0,1,2$ partons at next-to-leading order (NLO) merged with
a parton shower, shows generally good agreement, except at \PTll values
below 10\GeV both for the inclusive and one jet cross sections. 
This disagreement is enhanced for masses away from the \PZ
mass peak and is more pronounced for the higher dilepton masses, 
reaching 20\% for the highest mass bin.
\end{tolerant}

\begin{tolerant}{5000}
The \minnlo prediction provides the best global description of the data among the predictions
presented in this paper, both for the inclusive and the one jet cross sections. 
This approach, based on NNLO matrix element and \PYTHIA8 parton shower and MPI, describes well the large \PTll
cross sections and ratios, except for \PTll values above 400\GeV for dilepton masses around the \PZ mass peak. 
A good description of the medium and low \PTll cross sections is obtained using a 
modified primordial \kt parameter of the CP5 parton shower tune.
\end{tolerant}

\begin{tolerant}{5000}
\MGCAS predictions are based on Parton Branching TMDs obtained only from a 
fit to electron-proton deep
inelastic scattering measurements performed at HERA. These TMDs are merged with NLO matrix
element calculations. 
Low \PTll values are globally well described but with
too low cross sections at medium \PTll values. This discrepancy increases with increasing 
\Mll in a way similar to the \MGPE predictions. 
The high part of the \PTll distribution is not described by \cascade due to missing
higher fixed-order terms. 
The model can not describe the low \PTll region of the cross section in the presence of one jet 
due to the missing double parton scattering contributions. The recent inclusion of multi-jet merging allows a larger \PTll\  region to be described as well.  
\end{tolerant}

\begin{tolerant}{5000}
\artemide provides predictions based on TMDs extracted from previous measurements
including the Drell-Yan transverse momentum cross section at the LHC at the \PZ mass peak. 
By construction, the validity of \artemide predictions are limited to the range 
$\PTll < 0.2 \ \Mll$. In that range, they describe well the present measurements
up to the highest dilepton masses.
\end{tolerant}

\begin{tolerant}{5000}
The \GE prediction, combining resummation in the 0-jettiness variable $\tau_0$ (\GE-$\tau$)
and NNLO matrix element does not describe the measurement well for \PTll values below 40\GeV.
For the high \PTll region the inclusion of NNLO
in the matrix element provides a good description of the measured cross section.
The recent \GE prediction (\GE-\qt), using a \qt resummation, provides a much better
description of the measured
inclusive cross sections, describing very well the data in the full \PTll range except for middle \PTll
values in the lowest mass bin.
Both \GE approaches predict too hard \PTll spectra for the one jet cross sections.
\end{tolerant}

\begin{tolerant}{5000}
The ratio distributions presented in this paper confirm most of the observations based on the
comparison between the measurement and the predictions at the cross section level.
The observed scale dependence is well described by the different models.
Furthermore the partial cancellation of the uncertainties in the cross section ratios 
allows a higher level of precision to be reached for both the measurement and the predictions.
\end{tolerant}

\begin{tolerant}{5000}
 The present analysis shows the relevance of measuring the Drell--Yan cross section in a wide range in
dilepton masses to probe the interplay between the transverse momentum and the mass scales of the process.
Important theoretical efforts have been made during the last decade to improve the detailed
description of high energy processes involving multiple scales and partonic final states.
The understanding of the Drell--Yan process directly benefited from these developments.
The present paper shows that they individually describe the measurements well in the regions they were  
designed for. Nevertheless, no model is able to reproduce all dependencies over the complete covered range.   
Further progress might come from combining these approaches.
\end{tolerant}

\begin{acknowledgments}
We congratulate our colleagues in the CERN accelerator departments for the excellent performance of the LHC and thank the technical and administrative staffs at CERN and at other CMS institutes for their contributions to the success of the CMS effort. In addition, we gratefully acknowledge the computing centres and personnel of the Worldwide LHC Computing Grid and other centres for delivering so effectively the computing infrastructure essential to our analyses. Finally, we acknowledge the enduring support for the construction and operation of the LHC, the CMS detector, and the supporting computing infrastructure provided by the following funding agencies: BMBWF and FWF (Austria); FNRS and FWO (Belgium); CNPq, CAPES, FAPERJ, FAPERGS, and FAPESP (Brazil); MES and BNSF (Bulgaria); CERN; CAS, MoST, and NSFC (China); MINCIENCIAS (Colombia); MSES and CSF (Croatia); RIF (Cyprus); SENESCYT (Ecuador); MoER, ERC PUT and ERDF (Estonia); Academy of Finland, MEC, and HIP (Finland); CEA and CNRS/IN2P3 (France); BMBF, DFG, and HGF (Germany); GSRI (Greece); NKFIH (Hungary); DAE and DST (India); IPM (Iran); SFI (Ireland); INFN (Italy); MSIP and NRF (Republic of Korea); MES (Latvia); LAS (Lithuania); MOE and UM (Malaysia); BUAP, CINVESTAV, CONACYT, LNS, SEP, and UASLP-FAI (Mexico); MOS (Montenegro); MBIE (New Zealand); PAEC (Pakistan); MES and NSC (Poland); FCT (Portugal); J MESTD (Serbia); MCIN/AEI and PCTI (Spain); MOSTR (Sri Lanka); Swiss Funding Agencies (Switzerland); MST (Taipei); MHESI and NSTDA (Thailand); TUBITAK and TENMAK (Turkey); NASU (Ukraine); STFC (United Kingdom); DOE and NSF (USA).
    
\hyphenation{Rachada-pisek} Individuals have received support from the Marie-Curie programme and the European Research Council and Horizon 2020 Grant, contract Nos.\ 675440, 724704, 752730, 758316, 765710, 824093, 884104, and COST Action CA16108 (European Union); the Leventis Foundation; the Alfred P.\ Sloan Foundation; the Alexander von Humboldt Foundation; the Belgian Federal Science Policy Office; the Fonds pour la Formation \`a la Recherche dans l'Industrie et dans l'Agriculture (FRIA-Belgium); the Agentschap voor Innovatie door Wetenschap en Technologie (IWT-Belgium); the F.R.S.-FNRS and FWO (Belgium) under the ``Excellence of Science -- EOS" -- be.h project n.\ 30820817; the Beijing Municipal Science \& Technology Commission, No. Z191100007219010; the Ministry of Education, Youth and Sports (MEYS) of the Czech Republic; the Hellenic Foundation for Research and Innovation (HFRI), Project Number 2288 (Greece); the Deutsche Forschungsgemeinschaft (DFG), under Germany's Excellence Strategy -- EXC 2121 ``Quantum Universe" -- 390833306, and under project number 400140256 - GRK2497; the Hungarian Academy of Sciences, the New National Excellence Program - \'UNKP, the NKFIH research grants K 124845, K 124850, K 128713, K 128786, K 129058, K 131991, K 133046, K 138136, K 143460, K 143477, 2020-2.2.1-ED-2021-00181, and TKP2021-NKTA-64 (Hungary); the Council of Science and Industrial Research, India; the Latvian Council of Science; the Ministry of Education and Science, project no. 2022/WK/14, and the National Science Center, contracts Opus 2021/41/B/ST2/01369 and 2021/43/B/ST2/01552 (Poland); the Funda\c{c}\~ao para a Ci\^encia e a Tecnologia, grant CEECIND/01334/2018 (Portugal); the National Priorities Research Program by Qatar National Research Fund; MCIN/AEI/10.13039/501100011033, ERDF ``a way of making Europe", and the Programa Estatal de Fomento de la Investigaci{\'o}n Cient{\'i}fica y T{\'e}cnica de Excelencia Mar\'{\i}a de Maeztu, grant MDM-2017-0765 and Programa Severo Ochoa del Principado de Asturias (Spain); the Chulalongkorn Academic into Its 2nd Century Project Advancement Project, and the National Science, Research and Innovation Fund via the Program Management Unit for Human Resources \& Institutional Development, Research and Innovation, grant B05F650021 (Thailand); the Kavli Foundation; the Nvidia Corporation; the SuperMicro Corporation; the Welch Foundation, contract C-1845; and the Weston Havens Foundation (USA).
\end{acknowledgments}

\bibliography{auto_generated}   

\providecommand{\href}[2]{#2}\begingroup\raggedright\begin{thebibliography}{100}%
\makeatletter
\providecommand{\hrefCMSnoop }[0]{\@secondoftwo}%
\makeatother
\providecommand{\doi}{\texttt{doi:}\begingroup \urlstyle{tt}\Url}

\bibitem{Drell:1970wh}
\hrefCMSnoop {}{S.~D. Drell and T.-M. Yan, ``{Massive lepton pair production in
  hadron-hadron collisions at high energies}'',} \textit{ Phys. Rev. Lett.}
  \textbf{ 25} (1970) 316,
  \href{http://dx.doi.org/10.1103/PhysRevLett.25.316}{\doi{10.1103/PhysRevLett.25.316}}.
[Erratum: \DOI{10.1103/PhysRevLett.25.902.2}].

\bibitem{Dokshitzer:1978yd}
\hrefCMSnoop {}{Y.~L. Dokshitzer, D.~Diakonov, and S.~I. Troyan, ``{On the
  transverse momentum distribution of massive lepton pairs}'',} \textit{ Phys.
  Lett. B} \textbf{ 79} (1978) 269,
\href{http://dx.doi.org/10.1016/0370-2693(78)90240-X}{\doi{10.1016/0370-2693(78)90240-X}}.

\bibitem{Collins:1984kg}
\hrefCMSnoop {}{J.~C. Collins, D.~E. Soper, and G.~F. Sterman, ``{Transverse
  momentum distribution in Drell-Yan pair and W and Z boson production}'',}
  \textit{ Nucl. Phys. B} \textbf{ 250} (1985) 199,
\href{http://dx.doi.org/10.1016/0550-3213(85)90479-1}{\doi{10.1016/0550-3213(85)90479-1}}.

\bibitem{Hamberg:1990np}
\hrefCMSnoop {}{R.~Hamberg, W.~L. van Neerven, and T.~Matsuura, ``{A complete
  calculation of the order $\alpha_s^2$ correction to the Drell-Yan
  $K$-factor}'',} \textit{ Nucl. Phys. B} \textbf{ 359} (1991) 343,
  \href{http://dx.doi.org/10.1016/S0550-3213(02)00814-3}{\doi{10.1016/S0550-3213(02)00814-3}}.
[Erratum: \DOI{10.1016/0550-3213(91)90064-5}].

\bibitem{Catani:2009sm}
S.~Catani\hrefCMSnoop {}{ {et~al.}, ``{Vector boson production at hadron
  colliders: a fully exclusive QCD calculation at NNLO}'',} \textit{ Phys. Rev.
  Lett.} \textbf{ 103} (2009) 082001,
  \href{http://dx.doi.org/10.1103/PhysRevLett.103.082001}{\doi{10.1103/PhysRevLett.103.082001}},
\href{http://www.arXiv.org/abs/0903.2120}{\texttt{arXiv:0903.2120}}.

\bibitem{Catani:2007vq}
\hrefCMSnoop {}{S.~Catani and M.~Grazzini, ``{An NNLO subtraction formalism in
  hadron collisions and its application to Higgs boson production at the
  LHC}'',} \textit{ Phys. Rev. Lett.} \textbf{ 98} (2007) 222002,
  \href{http://dx.doi.org/10.1103/PhysRevLett.98.222002}{\doi{10.1103/PhysRevLett.98.222002}},
\href{http://www.arXiv.org/abs/hep-ph/0703012}{\texttt{arXiv:hep-ph/0703012}}.

\bibitem{Melnikov:2006kv}
\hrefCMSnoop {}{K.~Melnikov and F.~Petriello, ``{Electroweak gauge boson
  production at hadron colliders through $\mathcal O(\alpha_s^2)$}'',} \textit{
  Phys. Rev. D} \textbf{ 74} (2006) 114017,
  \href{http://dx.doi.org/10.1103/PhysRevD.74.114017}{\doi{10.1103/PhysRevD.74.114017}},
\href{http://www.arXiv.org/abs/hep-ph/0609070}{\texttt{arXiv:hep-ph/0609070}}.

\bibitem{Bacchetta:2017gcc}
A.~Bacchetta\hrefCMSnoop {}{ {et~al.}, ``{Extraction of partonic transverse
  momentum distributions from semi-inclusive deep-inelastic scattering,
  Drell-Yan and Z-boson production}'',} \textit{ JHEP} \textbf{ 06} (2017) 081,
  \href{http://dx.doi.org/10.1007/JHEP06(2017)081}{\doi{10.1007/JHEP06(2017)081}},
  \href{http://www.arXiv.org/abs/1703.10157}{\texttt{arXiv:1703.10157}}.
  [Erratum: JHEP 06, 051 (2019)].

\bibitem{Bacchetta:2019sam}
A.~Bacchetta\hrefCMSnoop {}{ {et~al.}, ``{Transverse-momentum-dependent parton
  distributions up to N$^{3}$LL from Drell-Yan data}'',} \textit{ JHEP}
  \textbf{ 07} (2020) 117,
  \href{http://dx.doi.org/10.1007/JHEP07(2020)117}{\doi{10.1007/JHEP07(2020)117}},
  \href{http://www.arXiv.org/abs/1912.07550}{\texttt{arXiv:1912.07550}}.

\bibitem{Camarda:2019zyx}
\hrefCMSnoop {}{S.~Camarda {et~al.}, ``{DYTurbo: Fast predictions for Drell-Yan
  processes}'',} \textit{ Eur. Phys. J. C} \textbf{ 80} (2020) 251,
  \href{http://dx.doi.org/10.1140/epjc/s10052-020-7757-5}{\doi{10.1140/epjc/s10052-020-7757-5}},
  \href{http://www.arXiv.org/abs/1910.07049}{\texttt{arXiv:1910.07049}}.
  [Erratum: Eur.Phys.J.C 80, 440 (2020)].

\bibitem{Bizon:2018foh}
W.~Bizo{\'n}\hrefCMSnoop {}{ {et~al.}, ``{Fiducial distributions in Higgs and
  Drell-Yan production at N$^{3}$LL+NNLO}'',} \textit{ JHEP} \textbf{ 12}
  (2018) 132,
  \href{http://dx.doi.org/10.1007/JHEP12(2018)132}{\doi{10.1007/JHEP12(2018)132}},
  \href{http://www.arXiv.org/abs/1805.05916}{\texttt{arXiv:1805.05916}}.

\bibitem{Ebert:2020dfc}
\hrefCMSnoop {}{M.~A. Ebert, J.~K.~L. Michel, I.~W. Stewart, and F.~J.
  Tackmann, ``{Drell-Yan $q_{T}$ resummation of fiducial power corrections at
  N$^{3}$LL}'',} \textit{ JHEP} \textbf{ 04} (2021) 102,
  \href{http://dx.doi.org/10.1007/JHEP04(2021)102}{\doi{10.1007/JHEP04(2021)102}},
  \href{http://www.arXiv.org/abs/2006.11382}{\texttt{arXiv:2006.11382}}.

\bibitem{Becher:2020ugp}
\hrefCMSnoop {}{T.~Becher and T.~Neumann, ``{Fiducial $q_T$ resummation of
  color-singlet processes at N$^3$LL+NNLO}'',} \textit{ JHEP} \textbf{ 03}
  (2021) 199,
  \href{http://dx.doi.org/10.1007/JHEP03(2021)199}{\doi{10.1007/JHEP03(2021)199}},
  \href{http://www.arXiv.org/abs/2009.11437}{\texttt{arXiv:2009.11437}}.

\bibitem{Hautmann:2017xtx}
F.~Hautmann\hrefCMSnoop {}{ {et~al.}, ``{Soft-gluon resolution scale in QCD
  evolution equations}'',} \textit{ Phys. Lett. B} \textbf{ 772} (2017) 446,
  \href{http://dx.doi.org/10.1016/j.physletb.2017.07.005}{\doi{10.1016/j.physletb.2017.07.005}},
  \href{http://www.arXiv.org/abs/1704.01757}{\texttt{arXiv:1704.01757}}.

\bibitem{Hautmann:2017fcj}
F.~Hautmann\hrefCMSnoop {}{ {et~al.}, ``{Collinear and TMD quark and gluon
  densities from parton branching solution of QCD evolution equations}'',}
  \textit{ JHEP} \textbf{ 01} (2018) 070,
  \href{http://dx.doi.org/10.1007/JHEP01(2018)070}{\doi{10.1007/JHEP01(2018)070}},
  \href{http://www.arXiv.org/abs/1708.03279}{\texttt{arXiv:1708.03279}}.

\bibitem{Banfi:2010cf}
A.~Banfi\hrefCMSnoop {}{ {et~al.}, ``{Optimisation of variables for studying
  dilepton transverse momentum distributions at hadron colliders}'',} \textit{
  Eur. Phys. J. C} \textbf{ 71} (2011) 1600,
  \href{http://dx.doi.org/10.1140/epjc/s10052-011-1600-y}{\doi{10.1140/epjc/s10052-011-1600-y}},
\href{http://www.arXiv.org/abs/1009.1580}{\texttt{arXiv:1009.1580}}.

\bibitem{Banfi:2012du}
\hrefCMSnoop {}{A.~Banfi, M.~Dasgupta, S.~Marzani, and L.~Tomlinson,
  ``{Predictions for Drell-Yan $\phi^*$ and $Q_T$ observables at the LHC}'',}
  \textit{ Phys. Lett. B} \textbf{ 715} (2012) 152,
  \href{http://dx.doi.org/10.1016/j.physletb.2012.07.035}{\doi{10.1016/j.physletb.2012.07.035}},
\href{http://www.arXiv.org/abs/1205.4760}{\texttt{arXiv:1205.4760}}.

\bibitem{Marzani:2013nza}
\hrefCMSnoop {}{S.~Marzani, ``{$Q_T$ and $\phi^*$ observables in Drell-Yan
  processes}'',} \textit{ Eur. Phys. J. Web Conf.} \textbf{ 49} (2013) 14007,
\href{http://dx.doi.org/10.1051/epjconf/20134914007}{\doi{10.1051/epjconf/20134914007}}.

\bibitem{Moureaux:2783681}
\href {https://cds.cern.ch/record/2783681}{L.~Moureaux, ``{Measurement of the
  transverse momentum of Drell-Yan lepton pairs over a wide mass range in
  proton-proton collisions at $\sqrt s = 13\:\mathrm{TeV}$ in CMS}''}.
\newblock PhD thesis, Universit{\'e} libre de Bruxelles, 2021.
\newblock Presented 24 Sep 2021.

\bibitem{CMS:2011aa}
\hrefCMSnoop {}{{CMS Collaboration}, ``{Measurement of the inclusive $W$ and
  $Z$ production cross sections in $\mathrm{pp}$ collisions at $\sqrt{s}=7$
  TeV}'',} \textit{ JHEP} \textbf{ 10} (2011) 132,
  \href{http://dx.doi.org/10.1007/JHEP10(2011)132}{\doi{10.1007/JHEP10(2011)132}},
  \href{http://www.arXiv.org/abs/1107.4789}{\texttt{arXiv:1107.4789}}.

\bibitem{Chatrchyan:2011cm}
\hrefCMSnoop {}{{CMS Collaboration}, ``{Measurement of the Drell-Yan cross
  section in $\mathrm{pp}$ collisions at $\sqrt{s}=7$ TeV}'',} \textit{ JHEP}
  \textbf{ 10} (2011) 007,
  \href{http://dx.doi.org/10.1007/JHEP10(2011)007}{\doi{10.1007/JHEP10(2011)007}},
  \href{http://www.arXiv.org/abs/1108.0566}{\texttt{arXiv:1108.0566}}.

\bibitem{Chatrchyan:2013tia}
\hrefCMSnoop {}{{CMS Collaboration}, ``{Measurement of the differential and
  double-differential Drell-Yan cross sections in proton-proton collisions at
  $\sqrt{s} =$ 7 TeV}'',} \textit{ JHEP} \textbf{ 12} (2013) 030,
  \href{http://dx.doi.org/10.1007/JHEP12(2013)030}{\doi{10.1007/JHEP12(2013)030}},
  \href{http://www.arXiv.org/abs/1310.7291}{\texttt{arXiv:1310.7291}}.

\bibitem{CMS:2014jea}
\hrefCMSnoop {}{{CMS Collaboration}, ``{Measurements of differential and
  double-differential Drell-Yan cross sections in proton-proton collisions at 8
  TeV}'',} \textit{ Eur. Phys. J. C} \textbf{ 75} (2015) 147,
  \href{http://dx.doi.org/10.1140/epjc/s10052-015-3364-2}{\doi{10.1140/epjc/s10052-015-3364-2}},
  \href{http://www.arXiv.org/abs/1412.1115}{\texttt{arXiv:1412.1115}}.

\bibitem{Khachatryan:2015oaa}
\hrefCMSnoop {}{{CMS Collaboration}, ``{Measurement of the Z boson differential
  cross section in transverse momentum and rapidity in
  proton\textendash{}proton collisions at 8 TeV}'',} \textit{ Phys. Lett. B}
  \textbf{ 749} (2015) 187,
  \href{http://dx.doi.org/10.1016/j.physletb.2015.07.065}{\doi{10.1016/j.physletb.2015.07.065}},
  \href{http://www.arXiv.org/abs/1504.03511}{\texttt{arXiv:1504.03511}}.

\bibitem{Sirunyan:2018owv}
\hrefCMSnoop {}{{CMS Collaboration}, ``{Measurement of the differential
  Drell-Yan cross section in proton-proton collisions at $ \sqrt{\mathrm{s}} $
  = 13 TeV}'',} \textit{ JHEP} \textbf{ 12} (2019) 059,
  \href{http://dx.doi.org/10.1007/JHEP12(2019)059}{\doi{10.1007/JHEP12(2019)059}},
  \href{http://www.arXiv.org/abs/1812.10529}{\texttt{arXiv:1812.10529}}.

\bibitem{Sirunyan:2019bzr}
\hrefCMSnoop {}{{CMS Collaboration}, ``{Measurements of differential Z boson
  production cross sections in proton-proton collisions at $ \sqrt{s} $ = 13
  TeV}'',} \textit{ JHEP} \textbf{ 12} (2019) 061,
  \href{http://dx.doi.org/10.1007/JHEP12(2019)061}{\doi{10.1007/JHEP12(2019)061}},
  \href{http://www.arXiv.org/abs/1909.04133}{\texttt{arXiv:1909.04133}}.

\bibitem{Aad:2011dm}
\hrefCMSnoop {}{{ATLAS Collaboration}, ``{Measurement of the inclusive $W^\pm$
  and $Z/\gamma^*$ cross sections in the electron and muon decay channels in
  $\mathrm{pp}$ collisions at $\sqrt{s}=7$ TeV with the ATLAS detector}'',}
  \textit{ Phys. Rev. D} \textbf{ 85} (2012) 072004,
  \href{http://dx.doi.org/10.1103/PhysRevD.85.072004}{\doi{10.1103/PhysRevD.85.072004}},
  \href{http://www.arXiv.org/abs/1109.5141}{\texttt{arXiv:1109.5141}}.

\bibitem{Aad:2013iua}
\hrefCMSnoop {}{{ATLAS Collaboration}, ``{Measurement of the high-mass
  Drell--Yan differential cross-section in pp collisions at $\sqrt s=7$ TeV
  with the ATLAS detector}'',} \textit{ Phys. Lett. B} \textbf{ 725} (2013)
  223,
  \href{http://dx.doi.org/10.1016/j.physletb.2013.07.049}{\doi{10.1016/j.physletb.2013.07.049}},
  \href{http://www.arXiv.org/abs/1305.4192}{\texttt{arXiv:1305.4192}}.

\bibitem{Aad:2014xaa}
\hrefCMSnoop {}{{ATLAS Collaboration}, ``{Measurement of the $Z/\gamma^*$ boson
  transverse momentum distribution in pp collisions at $\sqrt{s}$ = 7 TeV with
  the ATLAS detector}'',} \textit{ JHEP} \textbf{ 09} (2014) 145,
  \href{http://dx.doi.org/10.1007/JHEP09(2014)145}{\doi{10.1007/JHEP09(2014)145}},
  \href{http://www.arXiv.org/abs/1406.3660}{\texttt{arXiv:1406.3660}}.

\bibitem{Aad:2014qja}
\hrefCMSnoop {}{{ATLAS Collaboration}, ``{Measurement of the low-mass Drell-Yan
  differential cross section at $\sqrt{s}$ = 7 TeV using the ATLAS
  detector}'',} \textit{ JHEP} \textbf{ 06} (2014) 112,
  \href{http://dx.doi.org/10.1007/JHEP06(2014)112}{\doi{10.1007/JHEP06(2014)112}},
  \href{http://www.arXiv.org/abs/1404.1212}{\texttt{arXiv:1404.1212}}.

\bibitem{Aad:2015auj}
\hrefCMSnoop {}{{ATLAS Collaboration}, ``{Measurement of the transverse
  momentum and $\phi ^*_{\eta }$ distributions of Drell\textendash{}Yan lepton
  pairs in proton\textendash{}proton collisions at $\sqrt{s}=8$ TeV with the
  ATLAS detector}'',} \textit{ Eur. Phys. J. C} \textbf{ 76} (2016) 291,
  \href{http://dx.doi.org/10.1140/epjc/s10052-016-4070-4}{\doi{10.1140/epjc/s10052-016-4070-4}},
  \href{http://www.arXiv.org/abs/1512.02192}{\texttt{arXiv:1512.02192}}.

\bibitem{Aaboud:2016btc}
\hrefCMSnoop {}{{ATLAS Collaboration}, ``{Precision measurement and
  interpretation of inclusive $W^+$ , $W^-$ and $Z/\gamma ^*$ production cross
  sections with the ATLAS detector}'',} \textit{ Eur. Phys. J. C} \textbf{ 77}
  (2017) 367,
  \href{http://dx.doi.org/10.1140/epjc/s10052-017-4911-9}{\doi{10.1140/epjc/s10052-017-4911-9}},
  \href{http://www.arXiv.org/abs/1612.03016}{\texttt{arXiv:1612.03016}}.

\bibitem{Aad:2019wmn}
\hrefCMSnoop {}{{ATLAS Collaboration}, ``{Measurement of the transverse
  momentum distribution of Drell\textendash{}Yan lepton pairs in
  proton\textendash{}proton collisions at $\sqrt{s}=13$ TeV with the ATLAS
  detector}'',} \textit{ Eur. Phys. J. C} \textbf{ 80} (2020) 616,
  \href{http://dx.doi.org/10.1140/epjc/s10052-020-8001-z}{\doi{10.1140/epjc/s10052-020-8001-z}},
  \href{http://www.arXiv.org/abs/1912.02844}{\texttt{arXiv:1912.02844}}.

\bibitem{Aaij:2012vn}
\hrefCMSnoop {}{{LHCb Collaboration}, ``{Inclusive $W$ and $Z$ production in
  the forward region at $\sqrt{s} = 7$ TeV}'',} \textit{ JHEP} \textbf{ 06}
  (2012) 058,
  \href{http://dx.doi.org/10.1007/JHEP06(2012)058}{\doi{10.1007/JHEP06(2012)058}},
  \href{http://www.arXiv.org/abs/1204.1620}{\texttt{arXiv:1204.1620}}.

\bibitem{Aaij:2012mda}
\hrefCMSnoop {}{{LHCb Collaboration}, ``{Measurement of the cross-section for
  $Z \to e^+e^-$ production in $\mathrm{pp}$ collisions at $\sqrt{s}=7$
  TeV}'',} \textit{ JHEP} \textbf{ 02} (2013) 106,
  \href{http://dx.doi.org/10.1007/JHEP02(2013)106}{\doi{10.1007/JHEP02(2013)106}},
  \href{http://www.arXiv.org/abs/1212.4620}{\texttt{arXiv:1212.4620}}.

\bibitem{Aaij:2015gna}
\hrefCMSnoop {}{{LHCb Collaboration}, ``{Measurement of the forward $Z$ boson
  production cross-section in $\mathrm{pp}$ collisions at $\sqrt{s}=7$ TeV}'',}
  \textit{ JHEP} \textbf{ 08} (2015) 039,
  \href{http://dx.doi.org/10.1007/JHEP08(2015)039}{\doi{10.1007/JHEP08(2015)039}},
  \href{http://www.arXiv.org/abs/1505.07024}{\texttt{arXiv:1505.07024}}.

\bibitem{Aaij:2015zlq}
\hrefCMSnoop {}{{LHCb Collaboration}, ``{Measurement of forward W and Z boson
  production in $\mathrm{pp}$ collisions at $ \sqrt{s}=8 $ TeV}'',} \textit{
  JHEP} \textbf{ 01} (2016) 155,
  \href{http://dx.doi.org/10.1007/JHEP01(2016)155}{\doi{10.1007/JHEP01(2016)155}},
  \href{http://www.arXiv.org/abs/1511.08039}{\texttt{arXiv:1511.08039}}.

\bibitem{Aaij:2016mgv}
\hrefCMSnoop {}{{LHCb Collaboration}, ``{Measurement of the forward Z boson
  production cross-section in $\mathrm{pp}$ collisions at $\sqrt{s} = 13$
  TeV}'',} \textit{ JHEP} \textbf{ 09} (2016) 136,
  \href{http://dx.doi.org/10.1007/JHEP09(2016)136}{\doi{10.1007/JHEP09(2016)136}},
  \href{http://www.arXiv.org/abs/1607.06495}{\texttt{arXiv:1607.06495}}.

\bibitem{Affolder:1999jh}
\hrefCMSnoop {}{{CDF} Collaboration, ``{The transverse momentum and total cross
  section of $\rm{ e^+e^-}$ pairs in the Z boson region from $\Pp\bar{\Pp}$
  collisions at $\sqrt{s} = 1.8$ TeV}'',} \textit{ Phys. Rev. Lett.} \textbf{
  84} (2000) 845,
  \href{http://dx.doi.org/10.1103/PhysRevLett.84.845}{\doi{10.1103/PhysRevLett.84.845}},
\href{http://www.arXiv.org/abs/hep-ex/0001021}{\texttt{arXiv:hep-ex/0001021}}.

\bibitem{Abbott:1999yd}
\hrefCMSnoop {}{{D0} Collaboration, ``{Differential production cross section of
  Z bosons as a function of transverse momentum at $\sqrt{s} = 1.8$ TeV}'',}
  \textit{ Phys. Rev. Lett.} \textbf{ 84} (2000) 2792,
  \href{http://dx.doi.org/10.1103/PhysRevLett.84.2792}{\doi{10.1103/PhysRevLett.84.2792}},
\href{http://www.arXiv.org/abs/hep-ex/9909020}{\texttt{arXiv:hep-ex/9909020}}.

\bibitem{TevatronWZ:D0PhysRevLett2008_100}
\hrefCMSnoop {}{{D0} Collaboration, ``{Measurement of the shape of the boson
  transverse momentum distribution in $\Pp \bar{\Pp} \to \PZ / \gamma^{*} \to
  \Pe^+ \Pe^- + X$ events produced at $\sqrt{s}=1.96~\TeV$}'',} \textit{ Phys.
  Rev. Lett.} \textbf{ 100} (2008) 102002,
  \href{http://dx.doi.org/10.1103/PhysRevLett.100.102002}{\doi{10.1103/PhysRevLett.100.102002}},
\href{http://www.arXiv.org/abs/0712.0803}{\texttt{arXiv:0712.0803}}.

\bibitem{TevatronWZ:D0PhysLettB2010_693}
\hrefCMSnoop {}{{D0} Collaboration, ``{Measurement of the normalized
  $\PZ/\gamma^* \to \mu^+\mu^-$ transverse momentum distribution in
  $\Pp\bar{\Pp}$ collisions at $\sqrt{s}=1.96$ TeV}'',} \textit{ Phys. Lett. B}
  \textbf{ 693} (2010) 522,
  \href{http://dx.doi.org/10.1016/j.physletb.2010.09.012}{\doi{10.1016/j.physletb.2010.09.012}},
\href{http://www.arXiv.org/abs/1006.0618}{\texttt{arXiv:1006.0618}}.

\bibitem{TevatronWZ:D0PhysRevLett2011_106}
\hrefCMSnoop {}{{D0} Collaboration, ``{Precise study of the $\PZ/\gamma^*$
  boson transverse momentum distribution in $\Pp\PAp$ collisions using a novel
  technique}'',} \textit{ Phys. Rev. Lett.} \textbf{ 106} (2011) 122001,
  \href{http://dx.doi.org/10.1103/PhysRevLett.106.122001}{\doi{10.1103/PhysRevLett.106.122001}},
\href{http://www.arXiv.org/abs/1010.0262}{\texttt{arXiv:1010.0262}}.

\bibitem{CDF:2012brb}
\hrefCMSnoop {}{{CDF} Collaboration, ``{Transverse momentum cross section of
  $e^+e^-$ pairs in the $Z$-boson region from $p\bar{p}$ collisions at
  $\sqrt{s}=1.96$ TeV}'',} \textit{ Phys. Rev. D} \textbf{ 86} (2012) 052010,
  \href{http://dx.doi.org/10.1103/PhysRevD.86.052010}{\doi{10.1103/PhysRevD.86.052010}},
  \href{http://www.arXiv.org/abs/1207.7138}{\texttt{arXiv:1207.7138}}.

\bibitem{PhysRevD.91.072002}
\hrefCMSnoop {}{{D0} Collaboration, ``{Measurement of the
  ${\ensuremath{\phi}}_{\ensuremath{\eta}}^{*}$ distribution of muon pairs with
  masses between 30 and 500 GeV in $10.4\text{ }\text{
  }{\mathrm{fb}}^{\ensuremath{-}1}$ of $\Pp\bar{\Pp}$ collisions}'',} \textit{
  Phys. Rev. D} \textbf{ 91} (2015) 072002,
  \href{http://dx.doi.org/10.1103/PhysRevD.91.072002}{\doi{10.1103/PhysRevD.91.072002}},
  \href{http://www.arXiv.org/abs/1410.8052}{\texttt{arXiv:1410.8052}}.

\bibitem{Sirunyan:2017ulk}
\hrefCMSnoop {}{{CMS Collaboration}, ``{Particle-flow reconstruction and global
  event description with the CMS detector}'',} \textit{ JINST} \textbf{ 12}
  (2017) P10003,
  \href{http://dx.doi.org/10.1088/1748-0221/12/10/P10003}{\doi{10.1088/1748-0221/12/10/P10003}},
\href{http://www.arXiv.org/abs/1706.04965}{\texttt{arXiv:1706.04965}}.

\bibitem{Sirunyan:2020ycc}
\hrefCMSnoop {}{{CMS Collaboration}, ``{Electron and photon reconstruction and
  identification with the CMS experiment at the CERN LHC}'',} \textit{ JINST}
  \textbf{ 16} (2021) P05014,
  \href{http://dx.doi.org/10.1088/1748-0221/16/05/P05014}{\doi{10.1088/1748-0221/16/05/P05014}},
  \href{http://www.arXiv.org/abs/2012.06888}{\texttt{arXiv:2012.06888}}.

\bibitem{Sirunyan:2018fpa}
\hrefCMSnoop {}{{CMS Collaboration}, ``Performance of the {CMS} muon detector
  and muon reconstruction with proton-proton collisions at
  {$\sqrt{s}=13\TeV$}'',} \textit{ JINST} \textbf{ 13} (2018) P06015,
  \href{http://dx.doi.org/10.1088/1748-0221/13/06/P06015}{\doi{10.1088/1748-0221/13/06/P06015}},
\href{http://www.arXiv.org/abs/1804.04528}{\texttt{arXiv:1804.04528}}.

\bibitem{Bodek:2012id}
A.~Bodek\hrefCMSnoop {}{ {et~al.}, ``{Extracting muon momentum scale
  corrections for hadron collider experiments}'',} \textit{ Eur. Phys. J. C}
  \textbf{ 72} (2012) 2194,
  \href{http://dx.doi.org/10.1140/epjc/s10052-012-2194-8}{\doi{10.1140/epjc/s10052-012-2194-8}},
  \href{http://www.arXiv.org/abs/1208.3710}{\texttt{arXiv:1208.3710}}.

\bibitem{Cacciari:2008gp}
\hrefCMSnoop {}{M.~Cacciari, G.~P. Salam, and G.~Soyez, ``{The anti-$\kt$ jet
  clustering algorithm}'',} \textit{ JHEP} \textbf{ 04} (2008) 063,
  \href{http://dx.doi.org/10.1088/1126-6708/2008/04/063}{\doi{10.1088/1126-6708/2008/04/063}},
  \href{http://www.arXiv.org/abs/0802.1189}{\texttt{arXiv:0802.1189}}.

\bibitem{Cacciari:2011ma}
\hrefCMSnoop {}{M.~Cacciari, G.~P. Salam, and G.~Soyez, ``{FastJet user
  manual}'',} \textit{ Eur. Phys. J. C} \textbf{ 72} (2012) 1896,
  \href{http://dx.doi.org/10.1140/epjc/s10052-012-1896-2}{\doi{10.1140/epjc/s10052-012-1896-2}},
\href{http://www.arXiv.org/abs/1111.6097}{\texttt{arXiv:1111.6097}}.

\bibitem{CMS-TDR-15-02}
\href {http://cds.cern.ch/record/2020886}{{CMS Collaboration}, ``Technical
  proposal for the {Phase-II} upgrade of the {Compact Muon Solenoid}'',} CMS
  Technical Proposal CERN-LHCC-2015-010, CMS-TDR-15-02, 2015.

\bibitem{Sirunyan:2020zal}
\hrefCMSnoop {}{{CMS Collaboration}, ``{Performance of the CMS Level-1 trigger
  in proton-proton collisions at $\sqrt{s} = 13$\,TeV}'',} \textit{ JINST}
  \textbf{ 15} (2020) P10017,
  \href{http://dx.doi.org/10.1088/1748-0221/15/10/P10017}{\doi{10.1088/1748-0221/15/10/P10017}},
  \href{http://www.arXiv.org/abs/2006.10165}{\texttt{arXiv:2006.10165}}.

\bibitem{Khachatryan:2016bia}
\hrefCMSnoop {}{{CMS Collaboration}, ``{The CMS trigger system}'',} \textit{
  JINST} \textbf{ 12} (2017) P01020,
  \href{http://dx.doi.org/10.1088/1748-0221/12/01/P01020}{\doi{10.1088/1748-0221/12/01/P01020}},
\href{http://www.arXiv.org/abs/1609.02366}{\texttt{arXiv:1609.02366}}.

\bibitem{Chatrchyan:2008zzk}
\hrefCMSnoop {}{{CMS Collaboration}, ``The {CMS} experiment at the {CERN}
  {LHC}'',} \textit{ JINST} \textbf{ 3} (2008) S08004,
  \href{http://dx.doi.org/10.1088/1748-0221/3/08/S08004}{\doi{10.1088/1748-0221/3/08/S08004}}.

\bibitem{CMS-PAPERS-JME-18-001}
\hrefCMSnoop {}{{CMS Collaboration}, ``{Pileup mitigation at {CMS} in 13 {TeV}
  data}'',} \textit{ JINST} \textbf{ 15} (2020) P09018,
  \href{http://dx.doi.org/10.1088/1748-0221/15/09/P09018}{\doi{10.1088/1748-0221/15/09/P09018}},
  \href{http://www.arXiv.org/abs/2003.00503}{\texttt{arXiv:2003.00503}}.

\bibitem{Sirunyan:2017ezt}
\hrefCMSnoop {}{{CMS Collaboration}, ``{Identification of heavy-flavour jets
  with the CMS detector in pp collisions at 13 TeV}'',} \textit{ JINST}
  \textbf{ 13} (2018) P05011,
  \href{http://dx.doi.org/10.1088/1748-0221/13/05/P05011}{\doi{10.1088/1748-0221/13/05/P05011}},
\href{http://www.arXiv.org/abs/1712.07158}{\texttt{arXiv:1712.07158}}.

\bibitem{Alwall:2014hca}
J.~Alwall\hrefCMSnoop {}{ {et~al.}, ``{The automated computation of tree-level
  and next-to-leading order differential cross sections, and their matching to
  parton shower simulations}'',} \textit{ JHEP} \textbf{ 07} (2014) 079,
  \href{http://dx.doi.org/10.1007/JHEP07(2014)079}{\doi{10.1007/JHEP07(2014)079}},
\href{http://www.arXiv.org/abs/1405.0301}{\texttt{arXiv:1405.0301}}.

\bibitem{Frederix:2012ps}
\hrefCMSnoop {}{R.~Frederix and S.~Frixione, ``Merging meets matching in
  {MC@NLO}'',} \textit{ JHEP} \textbf{ 12} (2012) 061,
  \href{http://dx.doi.org/10.1007/JHEP12(2012)061}{\doi{10.1007/JHEP12(2012)061}},
\href{http://www.arXiv.org/abs/1209.6215}{\texttt{arXiv:1209.6215}}.

\bibitem{Sjostrand:2014zea}
T.~Sj{\"o}strand\hrefCMSnoop {}{ {et~al.}, ``An introduction to {PYTHIA
  8.2}'',} \textit{ Comput. Phys. Commun.} \textbf{ 191} (2015) 159,
  \href{http://dx.doi.org/10.1016/j.cpc.2015.01.024}{\doi{10.1016/j.cpc.2015.01.024}},
\href{http://www.arXiv.org/abs/1410.3012}{\texttt{arXiv:1410.3012}}.

\bibitem{Khachatryan:2015pea}
\hrefCMSnoop {}{{CMS Collaboration}, ``{Event generator tunes obtained from
  underlying event and multiparton scattering measurements}'',} \textit{ Eur.
  Phys. J. C} \textbf{ 76} (2016) 155,
  \href{http://dx.doi.org/10.1140/epjc/s10052-016-3988-x}{\doi{10.1140/epjc/s10052-016-3988-x}},
\href{http://www.arXiv.org/abs/1512.00815}{\texttt{arXiv:1512.00815}}.

\bibitem{Ball:2014uwa}
\hrefCMSnoop {}{{NNPDF} Collaboration, ``{Parton distributions for the LHC Run
  II}'',} \textit{ JHEP} \textbf{ 04} (2015) 040,
  \href{http://dx.doi.org/10.1007/JHEP04(2015)040}{\doi{10.1007/JHEP04(2015)040}},
\href{http://www.arXiv.org/abs/1410.8849}{\texttt{arXiv:1410.8849}}.

\bibitem{Nason:2004rx}
\hrefCMSnoop {}{P.~Nason, ``{A new method for combining NLO QCD with shower
  Monte Carlo algorithms}'',} \textit{ JHEP} \textbf{ 11} (2004) 040,
  \href{http://dx.doi.org/10.1088/1126-6708/2004/11/040}{\doi{10.1088/1126-6708/2004/11/040}},
\href{http://www.arXiv.org/abs/hep-ph/0409146}{\texttt{arXiv:hep-ph/0409146}}.

\bibitem{Frixione:2007vw}
\hrefCMSnoop {}{S.~Frixione, P.~Nason, and C.~Oleari, ``Matching {NLO} {QCD}
  computations with parton shower simulations: the {POWHEG} method'',} \textit{
  JHEP} \textbf{ 11} (2007) 070,
  \href{http://dx.doi.org/10.1088/1126-6708/2007/11/070}{\doi{10.1088/1126-6708/2007/11/070}},
\href{http://www.arXiv.org/abs/0709.2092}{\texttt{arXiv:0709.2092}}.

\bibitem{Alioli:2010xd}
\hrefCMSnoop {}{S.~Alioli, P.~Nason, C.~Oleari, and E.~Re, ``{A general
  framework for implementing NLO calculations in shower Monte Carlo programs:
  the POWHEG BOX}'',} \textit{ JHEP} \textbf{ 06} (2010) 043,
  \href{http://dx.doi.org/10.1007/JHEP06(2010)043}{\doi{10.1007/JHEP06(2010)043}},
\href{http://www.arXiv.org/abs/1002.2581}{\texttt{arXiv:1002.2581}}.

\bibitem{Frixione:2007nw}
\hrefCMSnoop {}{S.~Frixione, P.~Nason, and G.~Ridolfi, ``A positive-weight
  next-to-leading-order {M}onte {C}arlo for heavy flavour hadroproduction'',}
  \textit{ JHEP} \textbf{ 09} (2007) 126,
  \href{http://dx.doi.org/10.1088/1126-6708/2007/09/126}{\doi{10.1088/1126-6708/2007/09/126}},
\href{http://www.arXiv.org/abs/0707.3088}{\texttt{arXiv:0707.3088}}.

\bibitem{Nason:2013ydw}
\hrefCMSnoop {}{P.~Nason and G.~Zanderighi, ``{$\PW^+ \PW^-$}, {$\PW \cPZ$} and
  {$\cPZ \cPZ$} production in the {\sc powheg-box}-v2'',} \textit{ Eur. Phys.
  J. C} \textbf{ 74} (2014) 2702,
  \href{http://dx.doi.org/10.1140/epjc/s10052-013-2702-5}{\doi{10.1140/epjc/s10052-013-2702-5}},
\href{http://www.arXiv.org/abs/1311.1365}{\texttt{arXiv:1311.1365}}.

\bibitem{Vermaseren:1982cz}
\hrefCMSnoop {}{J.~A.~M. Vermaseren, ``Two photon processes at very
  high-energies'',} \textit{ Nucl. Phys. B} \textbf{ 229} (1983) 347,
  \href{http://dx.doi.org/10.1016/0550-3213(83)90336-X}{\doi{10.1016/0550-3213(83)90336-X}}.

\bibitem{Baranov:1991yq}
\hrefCMSnoop {}{S.~P. Baranov, O.~Duenger, H.~Shooshtari, and J.~A.~M.
  Vermaseren, ``{LPAIR: A generator for lepton pair production}'',} in \textit{
  {Workshop on Physics at HERA}}, p.~1478.
\newblock 1991.

\bibitem{Suri:1971yx}
\hrefCMSnoop {}{A.~Suri and D.~R. Yennie, ``{The space-time phenomenology of
  photon absorption and inelastic electron scattering}'',} \textit{ Annals
  Phys.} \textbf{ 72} (1972) 243,
  \href{http://dx.doi.org/10.1016/0003-4916(72)90242-4}{\doi{10.1016/0003-4916(72)90242-4}}.

\bibitem{Campbell:2015qma}
\hrefCMSnoop {}{J.~M. Campbell, R.~K. Ellis, and W.~T. Giele, ``{A
  multi-threaded version of MCFM}'',} \textit{ Eur. Phys. J. C} \textbf{ 75}
  (2015) 246,
  \href{http://dx.doi.org/10.1140/epjc/s10052-015-3461-2}{\doi{10.1140/epjc/s10052-015-3461-2}},
\href{http://www.arXiv.org/abs/1503.06182}{\texttt{arXiv:1503.06182}}.

\bibitem{Gehrmann:2014fva}
T.~Gehrmann\hrefCMSnoop {}{ {et~al.}, ``{$W^+W^-$ production at hadron
  colliders in next to next to leading order QCD}'',} \textit{ Phys. Rev.
  Lett.} \textbf{ 113} (2014) 212001,
  \href{http://dx.doi.org/10.1103/PhysRevLett.113.212001}{\doi{10.1103/PhysRevLett.113.212001}},
\href{http://www.arXiv.org/abs/1408.5243}{\texttt{arXiv:1408.5243}}.

\bibitem{Czakon:2011xx}
\hrefCMSnoop {}{M.~Czakon and A.~Mitov, ``{Top++: A program for the calculation
  of the top-pair cross-section at hadron colliders}'',} \textit{ Comput. Phys.
  Commun.} \textbf{ 185} (2014) 2930,
  \href{http://dx.doi.org/10.1016/j.cpc.2014.06.021}{\doi{10.1016/j.cpc.2014.06.021}},
\href{http://www.arXiv.org/abs/1112.5675}{\texttt{arXiv:1112.5675}}.

\bibitem{GEANT4:2002zbu}
\hrefCMSnoop {}{{\GEANTfour} Collaboration, ``{$\GEANTfour$}---a simulation
  toolkit'',} \textit{ Nucl. Instrum. Meth. A} \textbf{ 506} (2003) 250,
  \href{http://dx.doi.org/10.1016/S0168-9002(03)01368-8}{\doi{10.1016/S0168-9002(03)01368-8}}.

\bibitem{D'Agostini:1994zf}
\hrefCMSnoop {}{G.~D'Agostini, ``{A multidimensional unfolding method based on
  Bayes' theorem}'',} \textit{ Nucl. Instrum. Meth. A} \textbf{ 362} (1995)
  487,
\href{http://dx.doi.org/10.1016/0168-9002(95)00274-X}{\doi{10.1016/0168-9002(95)00274-X}}.

\bibitem{Sirunyan:2018cpw}
\hrefCMSnoop {}{{CMS Collaboration}, ``{Measurement of differential cross
  sections for Z boson production in association with jets in proton-proton
  collisions at $\sqrt{s} =$ 13 TeV}'',} \textit{ Eur. Phys. J. C} \textbf{ 78}
  (2018) 965,
  \href{http://dx.doi.org/10.1140/epjc/s10052-018-6373-0}{\doi{10.1140/epjc/s10052-018-6373-0}},
  \href{http://www.arXiv.org/abs/1804.05252}{\texttt{arXiv:1804.05252}}.

\bibitem{Khachatryan:2016crw}
\hrefCMSnoop {}{{CMS Collaboration}, ``{Measurements of differential production
  cross sections for a Z boson in association with jets in pp collisions at $
  \sqrt{s}=8 $ TeV}'',} \textit{ JHEP} \textbf{ 04} (2017) 022,
  \href{http://dx.doi.org/10.1007/JHEP04(2017)022}{\doi{10.1007/JHEP04(2017)022}},
  \href{http://www.arXiv.org/abs/1611.03844}{\texttt{arXiv:1611.03844}}.

\bibitem{CMS:2021xjt}
\hrefCMSnoop {}{{CMS Collaboration}, ``{Precision luminosity measurement in
  proton-proton collisions at $\sqrt{s} =$ 13 TeV in 2015 and 2016 at CMS}'',}
  \textit{ Eur. Phys. J. C} \textbf{ 81} (2021) 800,
  \href{http://dx.doi.org/10.1140/epjc/s10052-021-09538-2}{\doi{10.1140/epjc/s10052-021-09538-2}},
  \href{http://www.arXiv.org/abs/2104.01927}{\texttt{arXiv:2104.01927}}.

\bibitem{Manohar:2016nzj}
\hrefCMSnoop {}{A.~Manohar, P.~Nason, G.~P. Salam, and G.~Zanderighi, ``{How
  bright is the proton? A precise determination of the photon parton
  distribution function}'',} \textit{ Phys. Rev. Lett.} \textbf{ 117} (2016)
  242002,
  \href{http://dx.doi.org/10.1103/PhysRevLett.117.242002}{\doi{10.1103/PhysRevLett.117.242002}},
  \href{http://www.arXiv.org/abs/1607.04266}{\texttt{arXiv:1607.04266}}.

\bibitem{Manohar:2017eqh}
\hrefCMSnoop {}{A.~V. Manohar, P.~Nason, G.~P. Salam, and G.~Zanderighi, ``{The
  photon content of the proton}'',} \textit{ JHEP} \textbf{ 12} (2017) 046,
  \href{http://dx.doi.org/10.1007/JHEP12(2017)046}{\doi{10.1007/JHEP12(2017)046}},
  \href{http://www.arXiv.org/abs/1708.01256}{\texttt{arXiv:1708.01256}}.

\bibitem{Monni:2019whf}
P.~F. Monni\hrefCMSnoop {}{ {et~al.}, ``{MiNNLO$_{\text{PS}}$: a new method to
  match NNLO QCD to parton showers}'',} \textit{ JHEP} \textbf{ 05} (2020) 143,
  \href{http://dx.doi.org/10.1007/JHEP05(2020)143}{\doi{10.1007/JHEP05(2020)143}},
  \href{http://www.arXiv.org/abs/1908.06987}{\texttt{arXiv:1908.06987}}.

\bibitem{Monni:2020nks}
\hrefCMSnoop {}{P.~F. Monni, E.~Re, and M.~Wiesemann, ``{MiNNLO$_{\text {PS}}$:
  optimizing $2\rightarrow 1$ hadronic processes}'',} \textit{ Eur. Phys. J. C}
  \textbf{ 80} (2020) 1075,
  \href{http://dx.doi.org/10.1140/epjc/s10052-020-08658-5}{\doi{10.1140/epjc/s10052-020-08658-5}},
  \href{http://www.arXiv.org/abs/2006.04133}{\texttt{arXiv:2006.04133}}.

\bibitem{NNPDF:2017mvq}
\hrefCMSnoop {}{{NNPDF} Collaboration, ``{Parton distributions from
  high-precision collider data}'',} \textit{ Eur. Phys. J. C} \textbf{ 77}
  (2017) 663,
  \href{http://dx.doi.org/10.1140/epjc/s10052-017-5199-5}{\doi{10.1140/epjc/s10052-017-5199-5}},
  \href{http://www.arXiv.org/abs/1706.00428}{\texttt{arXiv:1706.00428}}.

\bibitem{CMS:2019csb}
\hrefCMSnoop {}{{CMS Collaboration}, ``{Extraction and validation of a new set
  of CMS PYTHIA8 tunes from underlying-event measurements}'',} \textit{ Eur.
  Phys. J. C} \textbf{ 80} (2020) 4,
  \href{http://dx.doi.org/10.1140/epjc/s10052-019-7499-4}{\doi{10.1140/epjc/s10052-019-7499-4}},
  \href{http://www.arXiv.org/abs/1903.12179}{\texttt{arXiv:1903.12179}}.

\bibitem{baranov2021cascade3}
S.~Baranov\hrefCMSnoop {}{ {et~al.}, ``{CASCADE3 A Monte Carlo event generator
  based on TMDs}'',} \textit{ Eur. Phys. J. C} \textbf{ 81} (2021) 425,
  \href{http://dx.doi.org/10.1140/epjc/s10052-021-09203-8}{\doi{10.1140/epjc/s10052-021-09203-8}},
  \href{http://www.arXiv.org/abs/2101.10221}{\texttt{arXiv:2101.10221}}.

\bibitem{Martinez:2019mwt}
A.~Bermudez~Martinez\hrefCMSnoop {}{ {et~al.}, ``{Production of Z-bosons in the
  parton branching method}'',} \textit{ Phys. Rev. D} \textbf{ 100} (2019)
  074027,
  \href{http://dx.doi.org/10.1103/PhysRevD.100.074027}{\doi{10.1103/PhysRevD.100.074027}},
  \href{http://www.arXiv.org/abs/1906.00919}{\texttt{arXiv:1906.00919}}.

\bibitem{PhysRevD.99.074008}
A.~Bermudez~Martinez\hrefCMSnoop {}{ {et~al.}, ``{Collinear and TMD parton
  densities from fits to precision DIS measurements in the parton branching
  method}'',} \textit{ Phys. Rev. D} \textbf{ 99} (2019) 074008,
  \href{http://dx.doi.org/10.1103/PhysRevD.99.074008}{\doi{10.1103/PhysRevD.99.074008}},
  \href{http://www.arXiv.org/abs/1804.11152}{\texttt{arXiv:1804.11152}}.

\bibitem{Sjostrand:2006za}
\hrefCMSnoop {}{T.~Sj{\"o}strand, S.~Mrenna, and P.~Z. Skands, ``{PYTHIA 6.4
  physics and manual}'',} \textit{ JHEP} \textbf{ 05} (2006) 026,
  \href{http://dx.doi.org/10.1088/1126-6708/2006/05/026}{\doi{10.1088/1126-6708/2006/05/026}},
  \href{http://www.arXiv.org/abs/hep-ph/0603175}{\texttt{arXiv:hep-ph/0603175}}.

\bibitem{Martinez:2021chk}
\hrefCMSnoop {}{A.~Bermudez~Martinez, F.~Hautmann, and M.~L. Mangano, ``{TMD
  evolution and multi-jet merging}'',} \textit{ Phys. Lett. B} \textbf{ 822}
  (2021) 136700,
  \href{http://dx.doi.org/10.1016/j.physletb.2021.136700}{\doi{10.1016/j.physletb.2021.136700}},
  \href{http://www.arXiv.org/abs/2107.01224}{\texttt{arXiv:2107.01224}}.

\bibitem{Scimemi:2017etj}
\hrefCMSnoop {}{I.~Scimemi and A.~Vladimirov, ``{Analysis of vector boson
  production within TMD factorization}'',} \textit{ Eur. Phys. J. C} \textbf{
  78} (2018) 89,
  \href{http://dx.doi.org/10.1140/epjc/s10052-018-5557-y}{\doi{10.1140/epjc/s10052-018-5557-y}},
  \href{http://www.arXiv.org/abs/1706.01473}{\texttt{arXiv:1706.01473}}.

\bibitem{Scimemi:2019cmh}
\hrefCMSnoop {}{I.~Scimemi and A.~Vladimirov, ``{Non-perturbative structure of
  semi-inclusive deep-inelastic and Drell-Yan scattering at small transverse
  momentum}'',} \textit{ JHEP} \textbf{ 06} (2020) 137,
  \href{http://dx.doi.org/10.1007/JHEP06(2020)137}{\doi{10.1007/JHEP06(2020)137}},
  \href{http://www.arXiv.org/abs/1912.06532}{\texttt{arXiv:1912.06532}}.

\bibitem{arTeMiDe-web}
\hrefCMSnoop {}{``arTeMiDe public repository''.}
  \url{https://github.com/VladimirovAlexey/artemide-public}, {2020}.

\bibitem{Alioli_2013}
S.~Alioli\hrefCMSnoop {}{ {et~al.}, ``Combining higher-order resummation with
  multiple {NLO} calculations and parton showers in {GENEVA}'',} \textit{ JHEP}
  \textbf{ 09} (2013) 120,
  \href{http://dx.doi.org/10.1007/jhep09(2013)120}{\doi{10.1007/jhep09(2013)120}},
  \href{http://www.arXiv.org/abs/1211.7049}{\texttt{arXiv:1211.7049}}.

\bibitem{Alioli:2015toa}
S.~Alioli\hrefCMSnoop {}{ {et~al.}, ``{Drell-Yan production at NNLL'+NNLO
  matched to parton showers}'',} \textit{ Phys. Rev. D} \textbf{ 92} (2015)
  094020,
  \href{http://dx.doi.org/10.1103/PhysRevD.92.094020}{\doi{10.1103/PhysRevD.92.094020}},
  \href{http://www.arXiv.org/abs/1508.01475}{\texttt{arXiv:1508.01475}}.

\bibitem{PhysRevD.104.094020}
S.~Alioli\hrefCMSnoop {}{ {et~al.}, ``{Matching NNLO predictions to parton
  showers using ${\mathrm{N}}^{3}\mathrm{LL}$ color-singlet transverse momentum
  resummation in GENEVA}'',} \textit{ Phys. Rev. D} \textbf{ 104} (2021)
  094020,
  \href{http://dx.doi.org/10.1103/PhysRevD.104.094020}{\doi{10.1103/PhysRevD.104.094020}},
  \href{http://www.arXiv.org/abs/2102.08390}{\texttt{arXiv:2102.08390}}.

\bibitem{PhysRevLett.105.092002}
\hrefCMSnoop {}{I.~W. Stewart, F.~J. Tackmann, and W.~J. Waalewijn,
  ``N-jettiness: An inclusive event shape to veto jets'',} \textit{ Phys. Rev.
  Lett.} \textbf{ 105} (2010) 092002,
  \href{http://dx.doi.org/10.1103/PhysRevLett.105.092002}{\doi{10.1103/PhysRevLett.105.092002}},
  \href{http://www.arXiv.org/abs/1004.2489}{\texttt{arXiv:1004.2489}}.

\bibitem{Monni:2016ktx}
\hrefCMSnoop {}{P.~F. Monni, E.~Re, and P.~Torrielli, ``{Higgs}
  transverse-momentum resummation in direct space'',} \textit{ Phys. Rev.
  Lett.} \textbf{ 116} (2016) 242001,
  \href{http://dx.doi.org/10.1103/PhysRevLett.116.242001}{\doi{10.1103/PhysRevLett.116.242001}},
  \href{http://www.arXiv.org/abs/1604.02191}{\texttt{arXiv:1604.02191}}.

\bibitem{Bizon:2017rah}
W.~Bizon\hrefCMSnoop {}{ {et~al.}, ``{Momentum-space resummation for transverse
  observables and the Higgs p$_{\perp}$ at N$^{3}$LL+NNLO}'',} \textit{ JHEP}
  \textbf{ 02} (2018) 108,
  \href{http://dx.doi.org/10.1007/JHEP02(2018)108}{\doi{10.1007/JHEP02(2018)108}},
  \href{http://www.arXiv.org/abs/1705.09127}{\texttt{arXiv:1705.09127}}.

\bibitem{Butterworth:2015oua}
\hrefCMSnoop {}{J.~Butterworth {et~al.}, ``{PDF4LHC} recommendations for {LHC
  Run II}'',} \textit{ J. Phys. G} \textbf{ 43} (2016) 023001,
  \href{http://dx.doi.org/10.1088/0954-3899/43/2/023001}{\doi{10.1088/0954-3899/43/2/023001}},
\href{http://www.arXiv.org/abs/1510.03865}{\texttt{arXiv:1510.03865}}.

\bibitem{hepdata}
\hrefCMSnoop {}{``{HEPD}ata record for this analysis'',} 2022.
\newblock
  \href{http://dx.doi.org/10.17182/hepdata.115656}{\doi{10.17182/hepdata.115656}}.

\end{thebibliography}\endgroup
\appendix
\numberwithin{figure}{section}
\section{Comparisons to other models\label{app:suppMat}}
\begin{figure*}[htbp!]
  \centering
  \includegraphics[width=0.48\textwidth]{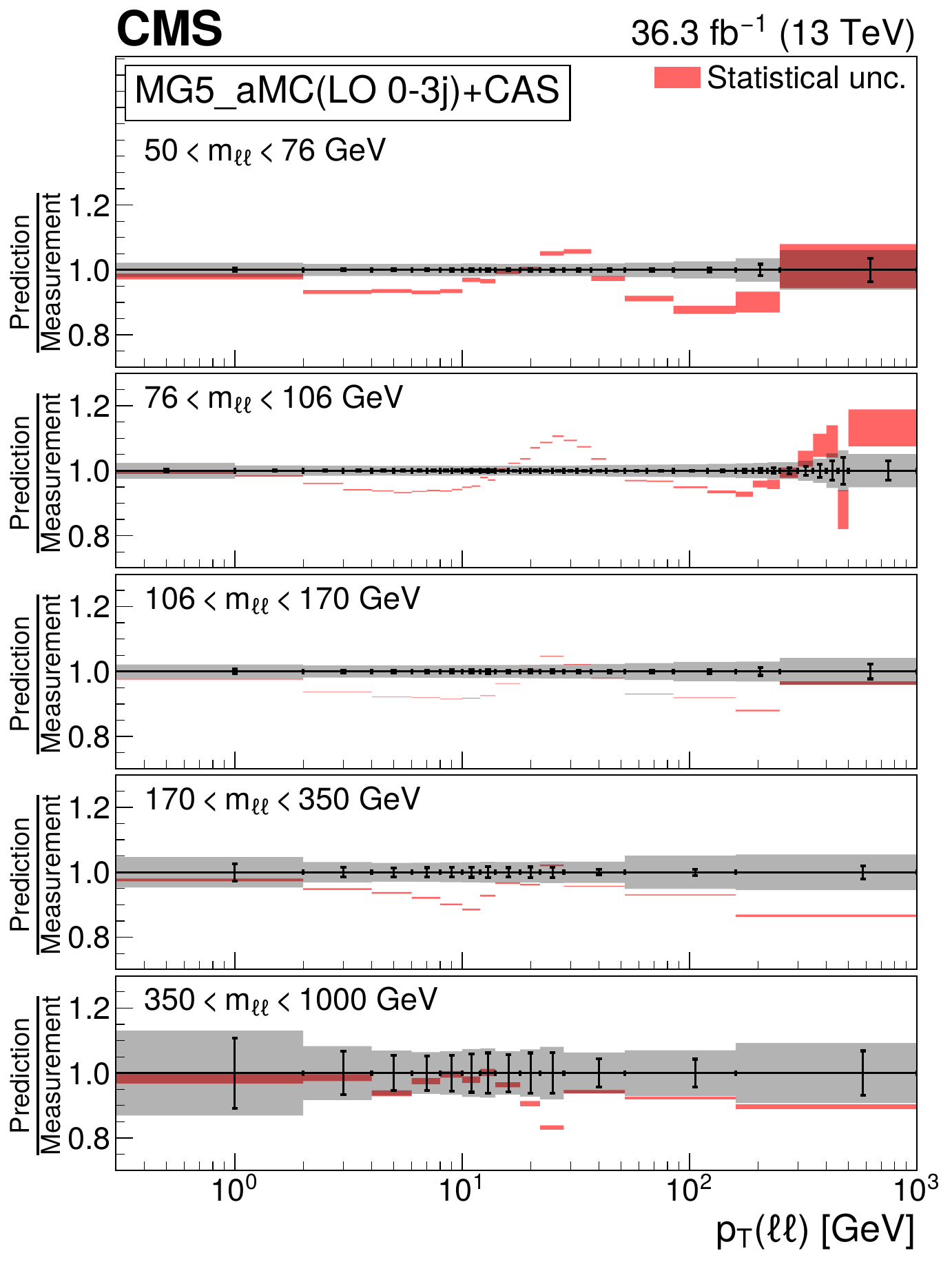}
  \includegraphics[width=0.48\textwidth]{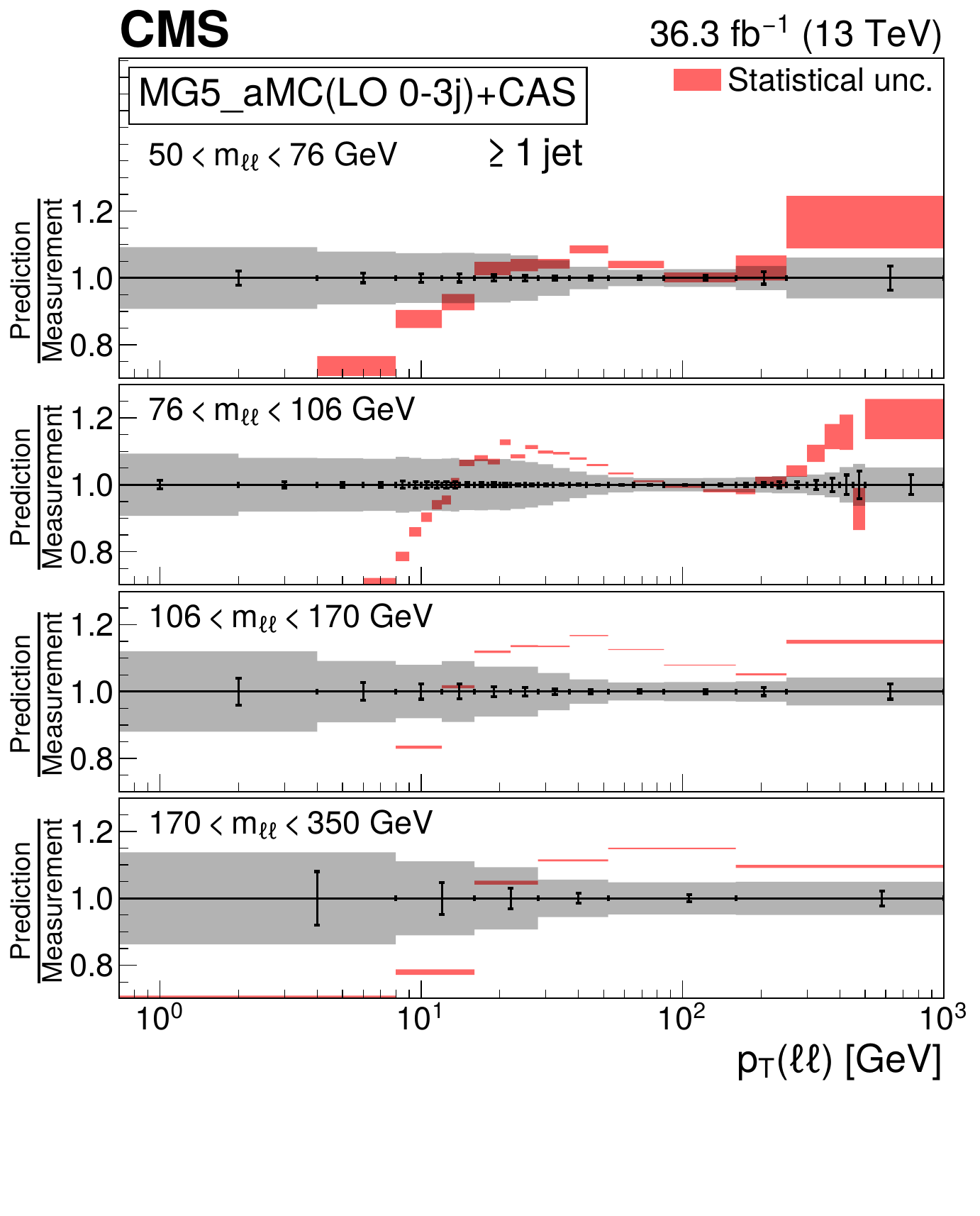}
  \includegraphics[width=0.48\textwidth]{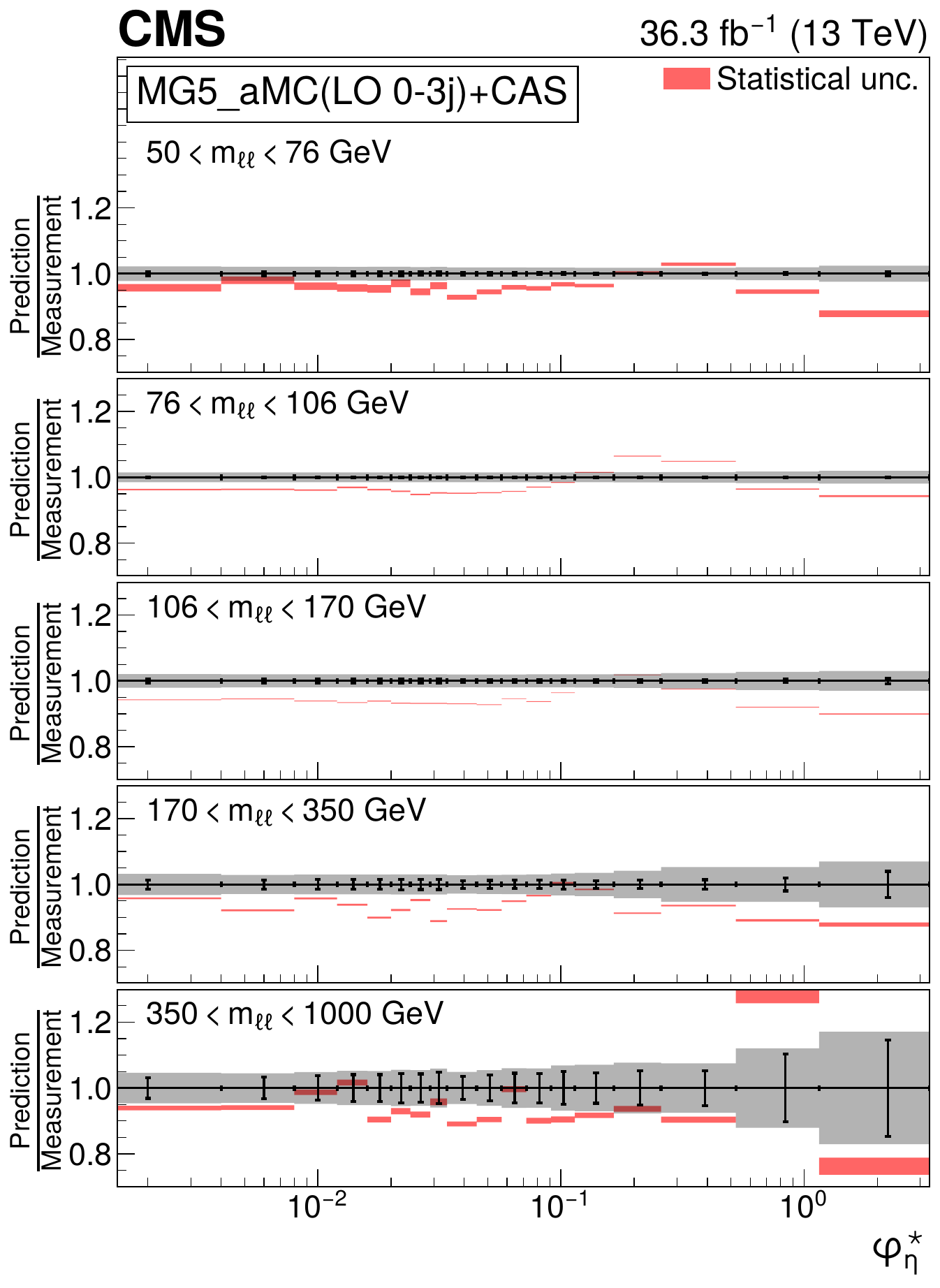}
  \caption{The ratio of \MGaMC (0, 1, 2 and 3 jets at LO)  + PB (\cascade) predictions to the measured differential cross sections
           in \PTll (upper left), in \PTll for the one or more jets case (upper right), and in \phistar (bottom)  are presented for various \Mll ranges.
           The error bars correspond to the statistical uncertainty of the measurement and the shaded bands to the total experimental uncertainty.
           The light color band around \cascade prediction corresponds to the statistical uncertainty of the simulation.
          }
 \label{fig:AppUnfCombPt0a}
\end{figure*}

\begin{figure*}[htbp!]
  \centering
  \includegraphics[width=0.48\textwidth]{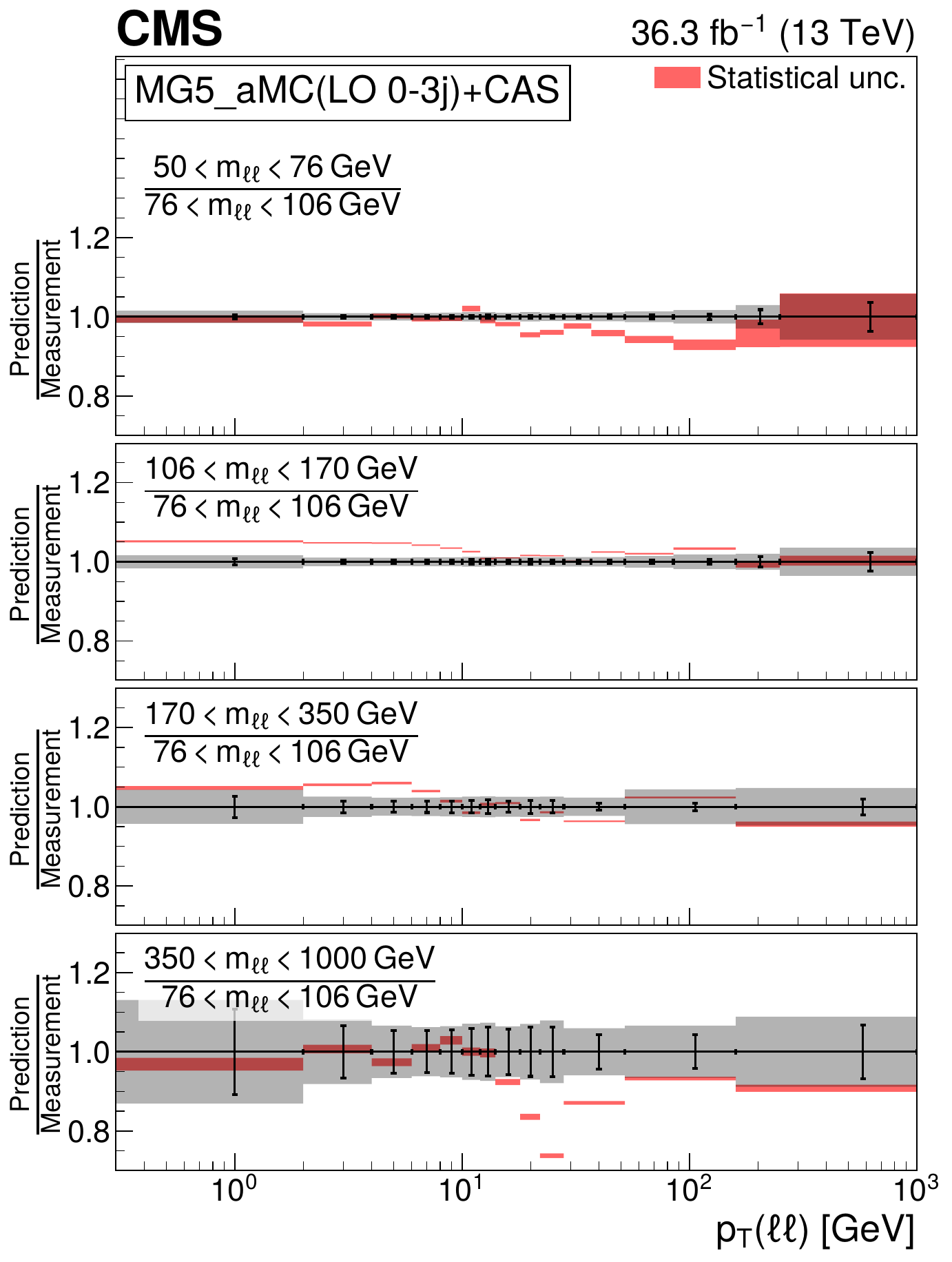}
  \includegraphics[width=0.48\textwidth]{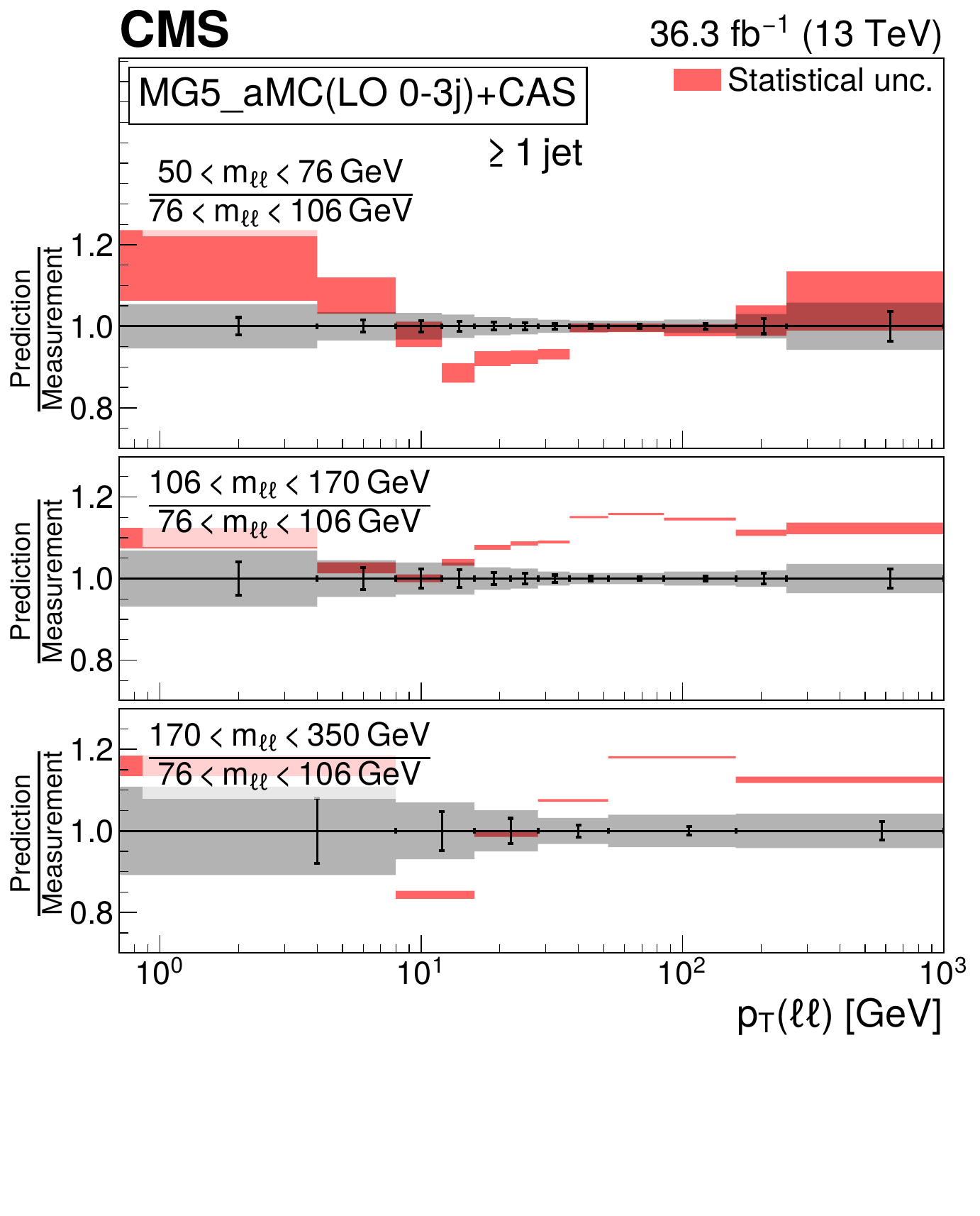}
  \includegraphics[width=0.48\textwidth]{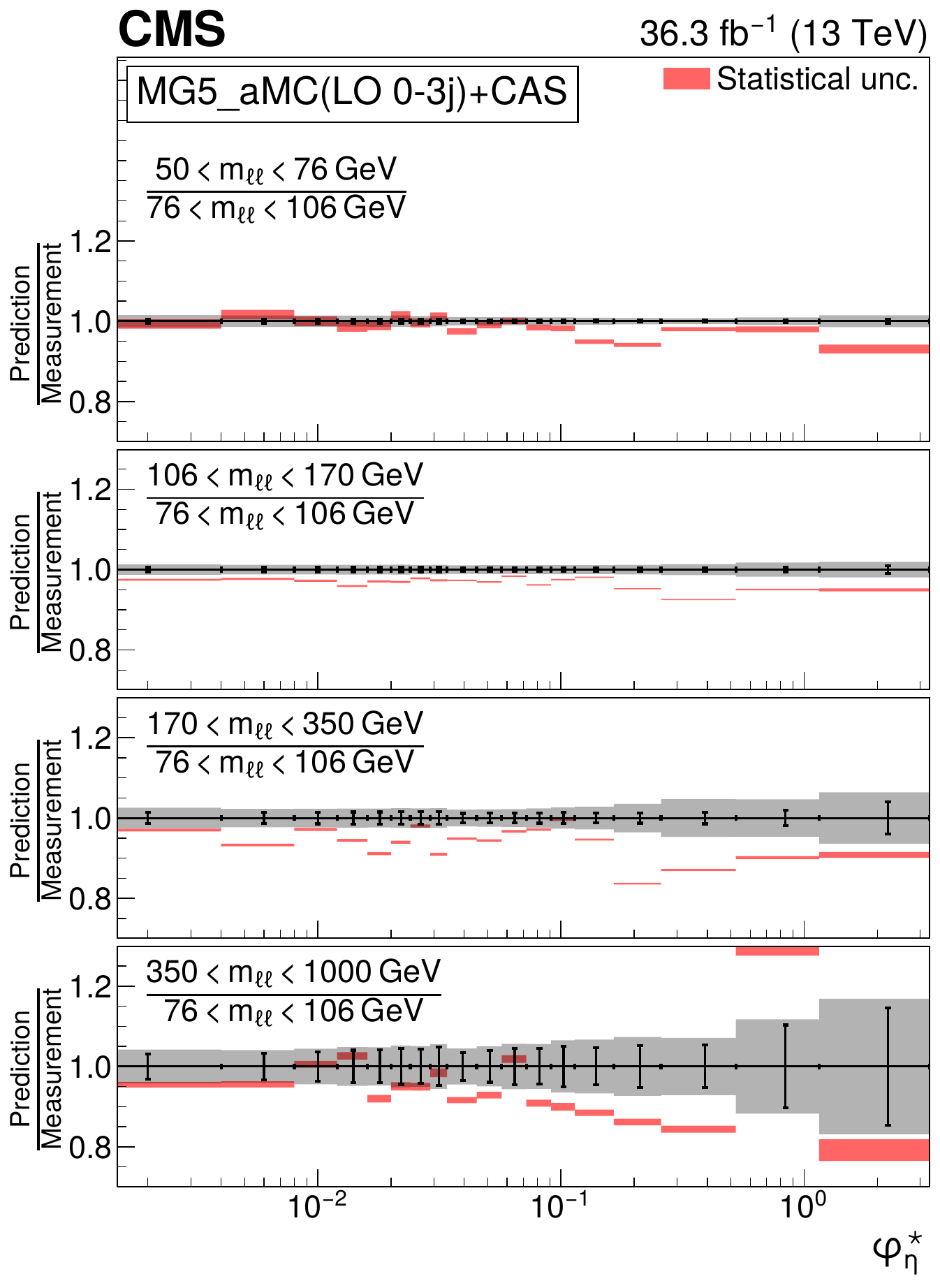}
  \caption{The ratio of \MGaMC (0, 1, 2 and 3 jets at LO)  + PB (\cascade) predictions to the ratios of differential cross sections for \Mll ranges with respect to the peak region 
           in \PTll (upper left), in \PTll for the one or more jets case (upper right), and in \phistar (bottom).
           The error bars correspond to the statistical uncertainty of the measurement and the shaded bands to the total experimental uncertainty.
           The light color band around \cascade prediction corresponds to the statistical uncertainty of the simulation.
          }
 \label{fig:AppUnfCombPt0Ratiosa}
\end{figure*}

In this section, comparisons of the obtained measurement results with predictions from a more recent parton branching (PB) TMD
method from \cascade are presented. The predictions are based on \MGaMC ME up to three partons at LO in QCD with multi-jet merging~\cite{Martinez:2021chk}. 
The ratio of the predictions over the data are presented in Fig.~\ref{fig:AppUnfCombPt0a}. 
The comparisons to predictions for the ratio of the cross sections for invariant masses outside the \PZ boson peak to the distribution within the \PZ boson
peak (\mrangeb) are shown in Fig.~\ref{fig:AppUnfCombPt0Ratiosa}.

\cleardoublepage \section{The CMS Collaboration \label{app:collab}}\begin{sloppypar}\hyphenpenalty=5000\widowpenalty=500\clubpenalty=5000
\cmsinstitute{Yerevan Physics Institute, Yerevan, Armenia}
{\tolerance=6000
A.~Tumasyan\cmsorcid{0009-0000-0684-6742}
\par}
\cmsinstitute{Institut f\"{u}r Hochenergiephysik, Vienna, Austria}
{\tolerance=6000
W.~Adam\cmsorcid{0000-0001-9099-4341}, J.W.~Andrejkovic, T.~Bergauer\cmsorcid{0000-0002-5786-0293}, S.~Chatterjee\cmsorcid{0000-0003-2660-0349}, M.~Dragicevic\cmsorcid{0000-0003-1967-6783}, A.~Escalante~Del~Valle\cmsorcid{0000-0002-9702-6359}, R.~Fr\"{u}hwirth\cmsAuthorMark{1}\cmsorcid{0000-0002-0054-3369}, M.~Jeitler\cmsAuthorMark{1}\cmsorcid{0000-0002-5141-9560}, N.~Krammer\cmsorcid{0000-0002-0548-0985}, L.~Lechner\cmsorcid{0000-0002-3065-1141}, D.~Liko\cmsorcid{0000-0002-3380-473X}, I.~Mikulec\cmsorcid{0000-0003-0385-2746}, P.~Paulitsch, F.M.~Pitters, J.~Schieck\cmsAuthorMark{1}\cmsorcid{0000-0002-1058-8093}, R.~Sch\"{o}fbeck\cmsorcid{0000-0002-2332-8784}, M.~Spanring\cmsorcid{0000-0001-6328-7887}, S.~Templ\cmsorcid{0000-0003-3137-5692}, W.~Waltenberger\cmsorcid{0000-0002-6215-7228}, C.-E.~Wulz\cmsAuthorMark{1}\cmsorcid{0000-0001-9226-5812}
% ...
\par}
\vskip\cmsinstskip
\dag:~Deceased\\
$^{1}$Also at TU Wien, Vienna, Austria\\
% ...
\end{sloppypar}
\end{document}